\documentclass[10pt,twocolumn,letterpaper]{article}
\usepackage[accsupp]{axessibility}
\usepackage{multirow}
\usepackage[pagenumbers]{cvpr} %

\newcommand{\oneline}[1]{%
  \newdimen{\namewidth}%
  \setlength{\namewidth}{\widthof{#1}}%
  \ifthenelse{\lengthtest{\namewidth < \textwidth}}%
  {#1}%
  {\resizebox{\textwidth}{!}{#1}}%
}

\newcommand{\nerds}{BrainNRDS\ } 
\newcommand{\nnerds}{BrainNRDS}

\definecolor{cvprblue}{rgb}{0.21,0.49,0.74}
\usepackage[pagebackref,breaklinks,colorlinks,allcolors=cvprblue]{hyperref}

\def\paperID{16804} %
\def\confName{CVPR}
\def\confYear{2025}

\title{Reanimating Images using Neural Representations of Dynamic Stimuli}

\author{%
  Jacob Yeung\thanks{Corresponding author: \texttt{jacobyeung@cmu.edu}}, 
  Andrew F. Luo, 
  Gabriel Sarch, 
  Margaret M. Henderson, 
  Deva Ramanan, 
  Michael J. Tarr \vspace{5pt}\\
  Carnegie Mellon University\vspace{-.2cm}
}

\begin{document}

\maketitle
\begin{abstract}
While computer vision models have made incredible strides in static image recognition, they still do not match human performance in tasks that require the understanding of complex, dynamic motion. This is notably true for real-world scenarios where embodied agents face complex and motion-rich environments. Our approach, BrainNRDS (\textbf{N}eural \textbf{R}epresentations of \textbf{D}ynamic \textbf{S}timuli), leverages state-of-the-art video diffusion models to decouple static image representation from motion generation, enabling us to utilize fMRI brain activity for a deeper understanding of human responses to dynamic visual stimuli. Conversely, we also demonstrate that information about the brain's representation of motion can enhance the prediction of optical flow in artificial systems. Our novel approach leads to four main findings: (1) Visual motion, represented as fine-grained, object-level resolution optical flow, can be decoded from brain activity generated by participants viewing video stimuli; (2) Video encoders outperform image-based models in predicting video-driven brain activity; (3) Brain-decoded motion signals enable realistic video reanimation based only on the initial frame of the video; and (4) We extend prior work to achieve full video decoding from video-driven brain activity. \nerds advances our understanding of how the brain represents spatial and temporal information in dynamic visual scenes. Our findings demonstrate the potential of combining brain imaging with video diffusion models for developing more robust and biologically-inspired computer vision systems. We show additional decoding and encoding examples on this site: \url{https://brain-nrds.github.io/}. 
\end{abstract}\vspace{-15pt}

\section{Introduction}

\begin{figure*}
    \centering
    \includegraphics[width=0.99\textwidth]{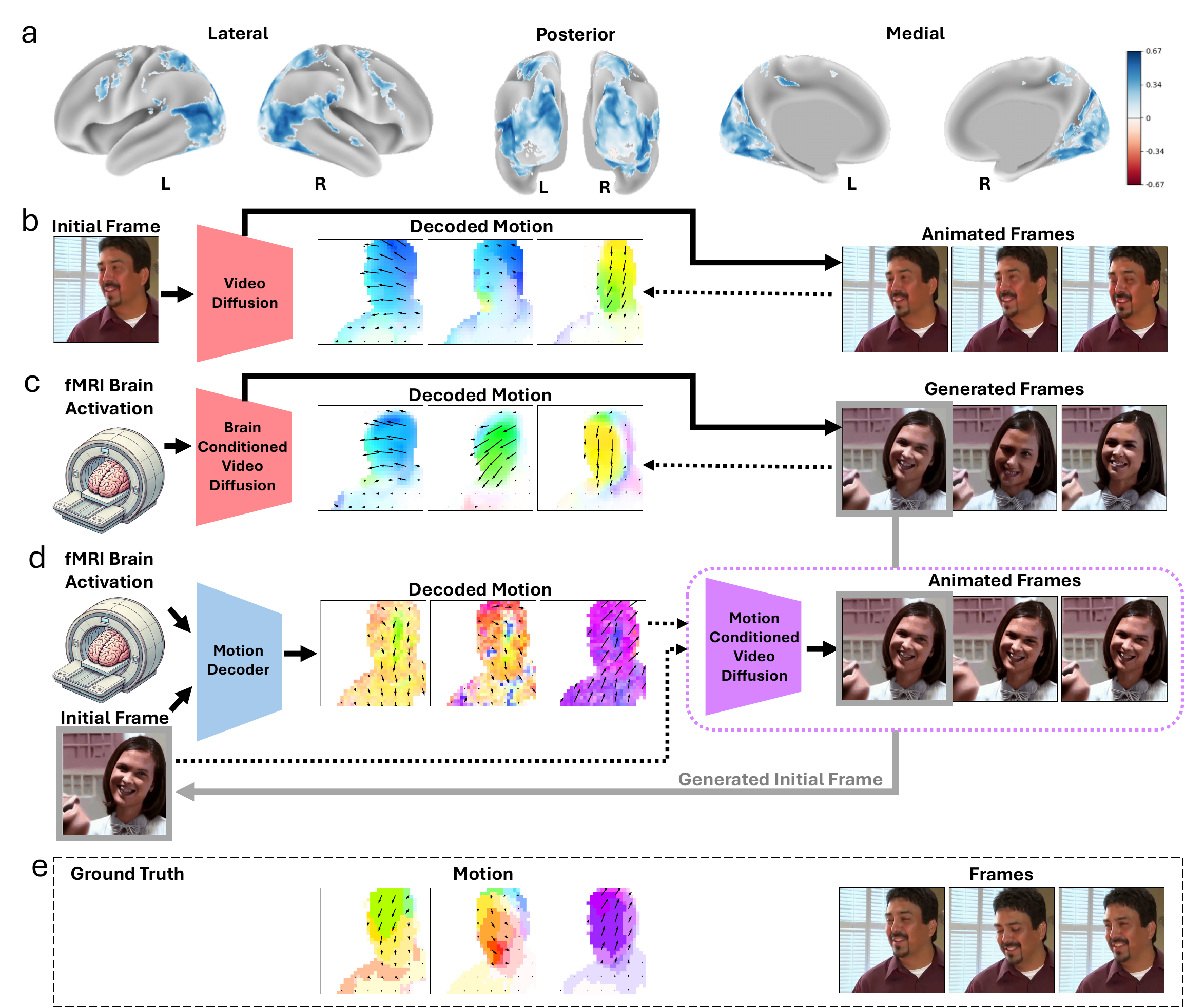} %
    
    \caption{{\bf Encoding and decoding video motion using brain activity.} {\bf (a)} fMRI brain activity can be predicted (using a Pearson coefficient) using off-the-shelf video encoders (VideoMAE~\cite{tong2022videomae}) extracted from the viewed video. 
    In the converse direction, we can generate video by decoding brain activity.
    \textbf{(b)} Many existing video diffusion models (e.g., SVD~\cite{blattmann2023stable}) generate a video by animating an initial frame. \textbf{(c)} This suggests that``brain-to-video" generation can be achieved by fine-tuning diffusers to condition on fMRI input (e.g., MindVideo~\cite{chen2023cinematic}).
    \textbf{(d)} In our approach we explicitly decouple the task of image and motion generation from brain activity. Given an initial video frame (which could be decoded from brain activity as in (c)) \textit{and} fMRI input%
    , we train a network to predict optical flow. We then animate the initial frame by feeding the predicted flow into an off-the-shelf motion-conditioned diffusion model (e.g., DragNUWA\cite{yin2023dragnuwa}). Our disentangled pipeline produces  more accurate brain-conditioned motion decodings than either (b) or (c). 
    \textbf{(e)} Ground truth optical flow and video.
    }
    
    \label{fig:teaser}
    \vspace{-.5cm}
\end{figure*}

Human vision excels at understanding complex, dynamic scenes, while computer vision offers powerful tools for data processing and analysis.
This work explores how insights from human brain activity, specifically in response to dynamic visual stimuli, can enhance the ability of computer vision systems to understand complex motion. Critically, as inherently embodied systems, our brains have evolved to process and represent the daily experience of naturally motion-rich environments in ways that are both semantically grounded and behaviorally-relevant. For example, a video of a person walking affords information not only about the visual features of the walker, but also about motion patterns, context, intent, and relationships between the different elements in the scene (e.g., sudden increase in gait implies urgency). This interplay between spatial and temporal information is impossible to capture from static images; accordingly, humans excel at inferring such high-level information from scene dynamics~\cite{Heider1944}.

Recent advances have improved our understanding of how visual information is represented in the brain through computer vision models~\cite{yamins2014performance,Wang2023machineintelligence,luo2024brainmappingdensefeatures}, yet a significant gap persists in explaining how the brain processes dynamic visual stimuli.
While some studies have started to explore the neural representation of dynamic visual stimuli~\cite{chen2023cinematic,lahner_modeling_2024,sun2024neurocinedecodingvividvideo}, there remains substantial potential to further harness modern computer vision techniques, especially video-based models.
There has also been little work demonstrating how insights from human vision can be leveraged to build better computer vision systems. Here we address both challenges, demonstrating that state-of-the-art video foundation models can be used to robustly predict human brain responses to dynamic visual inputs and that brain-decoded motion signals can help to reanimate videos based only on the first frame of the video. Critically, our framework outperforms prior work for both problems, achieving lower end point error (EPE) for motion decoding based on brain responses as compared to~\cite{chen2023cinematic} and better motion prediction based on initial static frames with brain data than without brain data.

In this paper, we present 
a framework for analyzing dynamic visual stimuli that explicitly separates the modeling of static and motion-related image representations as they are processed in the human brain: \textbf{N}eural \textbf{R}epresentations of \textbf{D}ynamic \textbf{S}timuli (``\nnerds"). By decoupling these elements,
\nerds offers a targeted approach to understanding the distinct neural pathways used to process and represent dynamic visual information in real-world scenes. Better specified models of dynamic visual processing are also critical for developing effective models of ecologically critical domains such as social interaction~\cite{Heider1944}, action representation, and dynamic causal structure~\cite{carpenter_fourteen_1998} -- none of which are easily extracted from static images.

\textbf{Contributions.} %
First, we demonstrate that visual motion information in the form of optical flow can be predicted (or decoded) from brain activity as measured by fMRI. %
Second, we show that this predicted motion can be used to realistically animate static images using a motion-conditioned video diffusion model in which the motion is driven by brain activity. Moreover, we demonstrate that brain activity can produce more reliable animations than image-conditioned video diffusion models trained on massively-large datasets. 
Third, we apply \nerds to existing methods that decode the viewed video by using their initial frame to do full video decoding from video-driven brain activity \textit{and} find that we more accurately predict the viewed motion than existing methods.
Finally, we show prediction in the reverse direction: recent video encoders (e.g. VideoMAE and VC-1~\cite{tong2022videomae,vc2023}) more effectively predict fMRI brain activity from videos compared to image encoders.
Overall these results demonstrate the effectiveness of \nerds as a novel, extensible framework for interpreting how the human brain processes dynamic visual information.

\textbf{Larger Objectives.} Our project also has several larger objectives. First, developing high-performing models for motion decoding based on neural data (Figure~\ref{fig:teaser}d) is an important precursor to building effective encoding models~\cite{naselaris2011encoding}. Second, the finding that VideoMAE is able to accurately predict brain responses (Figure~\ref{fig:teaser}a) is an important step towards building more general and informative video encoding models for brain prediction~\cite{Wang2023machineintelligence}). Encoding models are also important components in comparisons between model-based representations and empirical neural data, which can reveal key insights into which visual features the brain prioritizes and how they align with computational processing in interpretable artificial neural networks~\cite{sarch2023brain}. 
Third, we apply \nerds to existing methods which decode the viewed video to do full video decoding from video-driven brain activity~\cite{chen2023cinematic} \textit{and} find that we more accurately predict the viewed motion than existing methods.
Fourth, although comparison points are limited due to the underexploration of this research area, our finding that neural data improves video prediction performance better than other extant models is a simple demonstration of the point that latent information encoded in brain data may be useful for improving the performance of computer vision systems~\cite{Toneva_Improving}. This approach may prove to be especially powerful for domains that are inherently anchored in human behavior, for example, dynamic social interactions.

\section{Related Work}
\textbf{Static Visual Stimuli Decoding from fMRI.}
Reconstructing visual images from human brain activity, such as that measured by functional magnetic resonance imaging (fMRI), has significantly advanced through the use of deep learning models trained on natural image datasets \cite{yamins2014performance,horikawa2017generic,kietzmann2019recurrence,gucclu2015deep,groen2018distinct,wen2018neural,kell2018task,koumura2019cascaded,schrimpf2021neural,goldstein2022shared,caucheteux2022brains,schmitt2021predicting,sun2023contrast,luo2024brainmappingdensefeatures,yang2024braindecodesdeepnets,conwell_large-scale_2024,chen2024universaldimensionsvisualrepresentation} and  methods that leverage generative models to decode brain representations~\cite{takagi2022high,zeng2023controllable,chen2023cinematic,ozcelik2023brain,lin2022mind,lu2023minddiffuser,scotti2023reconstructing,koide2023mental,gu2022decoding,chen2023seeing,ferrante2023brain,liu2023brainclip,wang2024mindbridge,scotti2024mindeye2,kamitani2005decoding,dado_hyperrealistic_2022,doerig2022semantic,yang2024alignedcutvisualconceptsdiscovery,Du2022fmribci,efird2023identifying,efird2024s,gu2022neurogen,chen2014survey, DU2019948, brainsci12101394, RASTEGARNIA2023120395, 9349437, Prince2023.08.04.551888}. This progress is largely due to the availability of large, publicly accessible fMRI datasets with well-annotated stimuli, such as BOLD5000 \cite{chang2019bold5000} and the Natural Scenes Dataset (NSD) \cite{Allen2022}, along with multimodal generative models that have been pre-trained on very large datasets, such as Stable Diffusion \cite{rombach2022high}. 

\textbf{Dynamic Visual Stimuli Decoding from fMRI.}
While there has been extensive decoding work on image reconstruction from fMRI, there has been much less work on video reconstruction. Within the work on decoding motion from fMRI, much of prior work focuses on classifying global motion in artificial stimuli like random dots moving in a coherent direction \cite{maunsell1983functional, duffy1991sensitivity}. Recent work has also explored disentangling static and motion features using semantically meaningful dot movements \cite{Robert621}. We significantly advance this by decoding motion in complex, naturalistic visual stimuli at a more fine-grained, object-level resolution. This allows us to study how the brain represents motion in more ecologically valid scenarios. 

Our fMRI-based approach offers a whole-brain perspective compared to single-neuron recordings, which are limited in spatial coverage \cite{heeger1999motion, heuer2004optic, orban1992first, maunsell1983functional, duffy1991sensitivity,li2024enhance}. Despite the challenges of fMRI's slow temporal resolution, our results demonstrate successful decoding of motion information from complex, naturalistic stimuli at the object level. Thus, fMRI, despite its limitations, can enable valuable insights into the neural representation of dynamic information.

\begin{figure}[]
    \centering
    \includegraphics[width=0.5\textwidth]{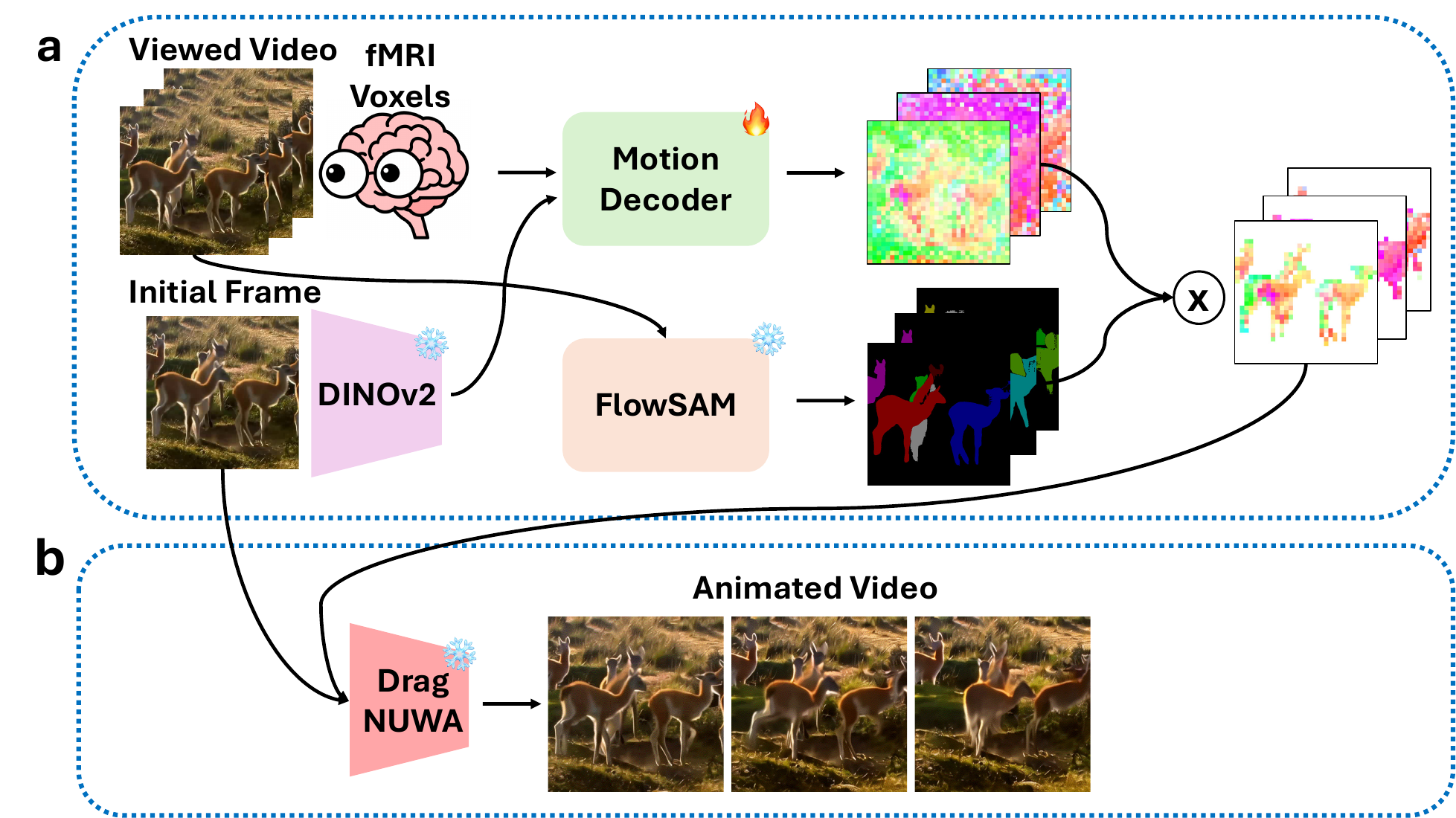}
    \caption{\textbf{\nerds pipeline for motion decoding and video generation.} \textbf{(a)} \nerds takes in neural data and image features from DINOv2~\cite{oquab2023dinov2}, extracted from the initial frame to predict consecutive future dense optical flow fields. Salient objects are masked out using FlowSAM~\cite{xie2024flowsam} to obtain the masked object flow. Snowflakes=frozen; flames=actively trained.
    \textbf{(b)} The initial frame is realistically animated using our predicted motion and a motion-conditioned video diffusion model, DragNUWA~\cite{yin2023dragnuwa}.}
    \label{fig:architecture}
    \vspace{-.5cm}
\end{figure}

Concurrent to motion decoding research, there is a parallel branch of research focusing on decoding the raw video from fMRI activity~\cite{wang2022reconstructing,kupershmidt2022pennyvisualthoughtsselfsupervised,chen2023cinematic, 6919270, gong2024neuroclipshighfidelitysmoothfmritovideo, 7056490}. Methods like MindVideo and NeuroClips~\cite{chen2023cinematic,gong2024neuroclipshighfidelitysmoothfmritovideo} train to align temporal semantics from fMRI and generate frames across time. Notably, our approach explicitly disentangles the representations of motion information (represented as optical flow) from static images in fMRI signals. This allows us to differentiate our present work from previous work by enabling interpretability. Specifically, we isolate and focus on decoding motion (as optical flow) from fMRI signals, as opposed to more difficult to interpret mixture of static image features and motion. This disentanglement allows us to learn an explicit representation of motion which can be directly compared to groundtruth and be more directly related to spatially localized neural mechanisms, as opposed to an entangled mixture of static image and dynamic features. In contrast, NeuroClips~\cite{gong2024neuroclipshighfidelitysmoothfmritovideo} decodes blurred video as a proxy for motion. Comparing the two approaches,
blurred video preserves rough scene composition, while optical flow explicitly encodes motion magnitude and direction, offering links to neural motion processing. At the same time, we ensure temporal consistency by integrating predicted optical flow with a motion-conditioned video diffusion model.

\section{Methods}

In order to understand how the brain processes dynamic visual stimuli, we focus on three directions. In the first, we examine the motion representation contained in the brain by decoding dense optical flow conditioned on the initial frame. In the second, we visualize the predicted motion by reconstructing the observed dynamic visual stimuli by leveraging a motion conditioned image diffusion model (DragNUWA). Finally, we examine which visual features, (static vs dynamic; semantic vs visual similarity) best explain brain activity on a voxel-wise level.

\subsection{Motion Prediction}
\label{subsection:motion prediction}

We extract the information relevant to visual motion in order to predict the optical flow in videos observed by participants. 
To enhance interpretability, we've architecturally separated static image features from dynamic features in \nnerds. By conditioning the model on the initial frame alongside neural data, we enable our model to focus on extracting dynamic visual information. This approach improves interpretability compared to models like MindVideo, which simultaneously predict both dynamic and static features. In contrast, our approach disentangles the specific contributions of dynamic components.

At a given fMRI scan at index $i$ we have fMRI voxels $B_i\in \mathbb{R}^n$ and corresponding video frames $\mathcal{I}_{i}\in \mathbb{R}^{T\times H\times W\times 3}$. 
We obtain optical flow fields $O_i\in \mathbb{R}^{(T-1)\times H\times W\times 2}$ from RAFT~\cite{teed2020raft}, which consist of $(T-1)\times H\times W$ optical flow vectors in $\mathbb{R}^2$. Prior work~\cite{walker2015dense} has found that reformulating optical flow prediction as classification improves performance. We take a similar approach and quantize the optical flow vectors into a codebook of $k$ clusters using k-means. We obtain image features with pretrained vision model $G$. Concretely, we learn a function $M_\theta$ that classifies the fMRI voxels $B_i$ and image features of the corresponding initial frame $G(\mathcal{I}_{i,1})$ to the quantized future predicted optical flows fields: $M_{\theta}([B_i, \mathcal{I}_{i,1}]) \rightarrow O_i$.

In order to focus our evaluation on salient objects in a scene, we apply masking using FlowSAM~\cite{xie2024flowsam} to track and identify salient objects.

\textbf{Visualizing the Predicted Motion.}
We can visualize the predicted motion by passing in the predicted optical flow field and the initial frame to a pretrained motion-conditioned video diffusion model, DragNUWA~\cite{yin2023dragnuwa}. This allows us to generate the dynamic vision stimuli (observed by the participants) or (driving the neural response).

\begin{figure}[h]
    \vspace{-.2cm}
    \centering
    \begin{subfigure}[b]{0.49\textwidth}
        \centering
        \includegraphics[width=0.9\textwidth]{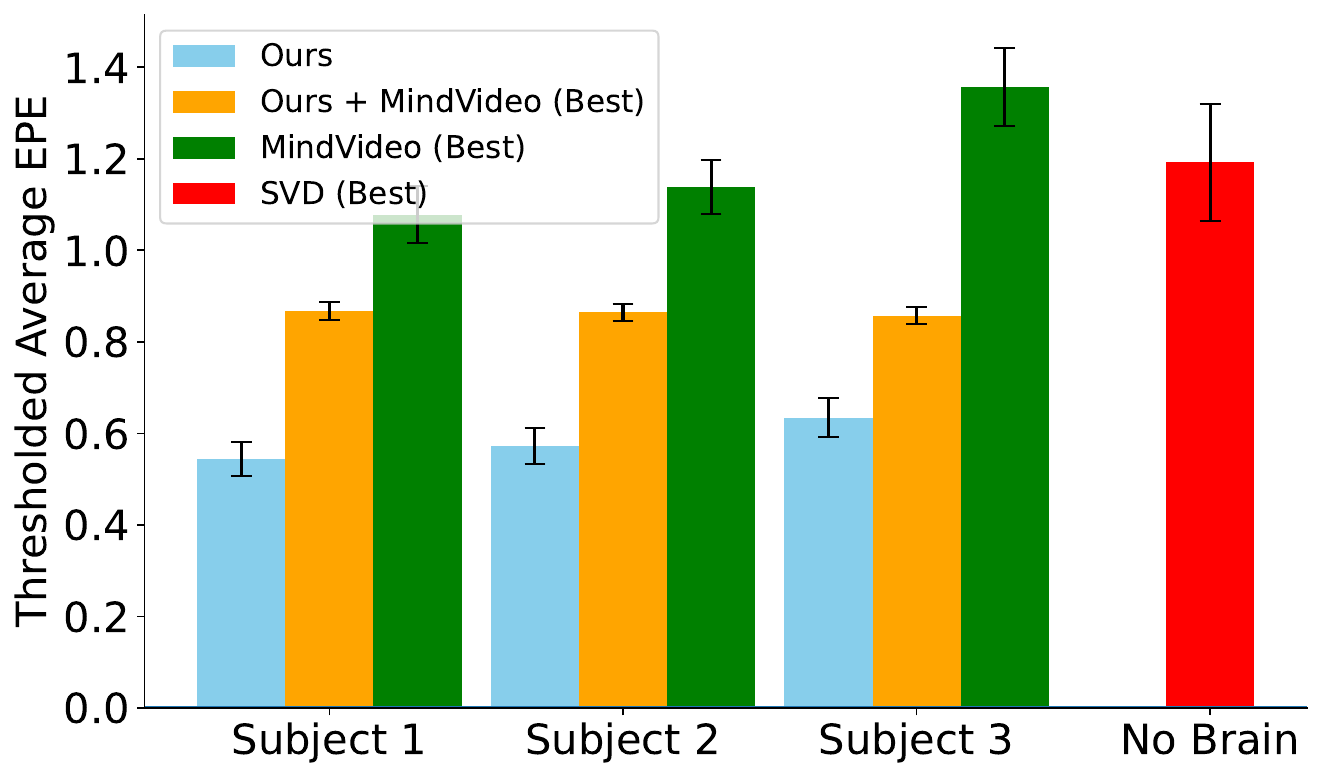}
        \label{fig:decoding_epe}
    \end{subfigure}
    \caption{\textbf{Quantitative motion decoding.} Optical flow predictors trained with neural data (Ours) are statistically better than both generative models trained without neural data (No Brain - Stable Video Diffusion (Best)~\cite{blattmann2023stable}) and generative models that fail to disentangle appearance and motion (MindVideo (Best)~\cite{chen2023cinematic}). We find that our method conditioned on the initial frame generated by MindVideo (Ours + MindVideo (Best)) better predicts the optical flow than MindVideo. %
    We average the end point error over the predicted and ground truth masked optical flow vectors. Similar to~\cite{eslami2024rethinkingraftefficientoptical}, we report end point error on pixels whose ground truth flow magnitudes exceed 1\% of the pixel width of the frames. Lower values are better. Paired $t$-tests comparing end point error for ``Ours'' versus SVD (Best) for each participant are as follows: S1: $p \le $ 1.197e-8; S2: $p \le $ 3.157e-8; S3: $p \le $ 9.140e-7. Error bars represent the standard error of the mean.}
    \label{fig:decoding quantitative}
    \vspace{-.2cm}
\end{figure}

\begin{figure*}[h]
    \centering
    
    \includegraphics[width=0.95\textwidth]{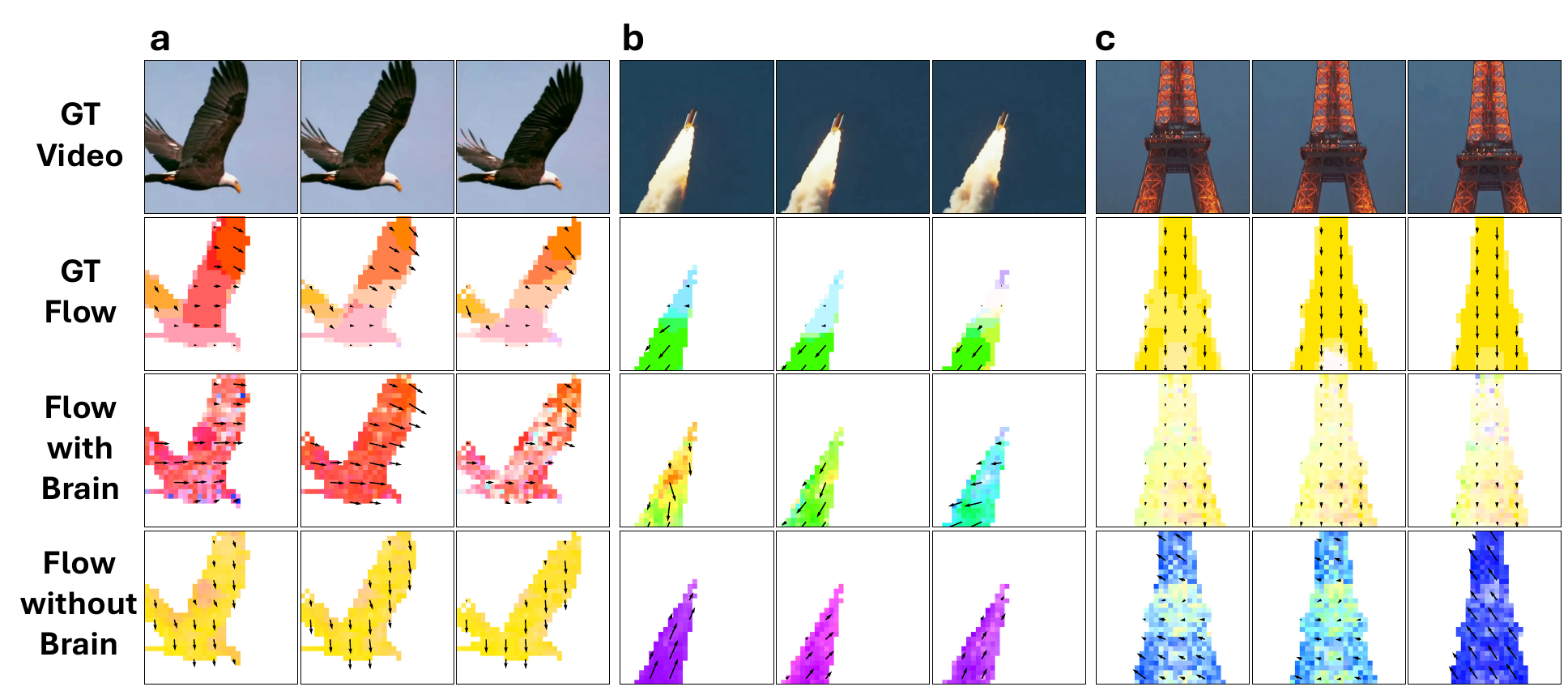} %
    \caption{\textbf{Motion predicted with and without neural data.} 
    Motion predicted using brain data and a static reference image is qualitatively both plausible and aligned with the original video compared to motion predicted from only the reference image using Stable Video Diffusion~\cite{blattmann2023stable}. Motion predicted from purely image features is plausible but not aligned with the original video. In this latter case, the model is hallucinating flows. 
    \textbf{(a)} An example video where there are multiple plausible actions. 
    \textbf{(b)} An example video where camera motion overrides the plausible motion. 
    \textbf{(c)} An example video where there is ambiguity of stationary objects due to camera motion. 
    }
    \label{fig:decoded examples wwo brain}
    \vspace{-.5cm}
\end{figure*}
\begin{table}
\begin{tabular}{c@{}c@{\hspace{1.5em}}@{}cc}
\toprule
\multirow{2}{*}{\vspace{-3pt}\small Method} & \multicolumn{1}{c}{\small Video-based \vspace{-3pt}} & \multicolumn{2}{c}{\small Frame-based} \\ 
\cmidrule(r){2-2} \cmidrule(l){3-4}\
 & \multicolumn{1}{c}{\footnotesize VideoMAE CosSim$\uparrow$} & \footnotesize CLIP CosSim$\uparrow$ & \footnotesize Pixel SSIM$\uparrow$ \\ \hline
\footnotesize MindVideo & \multicolumn{1}{c}{\footnotesize 0.742\scriptsize$\pm$0.006} & \footnotesize 0.879\scriptsize$\pm$0.004 & \footnotesize 0.171\scriptsize$\pm$0.02 \\
\footnotesize \textbf{Ours end-to-end} & \multicolumn{1}{c}{\footnotesize \textbf{0.769\scriptsize$\pm$0.006}} & \footnotesize \textbf{0.896\scriptsize$\pm$0.003} & \footnotesize \textbf{0.214\scriptsize$\pm$0.01} \\ \hline
\footnotesize Subject 1 & \multicolumn{1}{c}{\footnotesize 0.771\scriptsize$\pm$0.006} & \footnotesize 0.896\scriptsize$\pm$0.003 & \footnotesize 0.218\scriptsize$\pm$0.01 \\
\footnotesize Subject 2 & \multicolumn{1}{c}{\footnotesize 0.774\scriptsize$\pm$0.006} & \footnotesize 0.896\scriptsize$\pm$0.003 & \footnotesize 0.2\scriptsize$\pm$0.01 \\
\footnotesize Subject 3 & \multicolumn{1}{c}{\footnotesize 0.762\scriptsize$\pm$0.006} & \footnotesize 0.894\scriptsize$\pm$0.003 & \footnotesize 0.223\scriptsize$\pm$0.01\\
\cline{1-4}
\end{tabular}
\caption{\footnotesize\textbf{Video generation quantitative results. } 
Our end-to-end method uses the initial frame generated from fMRI by MindVideo and generates the video solely from fMRI without access to GT frames.
We report semantic fidelity metrics and SSIM on both video and frame level.
Results are significant ($p<$0.05 using two-sample t-test) compared to MindVideo.
\vspace{-25pt}}
\label{table:generation_quantitative}
\end{table}

\subsection{Visual Stimuli-To-Brain Encoding}
Here we describe a complementary approach to understanding dynamic visual stimuli processing in the brain. We consider models trained on different tasks:

\begin{enumerate}[wide]
    \item\noindent \textbf{Static Image:} These models are trained on static visual stimuli without guidance from dynamics. The features extracted can identify brain voxels that are semantically and spatially tuned.
    \begin{enumerate}
        \item \textbf{Supervised:} ResNet-50 is trained on ImageNet to categorize images into semantic classes~\cite{he2015resnet}.
        \item \textbf{Self-supervised:} DINOv1 and DINOv2 are vision transformer-based models trained to learn similar representations for visually similar patches of images without supervision on semantic similarity~\cite{caron2021dino, oquab2023dinov2}.
        \item \textbf{Semantically-supervised:} CLIP ViT-B/32 and CLIP ConvNeXt are trained to learn align semantics and images. CLIP ViT-B/32 uses a visual transformer architecture and CLIP ConvNeXt uses a convolutional backbone, sharing inductive biases with ResNet-50 but trained with a different objective. CLIP ViT-B/32 AF and CLIP ConvNeXt AF consist of features \textbf{A}veraged over \textbf{F}rames (AF) spanning the video~\cite{Radford2021clip,cherti2023reproducible}.
    \end{enumerate}
    \item\noindent  \textbf{Video:} These models are trained on dynamic visual stimuli. The features extracted can identify brain voxels that are spatially and dynamically tuned.
    \begin{enumerate}
        \item \textbf{Self-supervised:} VideoMAE, VideoMAE Large, Hiera Base Plus, and Hiera Huge are vision transformer-based architectures trained on Kinetics 400 with a masked autoencoder loss. The encoder learns embeddings used for downstream video tasks such as action recognition~\cite{tong2022videomae,ryali2023hiera}.
        \item \textbf{Semantically-supervised:} X-CLIP is a CLIP model fine-tuned with cross-frame attention to learn temporal relations~\cite{Ma2022XCLIP}.
    \end{enumerate}
    \item\noindent  \textbf{Embodied AI:} These models are trained to align representations of single frames across time for embodied AI visuomotor manipulation. The features extracted can identify brain voxels tuned to action representation.
    \begin{enumerate}
        \item \textbf{Self-supervised:} VIP is a ResNet-50 based architecture trained on Ego4D. VC-1 is a vision transformer-based architecture trained on seven human egocentric video datasets, including Ego4D. For both models, we also test an encoder trained on features averaged over frames spanning the video (VIP AF and VC-1 AF~\cite{ma2023vip, vc2023}).
        \item \textbf{Semantically-supervised:} R3M is a ResNet-50 based architecture trained on Ego4D to align frames that are close in time and semantically similar~\cite{nair2022r3m}.
    \end{enumerate}
\end{enumerate}

We train ridge regression models on visual features from different encoding models to predict voxel-wise responses.

\section{Experiments}
In~\ref{subsection:dataset}, we describe the dataset and preprocessing used. In~\ref{subsection:motion decoding} we explore motion decoding (reconstruction) conditioned on initial image and brain information using \nnerds, and augment quantitative analysis with visualizations derived from a motion-conditioned diffusion model. In~\ref{subsection:encoding}, we explore the performance of different visual backbones in predicting brain activations.

\subsection{Dataset}
\label{subsection:dataset}
We utilize the Dynamic Natural Vision dataset~\cite{wen_dynamic_vision}, a public, IRB approved fMRI dataset composed of brain responses from 3 participants viewing diverse naturalistic video clips. The dataset contains 11.5 hours of fMRI data, with the 3 participants viewing the same videos. The fMRI data was recorded using a 3T MRI system with a 2 second TR and the videos were presented at 30 FPS. 18 8-minute videos were seen twice and 5 8-minute videos were seen 10 times, including the train and test sets which have no overlapping videos. We use the same train/test split described in~\cite{wen_dynamic_vision}. The fMRI data is preprocessed and aligned with presented stimuli using the pipeline from the Human Connectome Project. We further normalized the fMRI activations so that each voxel's response was zero-mean and unit variance on a session basis. Then we average the response across repeated viewings of the same visual stimuli. 

In practice, we temporally downsample the flows to $c=3$ frames, selected in equal intervals over the corresponding 2 second fMRI brain scan, and aim to predict the $c$ quantized consecutive optical flow fields. We also spatially downsample the flow to a resolution of $32 \times 32$ and quantize the flows into 40 clusters. We extract image features from the first frame using DINOv2. During test time, we take a weighted-sum over all the classes per pixel, following~\cite{walker2015dense}. We train the model using cross entropy loss. 
We illustrate our model's pipeline in Figure~\ref{fig:architecture}. 

fMRI signal
peaks approximately 4 seconds after stimulus presentation and slowly decays due to the hemodynamic response. In preprocessing, the fMRI data is shifted to temporally align the peak of the response to the onset of its corresponding stimulus presentation. However,
fMRI scans 4 seconds(2 TRs) prior and after the current TR may share information with the current TR. Consequently, when predicting the motion in the video presented at TR=$i$, we concatenate data from TRs $[i-2, i-1, i, i+1, i+2]$. Video frames are generated conditioned on the predicted motion using DragNUWA~\cite{yin2023dragnuwa}.

\subsection{Motion Decoding}
\label{subsection:motion decoding}
\textbf{Architecture.} \nerds predicts motion labels by processing fMRI and image features. fMRI data is processed through an MLP then spatially broadcasted and concatenated over the image features. The concatenated features pass through three residual \texttt{1x1} convolutional blocks with dropout, while global features are extracted using \texttt{AdaptiveAvgPool2d} and concatenated back into the spatial features. A final \texttt{1x1} convolution classifies each spatial patch into the motion codebook.

\textbf{Quantitative Flow Evaluation.} We report evaluations of the decoded optical flow vectors on the end point error on the test set. We quantitatively compare the models trained on the 3 participants against baselines of: MindVideo (Best)~\cite{chen2023cinematic}, \nerds leveraging the initial frame generated from MindVideo (Ours + MindVideo (Best)), and Stable Video Diffusion (SVD)~\cite{blattmann2023stable} that does not use neural data and only the initial frame in Figure~\ref{fig:decoding quantitative}. We extract the optical flows from the videos generated by MindVideo and SVD using the same RAFT model we use to extract the ground truth flow. Since MindVideo and SVD are generative models that can produce multiple plausible outputs, we generate 100 samples per video for MindVideo and 10 samples for SVD (because the latter is able to condition on the true video's first frame) and select the sample with the lowest end point error for fair comparison. We evaluate \nerds using the true initial frame from the video (Ours) as well as the first frame generated by MindVideo (Ours + MindVideo(Best)). Since the SVD baseline (SVD (Best)) is not conditioned on neural data, it learns to predict plausible flows which do not necessarily align with the true flow. Indeed, we observe that our model trained on individual participant neural data performs significantly better on end point error compared to SVD (Best) (end point error for individual participants -- S1, S2, S3 -- respectively, S1: 0.543, S2: 0.572, S3: 0.634 vs. 1.192, all $p \ll 0.001$ using paired $t$-tests with Holm-Bonferroni correction).

\textbf{Reasoning With and Without fMRI.} We present examples and explain why
models trained on neural data outperform SVD
(Figure ~\ref{fig:decoded examples wwo brain}).
First, there exists ambiguity of action given static images - a frame may contain multiple possible actions. For instance, in Figure~\ref{fig:decoded examples wwo brain}a we show an eagle flying. Given the first frame, the eagle could be plausibly either flapping its wings or gliding. The model trained on neural data is able to identify that the eagle is gliding. However, the model without brain conditionining incorrectly predicts that the eagle is flapping its wings.

Second, even if the action is not ambiguous, there exists camera motion in the videos which can change the direction of the flow fields. For example, in Figure~\ref{fig:decoded examples wwo brain}b we show a space shuttle flying into the sky. In this case, with no camera motion the direction of motion is clearly upwards and to the right, which the model without brain conditioning correctly predicts. However, because the video is centered on the space shuttle with camera motion, the exhaust from the space shuttle appears to be going downwards and to the left. This causes the optical flow vectors to point downwards and to the left. Again, the model trained with neural data is able to correctly predict the motion of the object.

\begin{figure}[]
    \centering
    \includegraphics[width=0.47\textwidth]{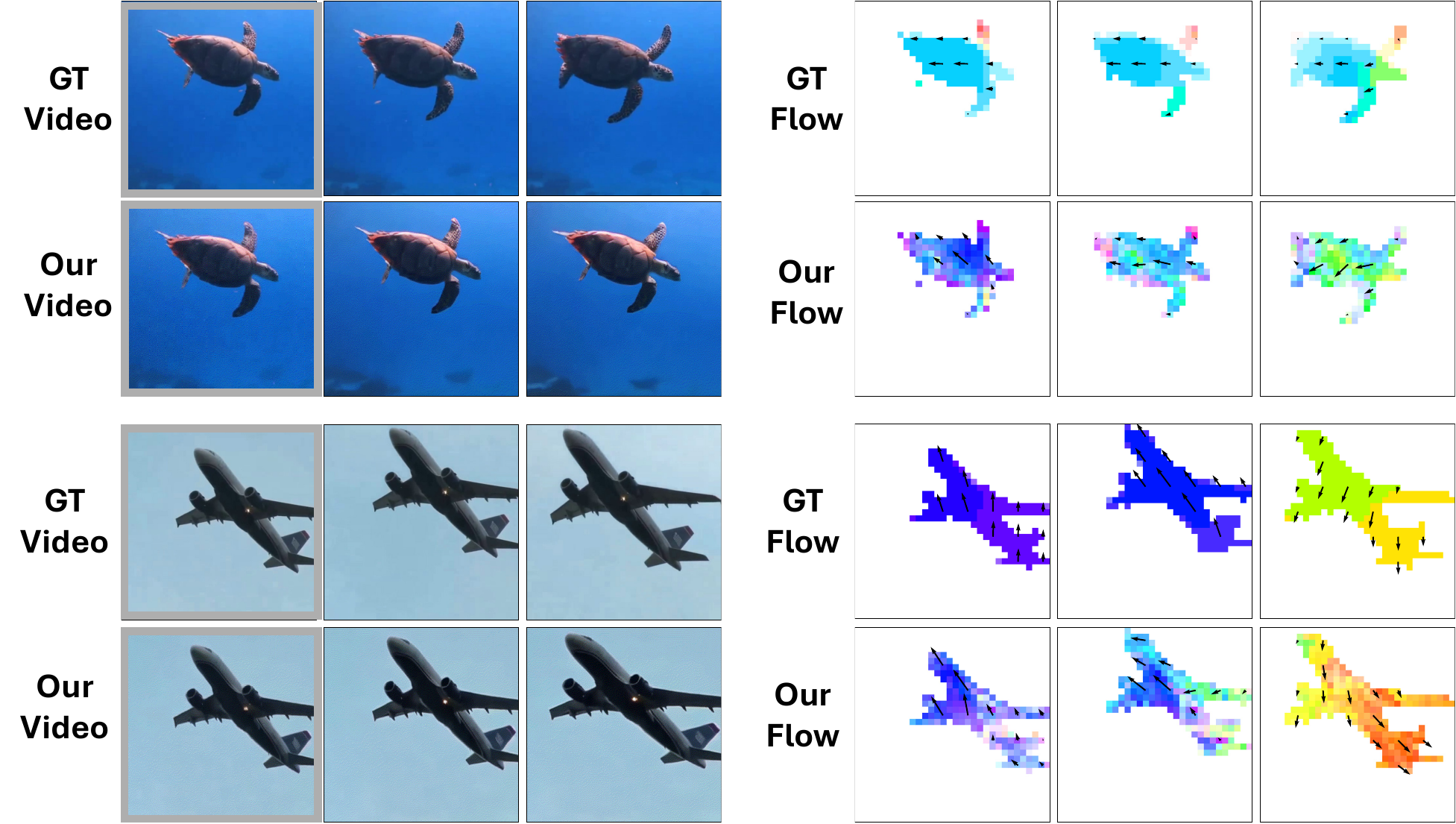}
    \caption{\textbf{Static image animation results.} Optical flow predicted from fMRI brain activity (``Our Flow'') and the ground truth initial frame (gray box) compared against the ground truth (``GT'') flow. Beneath the ground truth video frames, we show our results for animating the initial frame by combining the brain conditioned motion prediction with DragNUWA~\cite{yin2023dragnuwa} (``Our Video'').}
    \label{fig:decoded examples}
    \vspace{-.5cm}
\end{figure}

\begin{figure*}[]
    \centering
    \includegraphics[width=0.99\textwidth]{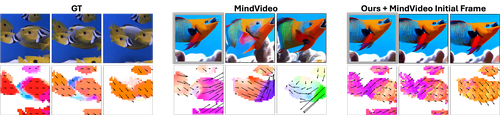} %
    \caption{\textbf{Extending the 
    method for video decoding.} Video decoding from fMRI is accomplished by leveraging the initial frame generated by MindVideo~\cite{chen2023cinematic} box). The ground truth flow (\textbf{bottom-left}) better aligns with our predicted motion (\textbf{bottom-right}) than with the motion predicted by MindVideo (\textbf{bottom-middle}).}
    \label{fig:decoding full video}
    \vspace{-0.5cm}
\end{figure*}

Third, there exists ambiguity of direction of motion in static images of stationary objects. For instance, if the first frame contains just a stationary portion of the Eiffel Tower, what movement should be predicted? One could choose to predict no motion, or arbitrarily predict a direction of camera motion. Indeed, we see in Figure~\ref{fig:decoded examples wwo brain}c that the model without brain conditioning chooses an arbitrary camera motion pointing leftwards. However, the model trained with neural data correctly predicts the true direction which is camera panning downwards.
We provide additional controls and quantitative results in the supplemental.

\textbf{Initial Frame Reanimation.} The results derived from these scenarios indicate that the neural data contains information about the correct movement which cannot be derived from just image features. 

We show motion decoding results along with the video reconstruction from DragNUWA (Figure~\ref{fig:decoded examples}). %
We find that the motion we predict from the fMRI brain activity is not only aligned with the ground truth motion, but also generates realistic videos. We further demonstrate that we are able to decode motion that is consistent and that suddenly changes direction, i.e. the airplane example in Figure
~\ref{fig:decoded examples} %
.

Because we disentangle the modeling of static image and motion representations, we can extend \nerds to other works that decode image frames from fMRI. We demonstrate examples of extending our method to initial frames decoded from MindVideo in Figure~\ref{fig:decoding full video} and find we decode motion more accurately (Figure~\ref{fig:decoding quantitative}). We provide additional video examples in the supplemental.

Generated videos are evaluated using initial frames decoded from MindVideo (Table~\ref{table:generation_quantitative}). Video-level evaluation is computed using the cosine similarity between averaged VideoMAE features of the ground truth and generated videos. Frame-level evaluation is computed using the cosine similarity between CLIP features of corresponding frames. SSIM is also reported. \nerds outperforms MindVideo across all metrics, demonstrating the advantage of separately modeling spatial and temporal 
information.

\subsection{Identifying Brain Regions Tuned to Dynamic Features}
\label{subsection:encoding}

Having demonstrated the ability to predict motion from fMRI brain activity, we identified the brain regions selective for dynamic visual features. To do so, we test various vision encoding models to determine which best predict the fMRI brain activity of participants viewing dynamic visual stimuli. Specifically, we extract visual features using these models and construct voxel-wise encoding models using ridge regression~\cite{Kay2008,naselaris2011encoding}. These voxel-wise encoding models predict the fMRI brain activity of participants watching dynamic visual stimuli based on the extracted features.

Importantly, we test models trained under different paradigms, such as those trained on static images or video input, and models that use visual or semantic contrastive losses. By analyzing the differences in brain prediction performance, we can identify brain regions more selective to dynamic visual features. 
\begin{figure}[]
    \centering
    \includegraphics[width=0.49\textwidth]{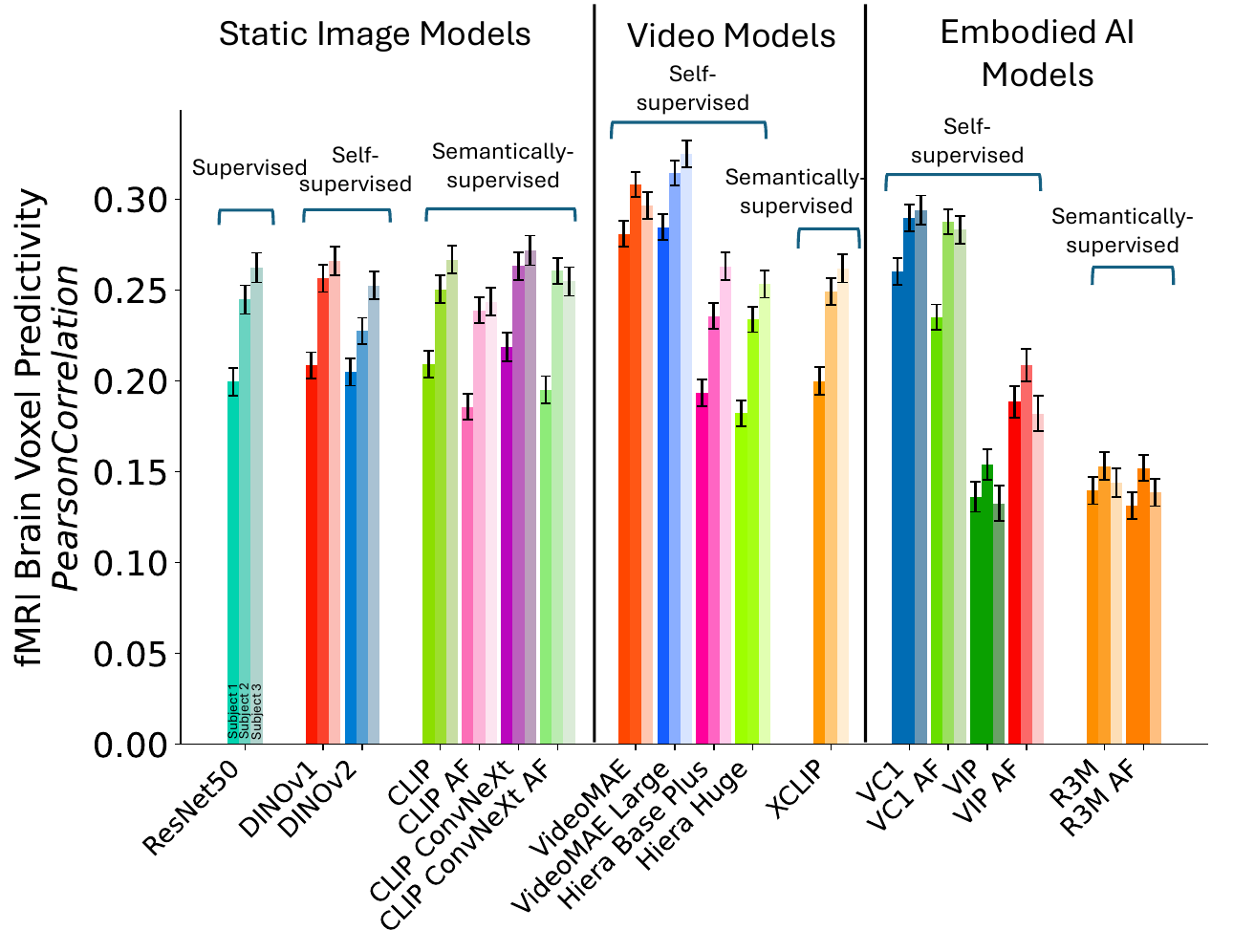} %
    \vspace{-14pt}\caption{\textbf{Brain Encoding Prediction Performance.} We analyze the ability of visual encoders (VideoMAE~\cite{tong2022videomae} and CLIP ConvNeXt) to predict (encode) fMRI brain activity from visual stimuli. Self-supervised video models, VideoMAE and VideoMAE Large, best predict fMRI brain activity. Details on which brain regions are best predicted by VideoMAE Large are shown in Figure~\ref{fig:encoding brain maps videomae large}. For static image models, semantically-supervised models such as CLIP ConvNeXt, perform best (consistent with~\cite{Wang2023machineintelligence}), while VC-1 performs best among embodied AI models. Figure~\ref{fig:encoding brain maps difference} compares the relative encoding performance of the best video (VideoMAE) and static image (CLIP ConvNeXt) models. Error bars represent the standard error of the mean.}
    \label{fig:encoding barplot}
    \vspace{-.65cm}
\end{figure}

We find that the best performing video encoding model, VideoMAE Large, outperforms the best static image encoding models at predicting fMRI brain activity (Pearson $r$'s for individual participants, S1: 0.285, S2: 0.314, S3: 0.324, all $p\ll0.001$ by a one-sample $t$-test). This suggests that fMRI data during video viewing contain dynamic information which is not captured in features from static images.

Within image encoding models, CLIP ConvNeXt predicts fMRI brain activity best across all participants (Pearson $r$'s for individual participants, S1: $0.219$, S2: $0.263$, S3: $0.272$, all $p\ll 0.001$), consistent with prior results~\cite{Wang2023machineintelligence} that alignment with language improves visual brain prediction. Averaging embeddings across frames from temporally unaligned encoding models does not enhance performance.

The best performing embodied AI vision encoding model, VC-1, outperforms the best static image encoding models at predicting brain activity (Pearson $r$'s for individual participants, S1: $0.260$, S2: $ 0.290$, S3: $0.294$, all $p\ll 0.001$). Interestingly, on average, VC-1 outperforms other embodied AI vision encoding models across a series of tasks~\cite{vc2023}. This suggests a possible avenue for ranking embodied AI vision encoding models on predicting fMRI brain activity. More generally, models which explicitly model temporal relationships are better at capturing neural responses to dynamic visual stimuli. 
We provide additional controls and quantitative results in the supplemental.

\begin{figure}[]
    \centering
    \includegraphics[width=0.45\textwidth]{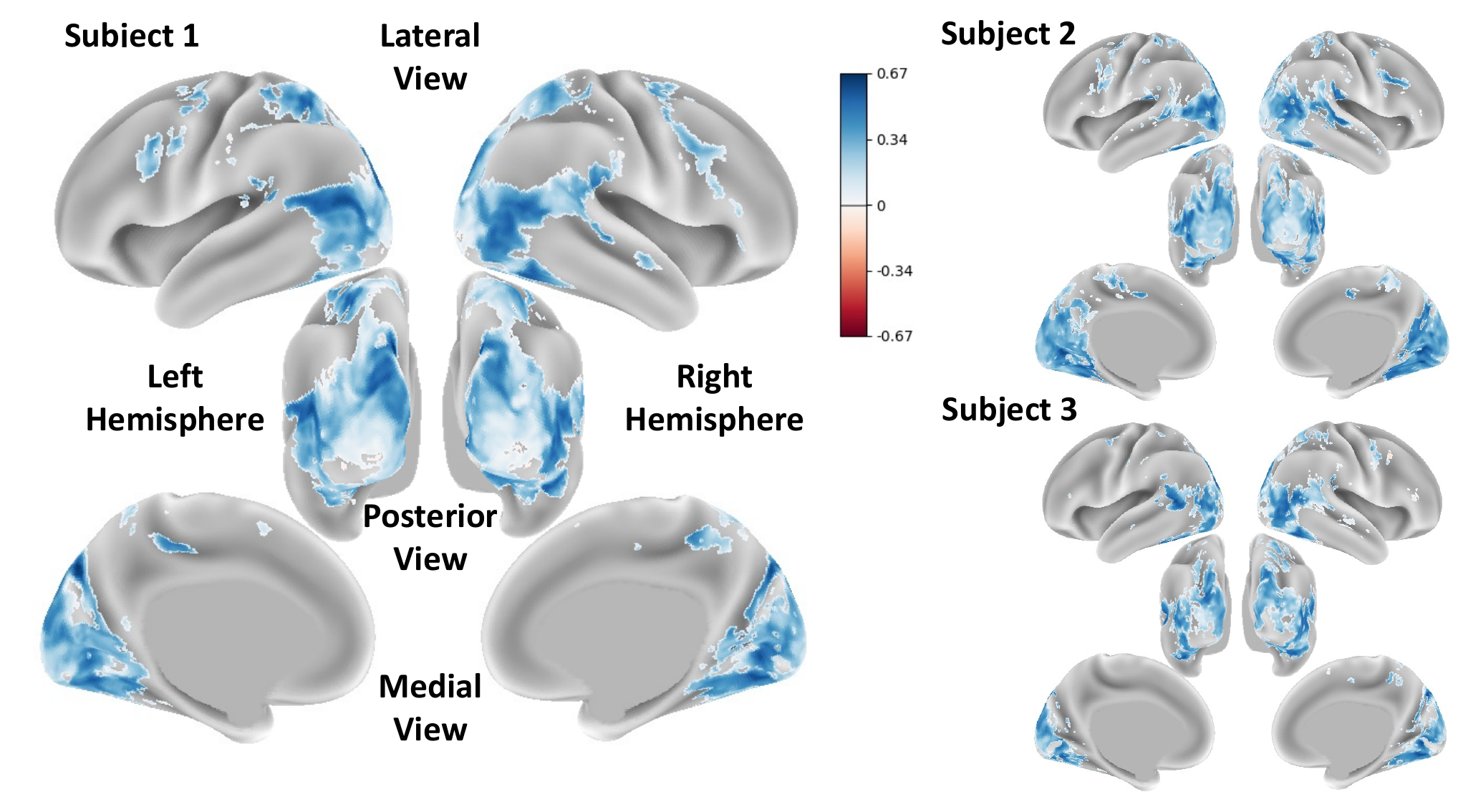}
    \caption{\textbf{VideoMAE voxel-wise prediction performance.} Voxel-wise fMRI voxel encoding accuracy of the best encoding model, VideoMAE Large, quantified as the Pearson correlation between measured and predicted responses. Refer to the supplemental %
    for a labeled fMRI data parcellation of brain regions.
    The majority of the voxels in visual cortex are predicted well by VideoMAE. VideoMAE predicts areas TPOJ1, STV, PGi, and 7PC (all subjects, Left Hemisphere, Lateral View) particularly well -- brain areas are implicated in processing visual and motor information and appear important for high-level brain functions, including motor planning and theory of mind~\cite{Baker2018c}. Thus, VideoMAE may be capturing dynamic information relevant for motor planning.}
    \label{fig:encoding brain maps videomae large}
    \vspace{-.65cm}
\end{figure}

\begin{figure}[]
    \centering
    \includegraphics[width=0.475\textwidth]{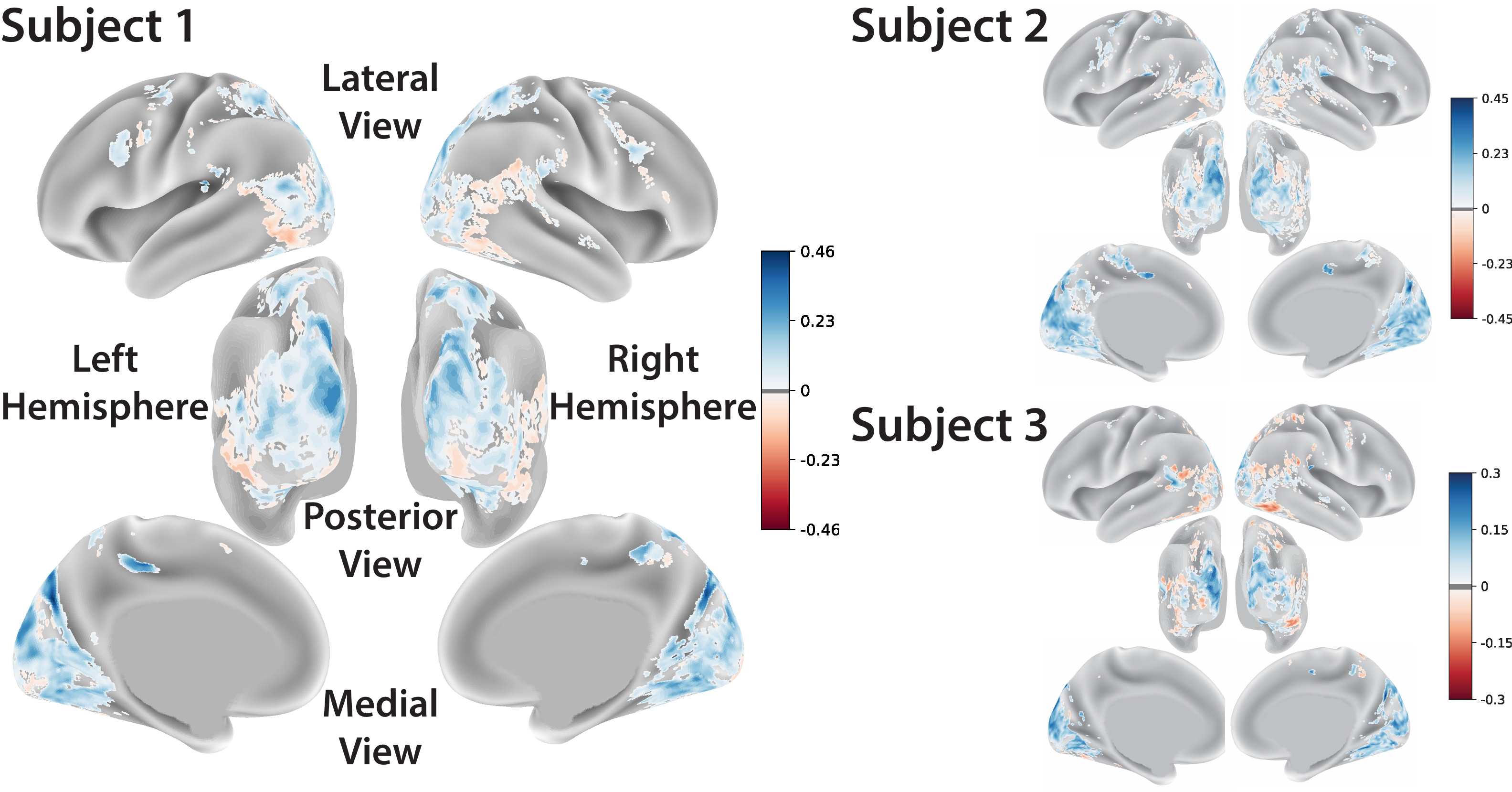}
    \caption{\textbf{Voxel-wise prediction difference between VideoMAE and CLIP ConvNeXt.} Voxel-wise  difference of fMRI voxel encoding accuracy between the best encoding model, VideoMAE Large, and the best performing static image encoding model, CLIP ConvNeXt, quantified as the Pearson correlation between measured and predicted responses. fMRI voxels where VideoMAE predicts better are in blue and where CLIP ConvNeXt predicts better are in red. %
    The majority of visual cortex is better predicted by VideoMAE. VideoMAE also predicts areas 5m, 5mv, 23c, 4, 24dd, and 24dv much better than CLIP ConvNeXt (S1 and S2, Left and Right Hemispheres, Medial View) -- all part of the somatosensory cortex which integrates visuomotor and somatosensory
    information for precise motor control and spatial awareness~\cite{Baker2018,Baker2018b}. Thus, VideoMAE may be capturing action representations which are less prevalent in static image models.}
    \label{fig:encoding brain maps difference}
    \vspace{-.5cm}
\end{figure}

\textbf{Interpreting Feature Performance.} We plot voxel-wise encoding performance of the best performing model, VideoMAE~\cite{tong2022videomae}, on an inflated cortical map to interpret the voxel-wise selectivity for dynamic visual stimuli features in Figure~\ref{fig:encoding brain maps videomae large}. Please refer to the caption for more detailed analysis.
We identify fMRI brain voxels that are motion selective by taking the voxel-wise encoding performance difference between the best encoding video model, VideoMAE Large, and the best encoding static image model, CLIP ConvNeXt, illustrated in Figure~\ref{fig:encoding brain maps difference}. We find results that suggest encoding visual stimuli with VideoMAE better predicts fMRI brain activity across visual cortex and in somatosensory cortex.
Voxel-wise encoding performance for each model is shown on inflated cortical maps in the supplementary information.

\section{Conclusion}
\textbf{Limitations and Future Work.}
\label{sec:conclusion limitations}
\nerds successfully decodes motion of salient objects from fMRI brain activity on the Dynamic Vision Dataset~\cite{wen_dynamic_vision}. However, the generalizability of our conclusions to different dataset distributions remains unclear. Future datasets with fine-grained motion can help determine the limits of motion decodability from fMRI brain activity. Furthermore, our subject-specific training approach could be improved by exploring cross-subject alignment methods, as has been shown to improve decoding performance for static stimuli~\cite{Huang2023functional,andreella2023functional}.

\textbf{Conclusion and Impact.}
\label{sec:broader impact}
To summarize, in this paper we propose to study neural representations of motion by disentangling the modeling of static image features and motion using \nnerds. We show that brain activity can be used to predict visual motion, demonstrate that our predicted motion can animate static images to reconstruct the observed dynamic visual stimuli, and identify regions in the brain that are tuned for dynamic stimuli. Our results lay the groundwork for better understanding how the brain represents visual motion.

\clearpage
\newpage
\bibliography{myref}

\begin{thebibliography}{95}
\providecommand{\natexlab}[1]{#1}
\providecommand{\url}[1]{\texttt{#1}}
\expandafter\ifx\csname urlstyle\endcsname\relax
  \providecommand{\doi}[1]{doi: #1}\else
  \providecommand{\doi}{doi: \begingroup \urlstyle{rm}\Url}\fi

\bibitem[Tong et~al.(2022)Tong, Song, Wang, and Wang]{tong2022videomae}
Zhan Tong, Yibing Song, Jue Wang, and Limin Wang.
\newblock Video{MAE}: Masked autoencoders are data-efficient learners for self-supervised video pre-training.
\newblock In Alice~H. Oh, Alekh Agarwal, Danielle Belgrave, and Kyunghyun Cho, editors, \emph{Advances in Neural Information Processing Systems}, 2022.

\bibitem[Blattmann et~al.(2023)Blattmann, Dockhorn, Kulal, Mendelevitch, Kilian, Lorenz, Levi, English, Voleti, Letts, et~al.]{blattmann2023stable}
Andreas Blattmann, Tim Dockhorn, Sumith Kulal, Daniel Mendelevitch, Maciej Kilian, Dominik Lorenz, Yam Levi, Zion English, Vikram Voleti, Adam Letts, et~al.
\newblock Stable video diffusion: Scaling latent video diffusion models to large datasets.
\newblock \emph{arXiv preprint arXiv:2311.15127}, 2023.

\bibitem[Chen et~al.(2023{\natexlab{a}})Chen, Qing, and Zhou]{chen2023cinematic}
Zijiao Chen, Jiaxin Qing, and Juan~Helen Zhou.
\newblock Cinematic mindscapes: High-quality video reconstruction from brain activity.
\newblock In \emph{Thirty-seventh Conference on Neural Information Processing Systems}, 2023{\natexlab{a}}.

\bibitem[Yin et~al.(2023)Yin, Wu, Liang, Shi, Li, Ming, and Duan]{yin2023dragnuwa}
Shengming Yin, Chenfei Wu, Jian Liang, Jie Shi, Houqiang Li, Gong Ming, and Nan Duan.
\newblock {DragNUWA}: Fine-grained control in video generation by integrating text, image, and trajectory.
\newblock 2023.

\bibitem[Heider and Simmel(1944)]{Heider1944}
F.~Heider and M.~Simmel.
\newblock An experimental study of apparent behaviour.
\newblock \emph{American Journal of Psychology}, 13, 1944.

\bibitem[Yamins et~al.(2014)Yamins, Hong, Cadieu, Solomon, Seibert, and DiCarlo]{yamins2014performance}
Daniel~LK Yamins, Ha~Hong, Charles~F Cadieu, Ethan~A Solomon, Darren Seibert, and James~J DiCarlo.
\newblock Performance-optimized hierarchical models predict neural responses in higher visual cortex.
\newblock \emph{Proceedings of the National Academy of Sciences}, 111\penalty0 (23):\penalty0 8619--8624, 2014.

\bibitem[Wang et~al.(2023)Wang, Kay, Naselaris, Tarr, and Wehbe]{Wang2023machineintelligence}
A.~Y. Wang, K.~Kay, T.~Naselaris, M.~J. Tarr, and L.~Wehbe.
\newblock Better models of human high-level visual cortex emerge from natural language supervision with a large and diverse dataset.
\newblock \emph{Nature Machine Intelligence}, 5\penalty0 (12):\penalty0 1415--1426, 2023.

\bibitem[Luo et~al.(2024)Luo, Yeung, Zawar, Dewan, Henderson, Wehbe, and Tarr]{luo2024brainmappingdensefeatures}
Andrew~F. Luo, Jacob Yeung, Rushikesh Zawar, Shaurya Dewan, Margaret~M. Henderson, Leila Wehbe, and Michael~J. Tarr.
\newblock Brain mapping with dense features: Grounding cortical semantic selectivity in natural images with vision transformers.
\newblock In \emph{Submitted to The Thirteenth International Conference on Learning Representations}, 2024.
\newblock under review.

\bibitem[Lahner et~al.(2024)Lahner, Dwivedi, Iamshchinina, Graumann, Lascelles, Roig, Gifford, Pan, Jin, Ratan~Murty, Kay, Oliva, and Cichy]{lahner_modeling_2024}
Benjamin Lahner, Kshitij Dwivedi, Polina Iamshchinina, Monika Graumann, Alex Lascelles, Gemma Roig, Alessandro~Thomas Gifford, Bowen Pan, SouYoung Jin, N.~Apurva Ratan~Murty, Kendrick Kay, Aude Oliva, and Radoslaw Cichy.
\newblock Modeling short visual events through the {BOLD} moments video {fMRI} dataset and metadata.
\newblock \emph{Nature Communications}, 15\penalty0 (1):\penalty0 6241, 2024.

\bibitem[Sun et~al.(2024)Sun, Li, Chen, and Moens]{sun2024neurocinedecodingvividvideo}
Jingyuan Sun, Mingxiao Li, Zijiao Chen, and Marie-Francine Moens.
\newblock {NeuroCine}: Decoding vivid video sequences from human brain activties, 2024.

\bibitem[Carpenter et~al.(1998)Carpenter, Akhtar, and Tomasello]{carpenter_fourteen_1998}
Malinda Carpenter, Nameera Akhtar, and Michael Tomasello.
\newblock Fourteen- through 18-month-old infants differentially imitate intentional and accidental actions.
\newblock \emph{Infant Behavior and Development}, 21\penalty0 (2):\penalty0 315--330, 1998.

\bibitem[Majumdar et~al.(2023)Majumdar, Yadav, Arnaud, Ma, Chen, Silwal, Jain, Berges, Wu, Vakil, Abbeel, Malik, Batra, Lin, Maksymets, Rajeswaran, and Meier]{vc2023}
Arjun Majumdar, Karmesh Yadav, Sergio Arnaud, Yecheng~Jason Ma, Claire Chen, Sneha Silwal, Aryan Jain, Vincent-Pierre Berges, Tingfan Wu, Jay Vakil, Pieter Abbeel, Jitendra Malik, Dhruv Batra, Yixin Lin, Oleksandr Maksymets, Aravind Rajeswaran, and Franziska Meier.
\newblock Where are we in the search for an artificial visual cortex for embodied intelligence?
\newblock In \emph{Thirty-seventh Conference on Neural Information Processing Systems}, 2023.

\bibitem[Naselaris et~al.(2011)Naselaris, Kay, Nishimoto, and Gallant]{naselaris2011encoding}
Thomas Naselaris, Kendrick~N Kay, Shinji Nishimoto, and Jack~L Gallant.
\newblock Encoding and decoding in {fMRI}.
\newblock \emph{Neuroimage}, 56\penalty0 (2):\penalty0 400--410, 2011.

\bibitem[Sarch et~al.(2023)Sarch, Tarr, Fragkiadaki, and Wehbe]{sarch2023brain}
Gabriel~Herbert Sarch, Michael~J. Tarr, Katerina Fragkiadaki, and Leila Wehbe.
\newblock Brain dissection: f{MRI}-trained networks reveal spatial selectivity in the processing of natural images.
\newblock In \emph{Thirty-seventh Conference on Neural Information Processing Systems}, 2023.

\bibitem[Toneva and Wehbe(2019)]{Toneva_Improving}
Mariya Toneva and Leila Wehbe.
\newblock \emph{Interpreting and improving natural-language processing (in machines) with natural language-processing (in the brain)}.
\newblock Curran Associates Inc., Red Hook, NY, USA, 2019.

\bibitem[Horikawa and Kamitani(2017)]{horikawa2017generic}
Tomoyasu Horikawa and Yukiyasu Kamitani.
\newblock Generic decoding of seen and imagined objects using hierarchical visual features.
\newblock \emph{Nature Communications}, 8\penalty0 (1):\penalty0 15037, 2017.

\bibitem[Kietzmann et~al.(2019)Kietzmann, Spoerer, S{\"o}rensen, Cichy, Hauk, and Kriegeskorte]{kietzmann2019recurrence}
Tim~C Kietzmann, Courtney~J Spoerer, Lynn~KA S{\"o}rensen, Radoslaw~M Cichy, Olaf Hauk, and Nikolaus Kriegeskorte.
\newblock Recurrence is required to capture the representational dynamics of the human visual system.
\newblock \emph{Proceedings of the National Academy of Sciences}, 116\penalty0 (43):\penalty0 21854--21863, 2019.

\bibitem[G{\"u}{\c{c}}l{\"u} and van Gerven(2015)]{gucclu2015deep}
Umut G{\"u}{\c{c}}l{\"u} and Marcel~AJ van Gerven.
\newblock Deep neural networks reveal a gradient in the complexity of neural representations across the ventral stream.
\newblock \emph{Journal of Neuroscience}, 35\penalty0 (27):\penalty0 10005--10014, 2015.

\bibitem[Groen et~al.(2018)Groen, Greene, Baldassano, Fei-Fei, Beck, and Baker]{groen2018distinct}
Iris~IA Groen, Michelle~R Greene, Christopher Baldassano, Li~Fei-Fei, Diane~M Beck, and Chris~I Baker.
\newblock Distinct contributions of functional and deep neural network features to representational similarity of scenes in human brain and behavior.
\newblock \emph{Elife}, 7, 2018.

\bibitem[Wen et~al.(2018)Wen, Shi, Zhang, Lu, Cao, and Liu]{wen2018neural}
Haiguang Wen, Junxing Shi, Yizhen Zhang, Kun-Han Lu, Jiayue Cao, and Zhongming Liu.
\newblock Neural encoding and decoding with deep learning for dynamic natural vision.
\newblock \emph{Cerebral Cortex}, 28\penalty0 (12):\penalty0 4136--4160, 2018.

\bibitem[Kell et~al.(2018)Kell, Yamins, Shook, Norman-Haignere, and McDermott]{kell2018task}
Alexander~JE Kell, Daniel~LK Yamins, Erica~N Shook, Sam~V Norman-Haignere, and Josh~H McDermott.
\newblock A task-optimized neural network replicates human auditory behavior, predicts brain responses, and reveals a cortical processing hierarchy.
\newblock \emph{Neuron}, 98\penalty0 (3):\penalty0 630--644, 2018.

\bibitem[Koumura et~al.(2019)Koumura, Terashima, and Furukawa]{koumura2019cascaded}
Takuya Koumura, Hiroki Terashima, and Shigeto Furukawa.
\newblock Cascaded tuning to amplitude modulation for natural sound recognition.
\newblock \emph{Journal of Neuroscience}, 39\penalty0 (28):\penalty0 5517--5533, 2019.

\bibitem[Schrimpf et~al.(2021)Schrimpf, Blank, Tuckute, Kauf, Hosseini, Kanwisher, Tenenbaum, and Fedorenko]{schrimpf2021neural}
Martin Schrimpf, Idan~Asher Blank, Greta Tuckute, Carina Kauf, Eghbal~A Hosseini, Nancy Kanwisher, Joshua~B Tenenbaum, and Evelina Fedorenko.
\newblock The neural architecture of language: Integrative modeling converges on predictive processing.
\newblock \emph{Proceedings of the National Academy of Sciences}, 118\penalty0 (45):\penalty0 e2105646118, 2021.

\bibitem[Goldstein et~al.(2022)Goldstein, Zada, Buchnik, Schain, Price, Aubrey, Nastase, Feder, Emanuel, Cohen, et~al.]{goldstein2022shared}
Ariel Goldstein, Zaid Zada, Eliav Buchnik, Mariano Schain, Amy Price, Bobbi Aubrey, Samuel~A Nastase, Amir Feder, Dotan Emanuel, Alon Cohen, et~al.
\newblock Shared computational principles for language processing in humans and deep language models.
\newblock \emph{Nature Neuroscience}, 25\penalty0 (3):\penalty0 369--380, 2022.

\bibitem[Caucheteux and King(2022)]{caucheteux2022brains}
Charlotte Caucheteux and Jean-R{\'e}mi King.
\newblock Brains and algorithms partially converge in natural language processing.
\newblock \emph{Communications Biology}, 5\penalty0 (1):\penalty0 1--10, 2022.

\bibitem[Schmitt et~al.(2021)Schmitt, Erb, Tune, Rysop, Hartwigsen, and Obleser]{schmitt2021predicting}
Lea-Maria Schmitt, Julia Erb, Sarah Tune, Anna~U Rysop, Gesa Hartwigsen, and Jonas Obleser.
\newblock Predicting speech from a cortical hierarchy of event-based time scales.
\newblock \emph{Science Advances}, 7\penalty0 (49):\penalty0 eabi6070, 2021.

\bibitem[Sun et~al.(2023)Sun, Li, Zhang, Moens, Chen, and Wang]{sun2023contrast}
Jingyuan Sun, Mingxiao Li, Yunhao Zhang, Marie-Francine Moens, Zijiao Chen, and Shaonan Wang.
\newblock Contrast, attend and diffuse to decode high-resolution images from brain activities.
\newblock In \emph{Thirty-seventh Conference on Neural Information Processing Systems}, 2023.

\bibitem[Yang et~al.(2024{\natexlab{a}})Yang, Gee, and Shi]{yang2024braindecodesdeepnets}
Huzheng Yang, James Gee, and Jianbo Shi.
\newblock Brain decodes deep nets.
\newblock In \emph{Proceedings of the IEEE/CVF Conference on Computer Vision and Pattern Recognition (CVPR)}, pages 23030--23040, 2024{\natexlab{a}}.

\bibitem[Conwell et~al.(2024)Conwell, Prince, Kay, Alvarez, and Konkle]{conwell_large-scale_2024}
Colin Conwell, Jacob~S. Prince, Kendrick~N. Kay, George~A. Alvarez, and Talia Konkle.
\newblock A large-scale examination of inductive biases shaping high-level visual representation in brains and machines.
\newblock \emph{Nature Communications}, 15\penalty0 (1):\penalty0 9383, 2024.

\bibitem[Chen and Bonner(2024)]{chen2024universaldimensionsvisualrepresentation}
Zirui Chen and Michael~F. Bonner.
\newblock Universal dimensions of visual representation, 2024.

\bibitem[Takagi and Nishimoto(2022)]{takagi2022high}
Yu~Takagi and Shinji Nishimoto.
\newblock High-resolution image reconstruction with latent diffusion models from human brain activity.
\newblock \emph{bioRxiv}, pages 2022--11, 2022.

\bibitem[Zeng et~al.(2024)Zeng, Li, Liu, Gao, Jiang, Tang, Hu, Liu, and Zhang]{zeng2023controllable}
Bohan Zeng, Shanglin Li, Xuhui Liu, Sicheng Gao, Xiaolong Jiang, Xu~Tang, Yao Hu, Jianzhuang Liu, and Baochang Zhang.
\newblock Controllable mind visual diffusion model.
\newblock \emph{Proceedings of the AAAI Conference on Artificial Intelligence}, 38\penalty0 (7):\penalty0 6935--6943, 2024.

\bibitem[Ozcelik and VanRullen(2023)]{ozcelik2023brain}
Furkan Ozcelik and Rufin VanRullen.
\newblock Natural scene reconstruction from {fMRI} signals using generative latent diffusion.
\newblock \emph{Scientific Reports}, 13\penalty0 (1):\penalty0 15666, September 2023.

\bibitem[Lin et~al.(2022)Lin, Sprague, and Singh]{lin2022mind}
Sikun Lin, Thomas~Christopher Sprague, and Ambuj Singh.
\newblock Mind reader: {R}econstructing complex images from brain activities.
\newblock In Alice~H. Oh, Alekh Agarwal, Danielle Belgrave, and Kyunghyun Cho, editors, \emph{Advances in Neural Information Processing Systems}, 2022.

\bibitem[Lu et~al.(2023)Lu, Du, Zhou, Wang, and He]{lu2023minddiffuser}
Yizhuo Lu, Changde Du, Qiongyi Zhou, Dianpeng Wang, and Huiguang He.
\newblock {MindDiffuser:} {C}ontrolled image reconstruction from human brain activity with semantic and structural diffusion.
\newblock In \emph{Proceedings of the 31st ACM International Conference on Multimedia}, page 5899–5908, New York, NY, USA, 2023. Association for Computing Machinery.

\bibitem[Scotti et~al.(2024{\natexlab{a}})Scotti, Banerjee, Goode, Shabalin, Nguyen, Cohen, Dempster, Verlinde, Yundler, Weisberg, Norman, and Abraham]{scotti2023reconstructing}
Paul~S. Scotti, Atmadeep Banerjee, Jimmie Goode, Stepan Shabalin, Alex Nguyen, Ethan Cohen, Aidan~J. Dempster, Nathalie Verlinde, Elad Yundler, David Weisberg, Kenneth~A. Norman, and Tanishq~Mathew Abraham.
\newblock Reconstructing the mind's eye: {fMRI}-to-image with contrastive learning and diffusion priors.
\newblock In \emph{Thirty-seventh Conference on Neural Information Processing Systems}, Red Hook, NY, USA, 2024{\natexlab{a}}. Curran Associates Inc.

\bibitem[Koide-Majima et~al.(2024)Koide-Majima, Nishimoto, and Majima]{koide2023mental}
Naoko Koide-Majima, Shinji Nishimoto, and Kei Majima.
\newblock Mental image reconstruction from human brain activity: {Neural} decoding of mental imagery via deep neural network-based {Bayesian} estimation.
\newblock \emph{Neural Networks}, 170:\penalty0 349--363, 2024.

\bibitem[Gu et~al.(2024)Gu, Jamison, Kuceyeski, and Sabuncu]{gu2022decoding}
Zijin Gu, Keith Jamison, Amy Kuceyeski, and Mert~R. Sabuncu.
\newblock Decoding natural image stimuli from fmri data with a surface-based convolutional network.
\newblock In Ipek Oguz, Jack Noble, Xiaoxiao Li, Martin Styner, Christian Baumgartner, Mirabela Rusu, Tobias Heinmann, Despina Kontos, Bennett Landman, and Benoit Dawant, editors, \emph{Medical Imaging with Deep Learning}, volume 227 of \emph{Proceedings of Machine Learning Research}, pages 107--118. PMLR, 2024.

\bibitem[Chen et~al.(2023{\natexlab{b}})Chen, Qing, Xiang, Yue, and Zhou]{chen2023seeing}
Zijiao Chen, Jiaxin Qing, Tiange Xiang, Wan~Lin Yue, and Juan~Helen Zhou.
\newblock Seeing beyond the brain: Conditional diffusion model with sparse masked modeling for vision decoding.
\newblock In \emph{Proceedings of the IEEE/CVF Conference on Computer Vision and Pattern Recognition}, pages 22710--22720, 2023{\natexlab{b}}.

\bibitem[Ferrante et~al.(2024)Ferrante, Boccato, Ozcelik, VanRullen, and Toschi]{ferrante2023brain}
Matteo Ferrante, Tommaso Boccato, Furkan Ozcelik, Rufin VanRullen, and Nicola Toschi.
\newblock Multimodal decoding of human brain activity into images and text.
\newblock In Marco Fumero, Emanuele Rodolá, Clementine Domine, Francesco Locatello, Karolina Dziugaite, and Caron Mathilde, editors, \emph{Proceedings of UniReps: the First Workshop on Unifying Representations in Neural Models}, volume 243 of \emph{Proceedings of Machine Learning Research}, pages 87--101. PMLR, 2024.

\bibitem[Liu et~al.(2023)Liu, Ma, Zhou, Zhu, and Zheng]{liu2023brainclip}
Yulong Liu, Yongqiang Ma, Wei Zhou, Guibo Zhu, and Nanning Zheng.
\newblock {BrainCLIP}: Bridging brain and visual-linguistic representation via clip for generic natural visual stimulus decoding from {fMRI}.
\newblock \emph{arXiv preprint arXiv:2302.12971}, 2023.

\bibitem[Wang et~al.(2024)Wang, Liu, Tan, and Wang]{wang2024mindbridge}
Shizun Wang, Songhua Liu, Zhenxiong Tan, and Xinchao Wang.
\newblock {MindBridge}: A cross-subject brain decoding framework.
\newblock In \emph{Proceedings of the IEEE/CVF Conference on Computer Vision and Pattern Recognition}, 2024.

\bibitem[Scotti et~al.(2024{\natexlab{b}})Scotti, Tripathy, Torrico, Kneeland, Chen, Narang, Santhirasegaran, Xu, Naselaris, Norman, and Abraham]{scotti2024mindeye2}
Paul~S. Scotti, Mihir Tripathy, Cesar Torrico, Reese Kneeland, Tong Chen, Ashutosh Narang, Charan Santhirasegaran, Jonathan Xu, Thomas Naselaris, Kenneth~A. Norman, and Tanishq~Mathew Abraham.
\newblock {MindEye2}: Shared-subject models enable f{MRI}-to-image with 1 hour of data.
\newblock In \emph{ICLR 2024 Workshop on Representational Alignment}, 2024{\natexlab{b}}.

\bibitem[Kamitani and Tong(2005)]{kamitani2005decoding}
Yukiyasu Kamitani and Frank Tong.
\newblock Decoding the visual and subjective contents of the human brain.
\newblock \emph{Nature Neuroscience}, 8\penalty0 (5):\penalty0 679--685, 2005.

\bibitem[Dado et~al.(2022)Dado, Güçlütürk, Ambrogioni, Ras, Bosch, van Gerven, and Güçlü]{dado_hyperrealistic_2022}
Thirza Dado, Yağmur Güçlütürk, Luca Ambrogioni, Gabriëlle Ras, Sander Bosch, Marcel van Gerven, and Umut Güçlü.
\newblock Hyperrealistic neural decoding for reconstructing faces from {fMRI} activations via the {GAN} latent space.
\newblock \emph{Scientific Reports}, 12\penalty0 (1):\penalty0 141, 2022.

\bibitem[Doerig et~al.(2022)Doerig, Kietzmann, Allen, Wu, Naselaris, Kay, and Charest]{doerig2022semantic}
Adrien Doerig, Tim~C Kietzmann, Emily Allen, Yihan Wu, Thomas Naselaris, Kendrick Kay, and Ian Charest.
\newblock Semantic scene descriptions as an objective of human vision.
\newblock \emph{arXiv preprint arXiv:2209.11737}, 2022.

\bibitem[Yang et~al.(2024{\natexlab{b}})Yang, Gee, and Shi]{yang2024alignedcutvisualconceptsdiscovery}
Huzheng Yang, James Gee, and Jianbo Shi.
\newblock {AlignedCut}: Visual concepts discovery on brain-guided universal feature space, 2024{\natexlab{b}}.

\bibitem[Du et~al.(2022)Du, Cheng, Duan, and Ning]{Du2022fmribci}
B.~Du, X.~Cheng, Y.~Duan, and H.~Ning.
\newblock {fMRI Brain Decoding and Its Applications in Brain-Computer Interface: A Survey}.
\newblock \emph{Brain Sciences}, 12\penalty0 (2):\penalty0 228, 2022.

\bibitem[Efird et~al.(2023)Efird, Murphy, Zylberberg, and Fyshe]{efird2023identifying}
Cory Efird, Alex Murphy, Joel Zylberberg, and Alona Fyshe.
\newblock Identifying shared decodable concepts in the human brain using image-language foundation models.
\newblock \emph{CoRR}, abs/2306.03375, 2023.

\bibitem[Efird et~al.(2024)Efird, Murphy, Zylberberg, and Fyshe]{efird2024s}
Cory Efird, Alex Murphy, Joel Zylberberg, and Alona Fyshe.
\newblock What's the opposite of a face? finding shared decodable concepts and their negations in the brain.
\newblock \emph{arXiv preprint arXiv:2405.17663}, 2024.

\bibitem[Gu et~al.(2022)Gu, Jamison, Khosla, Allen, Wu, St-Yves, Naselaris, Kay, Sabuncu, and Kuceyeski]{gu2022neurogen}
Zijin Gu, Keith~Wakefield Jamison, Meenakshi Khosla, Emily~J Allen, Yihan Wu, Ghislain St-Yves, Thomas Naselaris, Kendrick Kay, Mert~R Sabuncu, and Amy Kuceyeski.
\newblock {NeuroGen}: activation optimized image synthesis for discovery neuroscience.
\newblock \emph{NeuroImage}, 247:\penalty0 118812, 2022.

\bibitem[Chen et~al.(2014)Chen, Han, Hu, Jiang, Guo, and Liu]{chen2014survey}
Min Chen, Jiawei Han, Xiaoping Hu, Xia Jiang, Lei Guo, and Tianming Liu.
\newblock Survey of encoding and decoding of visual stimulus via fmri: an image analysis perspective.
\newblock \emph{Brain Imaging and Behavior}, 8\penalty0 (1):\penalty0 7--23, Mar 2014.
\newblock \doi{10.1007/s11682-013-9238-z}.

\bibitem[Du et~al.(2019)Du, Li, Huang, and He]{DU2019948}
Changde Du, Jinpeng Li, Lijie Huang, and Huiguang He.
\newblock Brain encoding and decoding in {fMRI} with bidirectional deep generative models.
\newblock \emph{Engineering}, 5\penalty0 (5):\penalty0 948--953, 2019.

\bibitem[Meng and Ge(2022)]{brainsci12101394}
Lu~Meng and Kang Ge.
\newblock Decoding visual {fMRI} stimuli from human brain based on graph convolutional neural network.
\newblock \emph{Brain Sciences}, 12\penalty0 (10), 2022.

\bibitem[Rastegarnia et~al.(2023)Rastegarnia, St-Laurent, DuPre, Pinsard, and Bellec]{RASTEGARNIA2023120395}
Shima Rastegarnia, Marie St-Laurent, Elizabeth DuPre, Basile Pinsard, and Pierre Bellec.
\newblock Brain decoding of the human connectome project tasks in a dense individual fmri dataset.
\newblock \emph{NeuroImage}, 283:\penalty0 120395, 2023.

\bibitem[Akamatsu et~al.(2021)Akamatsu, Harakawa, Ogawa, and Haseyama]{9349437}
Yusuke Akamatsu, Ryosuke Harakawa, Takahiro Ogawa, and Miki Haseyama.
\newblock Perceived image decoding from brain activity using shared information of multi-subject fmri data.
\newblock \emph{IEEE Access}, 9:\penalty0 26593--26606, 2021.
\newblock \doi{10.1109/ACCESS.2021.3057800}.

\bibitem[Prince et~al.(2024)Prince, Alvarez, and Konkle]{Prince2023.08.04.551888}
Jacob~S. Prince, George~A. Alvarez, and Talia Konkle.
\newblock Contrastive learning explains the emergence and function of visual category-selective regions.
\newblock \emph{Science Advances}, 10\penalty0 (39):\penalty0 eadl1776, 2024.

\bibitem[Chang et~al.(2019)Chang, Pyles, Marcus, Gupta, Tarr, and Aminoff]{chang2019bold5000}
Nadine Chang, John~A Pyles, Austin Marcus, Abhinav Gupta, Michael~J Tarr, and Elissa~M Aminoff.
\newblock Bold5000, a public {fMRI} dataset while viewing 5000 visual images.
\newblock \emph{Scientific Data}, 6\penalty0 (1):\penalty0 1--18, 2019.

\bibitem[Allen et~al.(2022)Allen, St-Yves, Wu, Breedlove, Prince, Dowdle, Nau, Caron, Pestilli, Charest, Hutchinson, Naselaris, and Kay]{Allen2022}
Emily~J. Allen, Ghislain St-Yves, Yihan Wu, Jesse~L. Breedlove, Jacob~S. Prince, Logan~T. Dowdle, Matthias Nau, Brad Caron, Franco Pestilli, Ian Charest, J.~Benjamin Hutchinson, Thomas Naselaris, and Kendrick Kay.
\newblock A massive 7t {fMRI} dataset to bridge cognitive neuroscience and artificial intelligence.
\newblock \emph{Nature Neuroscience}, 25:\penalty0 116--126, 2022.

\bibitem[Rombach et~al.(2022)Rombach, Blattmann, Lorenz, Esser, and Ommer]{rombach2022high}
Robin Rombach, Andreas Blattmann, Dominik Lorenz, Patrick Esser, and Bj{\"o}rn Ommer.
\newblock High-resolution image synthesis with latent diffusion models.
\newblock In \emph{Proceedings of the IEEE/CVF Conference on Computer Vision and Pattern Recognition}, pages 10684--10695, 2022.

\bibitem[Maunsell and Van~Essen(1983)]{maunsell1983functional}
John~HR Maunsell and David~C Van~Essen.
\newblock Functional properties of neurons in middle temporal visual area of the macaque monkey. i. selectivity for stimulus direction, speed, and orientation.
\newblock \emph{Journal of Neurophysiology}, 49\penalty0 (5):\penalty0 1127--1147, 1983.

\bibitem[Duffy and Wurtz(1991)]{duffy1991sensitivity}
Charles~J Duffy and Robert~H Wurtz.
\newblock Sensitivity of mst neurons to optic flow stimuli. i. a continuum of response selectivity to large-field stimuli.
\newblock \emph{Journal of Neurophysiology}, 65\penalty0 (6):\penalty0 1329--1345, 1991.

\bibitem[Robert et~al.(2023)Robert, Ungerleider, and Vaziri-Pashkam]{Robert621}
Sophia Robert, Leslie~G. Ungerleider, and Maryam Vaziri-Pashkam.
\newblock Disentangling object category representations driven by dynamic and static visual input.
\newblock \emph{Journal of Neuroscience}, 43\penalty0 (4):\penalty0 621--634, 2023.

\bibitem[Heeger et~al.(1999)Heeger, Boynton, Demb, Seidemann, and Newsome]{heeger1999motion}
David~J Heeger, Geoffrey~M Boynton, Josh~B Demb, Eyal Seidemann, and William~T Newsome.
\newblock Motion opponency in visual cortex.
\newblock \emph{Journal of Neuroscience}, 19\penalty0 (16):\penalty0 7162--7174, 1999.

\bibitem[Heuer and Britten(2004)]{heuer2004optic}
Hilary~W Heuer and Kenneth~H Britten.
\newblock Optic flow signals in extrastriate area mst: comparison of perceptual and neuronal sensitivity.
\newblock \emph{Journal of Neurophysiology}, 91\penalty0 (3):\penalty0 1314--1326, 2004.
\newblock Epub 2003 Oct 8.

\bibitem[Orban et~al.(1992)Orban, Lagae, Verri, Raiguel, Xiao, Maes, and Torre]{orban1992first}
Guy~A Orban, Luc Lagae, Alessandro Verri, Suzanne Raiguel, Dong Xiao, Hans Maes, and Vincent Torre.
\newblock First-order analysis of optical flow in monkey brain.
\newblock \emph{Proceedings of the National Academy of Sciences of the United States of America}, 89\penalty0 (7):\penalty0 2595--2599, 1992.

\bibitem[Li et~al.(2024)Li, Qian, Wang, Huo, Xue, Fu, and Feng]{li2024enhance}
Chong Li, Xuelin Qian, Yun Wang, Jingyang Huo, Xiangyang Xue, Yanwei Fu, and Jianfeng Feng.
\newblock Enhancing cross-subject {fMRI}-to-video decoding with global-local functional alignment.
\newblock pages 353--369, 11 2024.

\bibitem[Oquab et~al.(2024)Oquab, Darcet, Moutakanni, Vo, Szafraniec, Khalidov, Fernandez, HAZIZA, Massa, El-Nouby, Assran, Ballas, Galuba, Howes, Huang, Li, Misra, Rabbat, Sharma, Synnaeve, Xu, Jegou, Mairal, Labatut, Joulin, and Bojanowski]{oquab2023dinov2}
Maxime Oquab, Timoth{\'e}e Darcet, Th{\'e}o Moutakanni, Huy~V. Vo, Marc Szafraniec, Vasil Khalidov, Pierre Fernandez, Daniel HAZIZA, Francisco Massa, Alaaeldin El-Nouby, Mido Assran, Nicolas Ballas, Wojciech Galuba, Russell Howes, Po-Yao Huang, Shang-Wen Li, Ishan Misra, Michael Rabbat, Vasu Sharma, Gabriel Synnaeve, Hu~Xu, Herve Jegou, Julien Mairal, Patrick Labatut, Armand Joulin, and Piotr Bojanowski.
\newblock {DINO}v2: Learning robust visual features without supervision.
\newblock \emph{Transactions on Machine Learning Research}, 2024.

\bibitem[Xie et~al.(2024)Xie, Yang, Xie, and Zisserman]{xie2024flowsam}
Junyu Xie, Charig Yang, Weidi Xie, and Andrew Zisserman.
\newblock Moving object segmentation: All you need is {SAM} (and flow).
\newblock \emph{CoRR}, abs/2404.12389, 2024.

\bibitem[Wang et~al.(2022)Wang, Yan, Huang, Li, Wang, Fan, Sheng, Liu, Li, and Chen]{wang2022reconstructing}
Cheng Wang, Haonan Yan, Wei Huang, Jiaqi Li, Ying Wang, Yunzhe~S Fan, Wei Sheng, Ting Liu, Rong Li, and Haibo Chen.
\newblock Reconstructing rapid natural vision with fmri-conditional video generative adversarial network.
\newblock \emph{Cerebral Cortex}, 32\penalty0 (20):\penalty0 4502--4511, 2022.

\bibitem[Kupershmidt et~al.(2022)Kupershmidt, Beliy, Gaziv, and Irani]{kupershmidt2022pennyvisualthoughtsselfsupervised}
Ganit Kupershmidt, Roman Beliy, Guy Gaziv, and Michal Irani.
\newblock A penny for your (visual) thoughts: Self-supervised reconstruction of natural movies from brain activity.
\newblock 2022.

\bibitem[Han et~al.(2015)Han, Chen, Shao, Hu, Han, and Liu]{6919270}
Junwei Han, Changyuan Chen, Ling Shao, Xintao Hu, Jungong Han, and Tianming Liu.
\newblock Learning computational models of video memorability from fmri brain imaging.
\newblock \emph{IEEE Transactions on Cybernetics}, 45\penalty0 (8):\penalty0 1692--1703, 2015.

\bibitem[Gong et~al.(2024)Gong, Bao, Zhang, Wan, Miao, Wang, Zhu, Wang, Xu, Hu, Liu, and Zhang]{gong2024neuroclipshighfidelitysmoothfmritovideo}
Zixuan Gong, Guangyin Bao, Qi~Zhang, Zhongwei Wan, Duoqian Miao, Shoujin Wang, Lei Zhu, Changwei Wang, Rongtao Xu, Liang Hu, Ke~Liu, and Yu~Zhang.
\newblock {NeuroClips}: Towards high-fidelity and smooth f{MRI}-to-video reconstruction.
\newblock In \emph{The Thirty-eighth Annual Conference on Neural Information Processing Systems}, 2024.

\bibitem[Hu et~al.(2015)Hu, Lv, Cheng, Lv, Guo, Han, and Liu]{7056490}
Xintao Hu, Cheng Lv, Gong Cheng, Jinglei Lv, Lei Guo, Junwei Han, and Tianming Liu.
\newblock Sparsity-constrained fmri decoding of visual saliency in naturalistic video streams.
\newblock \emph{IEEE Transactions on Autonomous Mental Development}, 7\penalty0 (2):\penalty0 65--75, 2015.

\bibitem[Teed and Deng(2020)]{teed2020raft}
Zachary Teed and Jia Deng.
\newblock {RAFT:} recurrent all-pairs field transforms for optical flow.
\newblock In \emph{Computer Vision – ECCV 2020: 16th European Conference, Glasgow, UK, August 23–28, 2020, Proceedings, Part II}, page 402–419, Berlin, Heidelberg, 2020. Springer-Verlag.

\bibitem[Walker et~al.(2015)Walker, Gupta, and Hebert]{walker2015dense}
Jacob Walker, Abhinav Gupta, and Martial Hebert.
\newblock Dense optical flow prediction from a static image.
\newblock In \emph{2015 IEEE International Conference on Computer Vision (ICCV)}, pages 2443--2451, 2015.

\bibitem[Eslami et~al.(2024)Eslami, Arefi, Mansourian, and Kasaei]{eslami2024rethinkingraftefficientoptical}
Navid Eslami, Farnoosh Arefi, Amir~M. Mansourian, and Shohreh Kasaei.
\newblock Rethinking {RAFT} for efficient optical flow.
\newblock \emph{CoRR}, abs/2401.00833, 2024.

\bibitem[He et~al.(2016)He, Zhang, Ren, and Sun]{he2015resnet}
Kaiming He, Xiangyu Zhang, Shaoqing Ren, and Jian Sun.
\newblock Deep residual learning for image recognition.
\newblock In \emph{2016 IEEE Conference on Computer Vision and Pattern Recognition (CVPR)}, pages 770--778, 2016.

\bibitem[Caron et~al.(2021)Caron, Touvron, Misra, J\'egou, Mairal, Bojanowski, and Joulin]{caron2021dino}
Mathilde Caron, Hugo Touvron, Ishan Misra, Herv\'e J\'egou, Julien Mairal, Piotr Bojanowski, and Armand Joulin.
\newblock Emerging properties in self-supervised vision transformers.
\newblock In \emph{Proceedings of the International Conference on Computer Vision (ICCV)}, 2021.

\bibitem[Radford et~al.(2021)Radford, Kim, Hallacy, Ramesh, Goh, Agarwal, Sastry, Askell, Mishkin, Clark, Krueger, and Sutskever]{Radford2021clip}
Alec Radford, Jong~Wook Kim, Chris Hallacy, A.~Ramesh, Gabriel Goh, Sandhini Agarwal, Girish Sastry, Amanda Askell, Pamela Mishkin, Jack Clark, Gretchen Krueger, and Ilya Sutskever.
\newblock Learning transferable visual models from natural language supervision.
\newblock In \emph{ICML}, 2021.

\bibitem[Cherti et~al.(2023)Cherti, Beaumont, Wightman, Wortsman, Ilharco, Gordon, Schuhmann, Schmidt, and Jitsev]{cherti2023reproducible}
Mehdi Cherti, Romain Beaumont, Ross Wightman, Mitchell Wortsman, Gabriel Ilharco, Cade Gordon, Christoph Schuhmann, Ludwig Schmidt, and Jenia Jitsev.
\newblock Reproducible scaling laws for contrastive language-image learning.
\newblock In \emph{Proceedings of the IEEE/CVF Conference on Computer Vision and Pattern Recognition}, pages 2818--2829, 2023.

\bibitem[Ryali et~al.(2023)Ryali, Hu, Bolya, Wei, Fan, Huang, Aggarwal, Chowdhury, Poursaeed, Hoffman, Malik, Li, and Feichtenhofer]{ryali2023hiera}
Chaitanya Ryali, Yuan-Ting Hu, Daniel Bolya, Chen Wei, Haoqi Fan, Po-Yao Huang, Vaibhav Aggarwal, Arkabandhu Chowdhury, Omid Poursaeed, Judy Hoffman, Jitendra Malik, Yanghao Li, and Christoph Feichtenhofer.
\newblock Hiera: A hierarchical vision transformer without the bells-and-whistles.
\newblock \emph{ICML}, 2023.

\bibitem[Ma et~al.(2022{\natexlab{a}})Ma, Xu, Sun, Yan, Zhang, and Ji]{Ma2022XCLIP}
Yiwei Ma, Guohai Xu, Xiaoshuai Sun, Ming Yan, Ji~Zhang, and Rongrong Ji.
\newblock {X-CLIP:}: End-to-end multi-grained contrastive learning for video-text retrieval.
\newblock \emph{arXiv preprint arXiv:2207.07285}, 2022{\natexlab{a}}.

\bibitem[Ma et~al.(2022{\natexlab{b}})Ma, Sodhani, Jayaraman, Bastani, Kumar, and Zhang]{ma2023vip}
Yecheng~Jason Ma, Shagun Sodhani, Dinesh Jayaraman, Osbert Bastani, Vikash Kumar, and Amy Zhang.
\newblock {VIP}: Towards universal visual reward and representation via value-implicit pre-training.
\newblock In \emph{5th Robot Learning Workshop: Trustworthy Robotics @ Thirty-sixth Conference on Neural Information Processing Systems}, 2022{\natexlab{b}}.

\bibitem[Nair et~al.(2023)Nair, Rajeswaran, Kumar, Finn, and Gupta]{nair2022r3m}
Suraj Nair, Aravind Rajeswaran, Vikash Kumar, Chelsea Finn, and Abhinav Gupta.
\newblock {R3M}: A universal visual representation for robot manipulation.
\newblock In Karen Liu, Dana Kulic, and Jeff Ichnowski, editors, \emph{Proceedings of The 6th Conference on Robot Learning}, volume 205 of \emph{Proceedings of Machine Learning Research}, pages 892--909. PMLR, 2023.

\bibitem[Wen et~al.(2017)Wen, Shi, Zhang, Lu, Cao, and Liu]{wen_dynamic_vision}
Haiguang Wen, Junxing Shi, Yizhen Zhang, Kun-Han Lu, Jiayue Cao, and Zhongming Liu.
\newblock {Neural Encoding and Decoding with Deep Learning for Dynamic Natural Vision}.
\newblock \emph{Cerebral Cortex}, 28\penalty0 (12):\penalty0 4136--4160, 2017.

\bibitem[Kay et~al.(2008)Kay, Naselaris, Prenger, and Gallant]{Kay2008}
Kendrick~N Kay, Thomas Naselaris, Ryan~J Prenger, and Jack~L Gallant.
\newblock Identifying natural images from human brain activity.
\newblock \emph{Nature}, 452\penalty0 (7185):\penalty0 352--355, 2008.
\newblock Epub 2008 Mar 5.

\bibitem[Baker et~al.(2018{\natexlab{a}})Baker, Burks, Briggs, Conner, Glenn, Taylor, Sali, McCoy, Battiste, O'Donoghue, and Sughrue]{Baker2018c}
C.~M. Baker, J.~D. Burks, R.~G. Briggs, A.~K. Conner, C.~A. Glenn, K.~N. Taylor, G.~Sali, T.~M. McCoy, J.~D. Battiste, D.~L. O'Donoghue, and M.~E. Sughrue.
\newblock A connectomic atlas of the human cerebrum-chapter 7: The lateral parietal lobe.
\newblock \emph{Operative Neurosurgery (Hagerstown)}, 15\penalty0 (suppl\_1):\penalty0 S295--S349, 2018{\natexlab{a}}.

\bibitem[Baker et~al.(2018{\natexlab{b}})Baker, Burks, Briggs, Sheets, Conner, Glenn, Sali, McCoy, Battiste, O'Donoghue, and Sughrue]{Baker2018}
C.~M. Baker, J.~D. Burks, R.~G. Briggs, J.~R. Sheets, A.~K. Conner, C.~A. Glenn, G.~Sali, T.~M. McCoy, J.~D. Battiste, D.~L. O'Donoghue, and M.~E. Sughrue.
\newblock A connectomic atlas of the human cerebrum-chapter 3: The motor, premotor, and sensory cortices.
\newblock \emph{Operative Neurosurgery (Hagerstown)}, 15\penalty0 (suppl\_1):\penalty0 S75--S121, 2018{\natexlab{b}}.

\bibitem[Baker et~al.(2018{\natexlab{c}})Baker, Burks, Briggs, Stafford, Conner, Glenn, Sali, McCoy, Battiste, O'Donoghue, and Sughrue]{Baker2018b}
C.~M. Baker, J.~D. Burks, R.~G. Briggs, J.~Stafford, A.~K. Conner, C.~A. Glenn, G.~Sali, T.~M. McCoy, J.~D. Battiste, D.~L. O'Donoghue, and M.~E. Sughrue.
\newblock A connectomic atlas of the human cerebrum-chapter 4: The medial frontal lobe, anterior cingulate gyrus, and orbitofrontal cortex.
\newblock \emph{Operative Neurosurgery (Hagerstown)}, 15\penalty0 (suppl\_1):\penalty0 S122--S174, 2018{\natexlab{c}}.

\bibitem[Huang et~al.(2023)Huang, Sun, Yousefnezhad, Wang, and Zhang]{Huang2023functional}
S.~Huang, L.~Sun, M.~Yousefnezhad, M.~Wang, and D.~Zhang.
\newblock Functional alignment-auxiliary generative adversarial network-based visual stimuli reconstruction via multi-subject {fMRI}.
\newblock \emph{IEEE Transactions on Neural Systems and Rehabilitation Engineering}, 31:\penalty0 2715--2725, 2023.
\newblock Epub 2023 Jun 20.

\bibitem[Andreella et~al.(2023)Andreella, Finos, and Lindquist]{andreella2023functional}
Angela Andreella, Livio Finos, and Martin~A Lindquist.
\newblock Enhanced hyperalignment via spatial prior information.
\newblock \emph{Human Brain Mapping}, 44\penalty0 (4):\penalty0 1725--1740, 2023.

\bibitem[Glasser et~al.(2016)Glasser, Coalson, Robinson, Hacker, Harwell, Yacoub, Ugurbil, Andersson, Beckmann, Jenkinson, et~al.]{glasser2016multi}
Matthew~F Glasser, Timothy~S Coalson, Emma~C Robinson, Carl~D Hacker, John Harwell, Essa Yacoub, Kamil Ugurbil, Jesper Andersson, Christian~F Beckmann, Mark Jenkinson, et~al.
\newblock A multi-modal parcellation of human cerebral cortex.
\newblock \emph{Nature}, 536\penalty0 (7615):\penalty0 171--178, 2016.

\bibitem[Huang et~al.(2022)Huang, Rolls, Feng, and Lin]{huang_extended_2022}
Chu-Chung Huang, Edmund~T. Rolls, Jianfeng Feng, and Ching-Po Lin.
\newblock An extended {Human} {Connectome} {Project} multimodal parcellation atlas of the human cortex and subcortical areas.
\newblock \emph{Brain Structure and Function}, 227\penalty0 (3):\penalty0 763--778, 2022.

\bibitem[Rolls et~al.(2022)Rolls, Deco, Huang, and Feng]{hcp_parcellation_map}
Edmund~T. Rolls, Gustavo Deco, Chu-Chung Huang, and Jianfeng Feng.
\newblock The human language effective connectome.
\newblock \emph{NeuroImage}, 258:\penalty0 119352, 2022.

\end{thebibliography}
\onecolumn
\appendix

\setcounter{section}{0}

\section{Supplemental}
\renewcommand\thefigure{S\arabic{figure}}    
\setcounter{figure}{0}
\renewcommand\thetable{S\arabic{table}}   
\setcounter{table}{0}
\textbf{\large Sections}
\begin{enumerate}
    \item HCP Parcellation Map~(\ref{section: supplemental hcp})
    \item Additional Dataset Information~(\ref{section: supplemental additional dataset information})
    \item Decoding Models: Visualizations~(\ref{section: supplemental decoding models})
    \item Encoding Models: Controls and Baselines ~(\ref{section: supplemental encoding models controls})
    \item Encoding Models: Voxel-wise Prediction Performance on Inflated Cortex ~(\ref{section: supplemental encoding models})
\end{enumerate}
\clearpage

\begin{figure*}[h]
    \centering
    \includegraphics[width=0.65\textwidth]{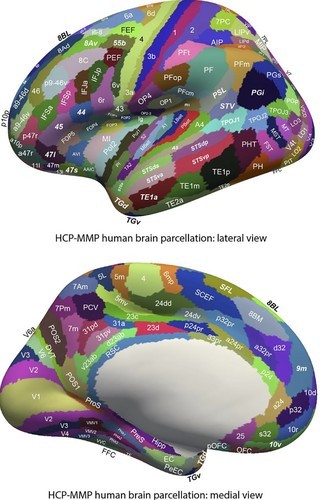}
    \caption{\textbf{HCP Parcellation Map.} The human cortex parceled into regions of interest as defined by the HCP-MMP~\cite{glasser2016multi} atlas and HCPex~\cite{huang_extended_2022}. Regions were defined through multimodal imaging and take into account neural structure, connectivity, and task-related selectivity. This figure is a reproduction of a subset of Figure 1 from ~\citet{hcp_parcellation_map}.
    }
    \label{fig:hcp parcellation}
\end{figure*}

\newcommand{\brainmaptext}[1]{\textbf{#1 Voxel-wise Prediction Performance.} Voxel-wise fMRI encoding accuracy of #1 quantified as the Pearson correlation between measured and predicted responses. Refer to Figure~\ref{fig:hcp parcellation} for a labeled parcellation of the human cortex.}

\subsection{HCP Parcellation Map}
\label{section: supplemental hcp}

\subsection{Additional Dataset Information}
\label{section: supplemental additional dataset information}
We use the Dynamic Vision Dataset collected by~\citet{wen_dynamic_vision}. The dataset is released under a CC0 1.0 universal license and can be found at \url{https://purr.purdue.edu/publications/2805/1}.

\subsection{Decoding Models: Visualizations}
\label{section: supplemental decoding models}
Our website \url{https://brain-nrds.github.io/} has multiple reanimation examples as videos. 
For each example, we show the ground truth video, the reanimated video using the flow predicted from the ground truth initial frame, and the reanimated video using the flow predicted from the initial frame generated by MindVideo~\cite{chen2023cinematic}. The MindVideo initial frame represents the diffusion image as predicted by MindVideo from fMRI data. As such, this frame is often quite different in content from the ground truth, although the objects in each are in similar locations. However, the motion predicted by our model is still consistent with ground truth motion, for example, in the first video the jellyfish retracts backwards at the end clip. The same backwards motion can be seen in the reanimated videos for with both the ground truth initial frame and the MindVideo generated initial frame (which happens to be a boat). This establishes that the motion visualized using DragNUWA~\cite{yin2023dragnuwa} is not derived solely from the diffusion model, but rather, incorporates our predicted motion.

In addition, we display example generated video frames below. Within each figure the first row is the ground truth frame, the second row is our reanimated video using the flow predicted from the ground truth initial frame, and the final row is the reanimated video using the flow predicted from the initial frame generated by MindVideo.

\begin{figure}[h]
    \centering
    \includegraphics[width=0.97\textwidth]{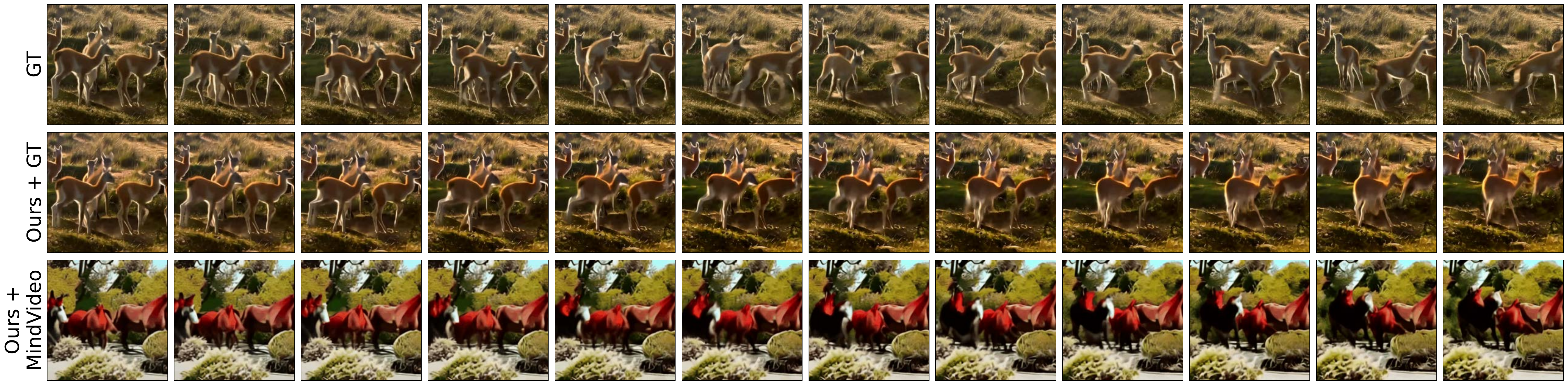}
    \caption{\textbf{Static image animation results.} Here we see an example of deer running to the right in the ground truth video (\textbf{first row}, ``GT"). We show an example of animating the initial frame of the ground truth video by combining the brain conditioned motion prediction with DragNUWA~\cite{yin2023dragnuwa} (\textbf{second row}, ``Ours + GT"). We show an example of animating the initial frame obtained from fMRI data using MindVideo by combining the brain conditioned motion prediction with DragNUWA (\textbf{third row}, ``Ours + MindVideo") and observe that the creatures run to the right.}
    \label{fig:supplemental decoded example deer}
    \vspace{-.5cm}
\end{figure}

\begin{figure}[h]
    \centering
    \includegraphics[width=0.97\textwidth]{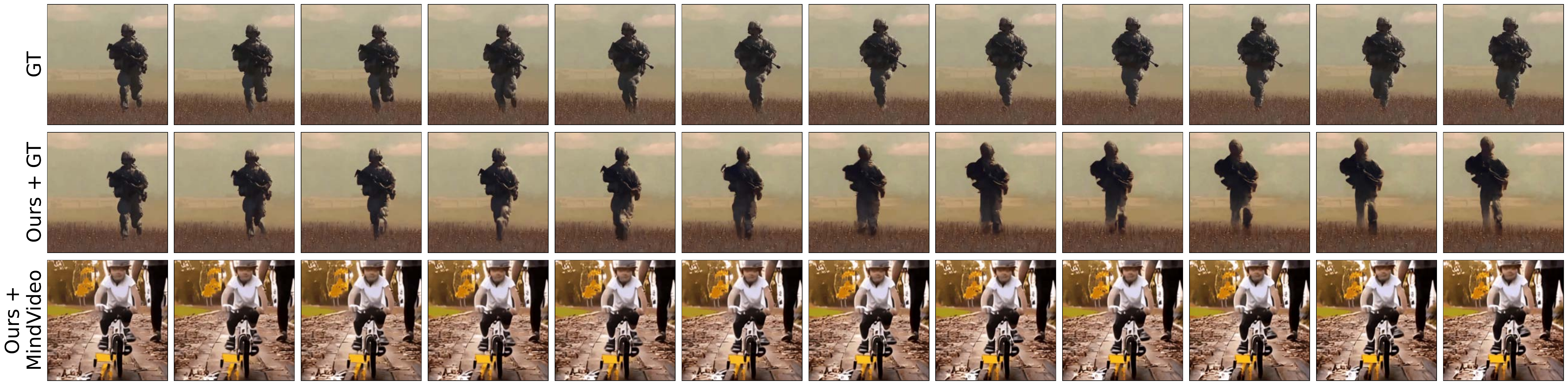}
    \caption{\textbf{Static image animation results.} Here we see an example of a soldier running to the left in the ground truth video (\textbf{first row}, ``GT"). We show an example of animating the initial frame of the ground truth video by combining the brain conditioned motion prediction with DragNUWA~\cite{yin2023dragnuwa} (\textbf{second row}, ``Ours + GT"). We show an example of animating the initial frame obtained from fMRI data using MindVideo by combining the brain conditioned motion prediction with DragNUWA (\textbf{third row}, ``Ours + MindVideo") and observe that the child pedals its feet in a similar motion to that of the soldier running.}
    \label{fig:supplemental decoded example soldier}
    \vspace{-.5cm}
\end{figure}

\newpage
\subsection{Encoding Models: Controls and Baselines}
\vspace{-1em}
\label{section: supplemental encoding models controls}
\begin{figure}[h]
    \centering
    \begin{subfigure}[b]{0.55\textwidth}
        \centering
        \includegraphics[width=1.0\textwidth]{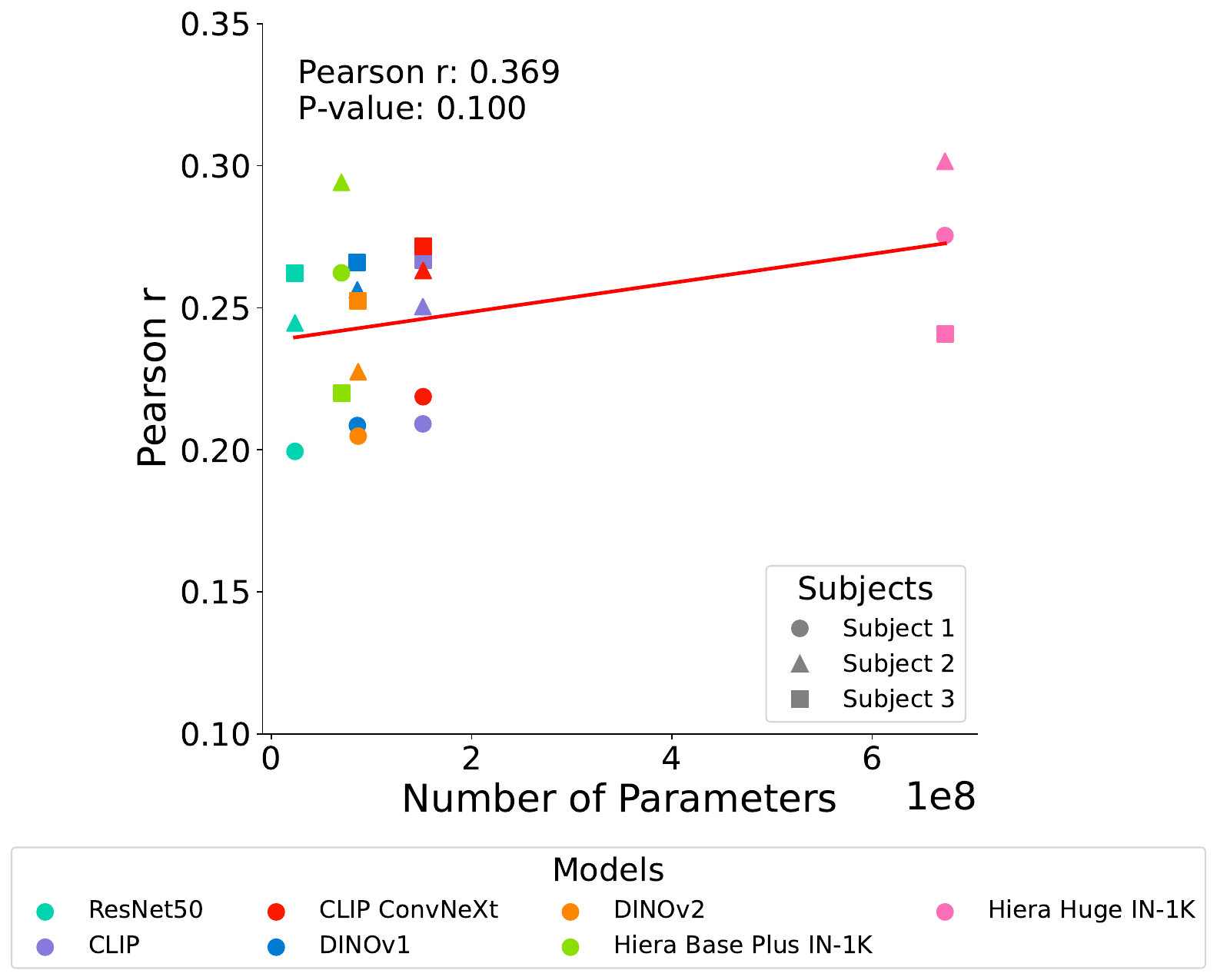}
        \caption{\textbf{Image Models} }
        \label{fig:decoding_cossim}
    \end{subfigure}\hfill

    \vspace{1em}
    
    \begin{subfigure}[b]{0.51\textwidth}
        \centering
        \includegraphics[width=1.0\textwidth]{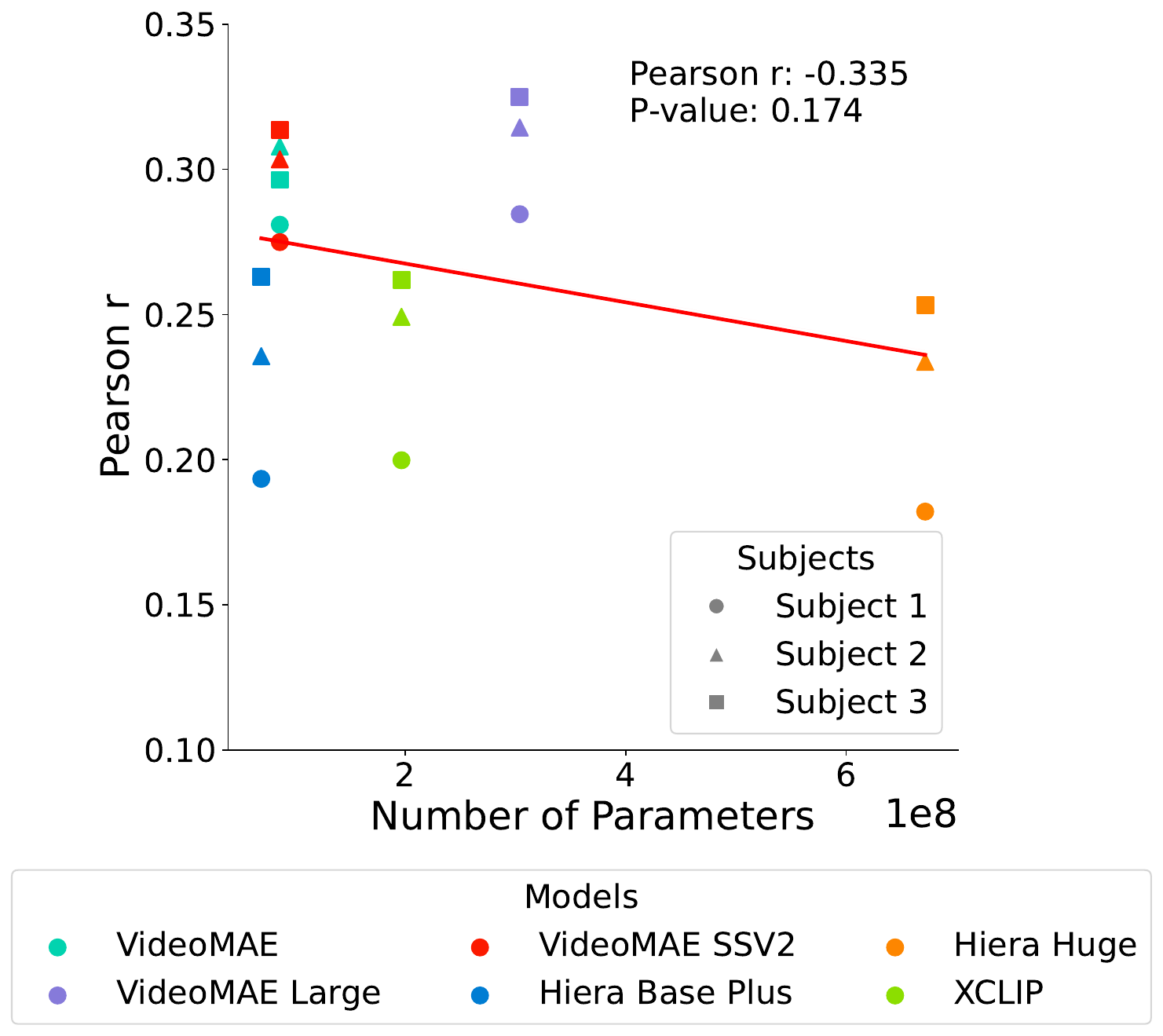}
        \caption{\textbf{Video Models} }
        \label{fig:decoding_cossim}
    \end{subfigure}\hfill
    \begin{subfigure}[b]{0.43\textwidth}
        \centering
        \includegraphics[width=0.9\textwidth]{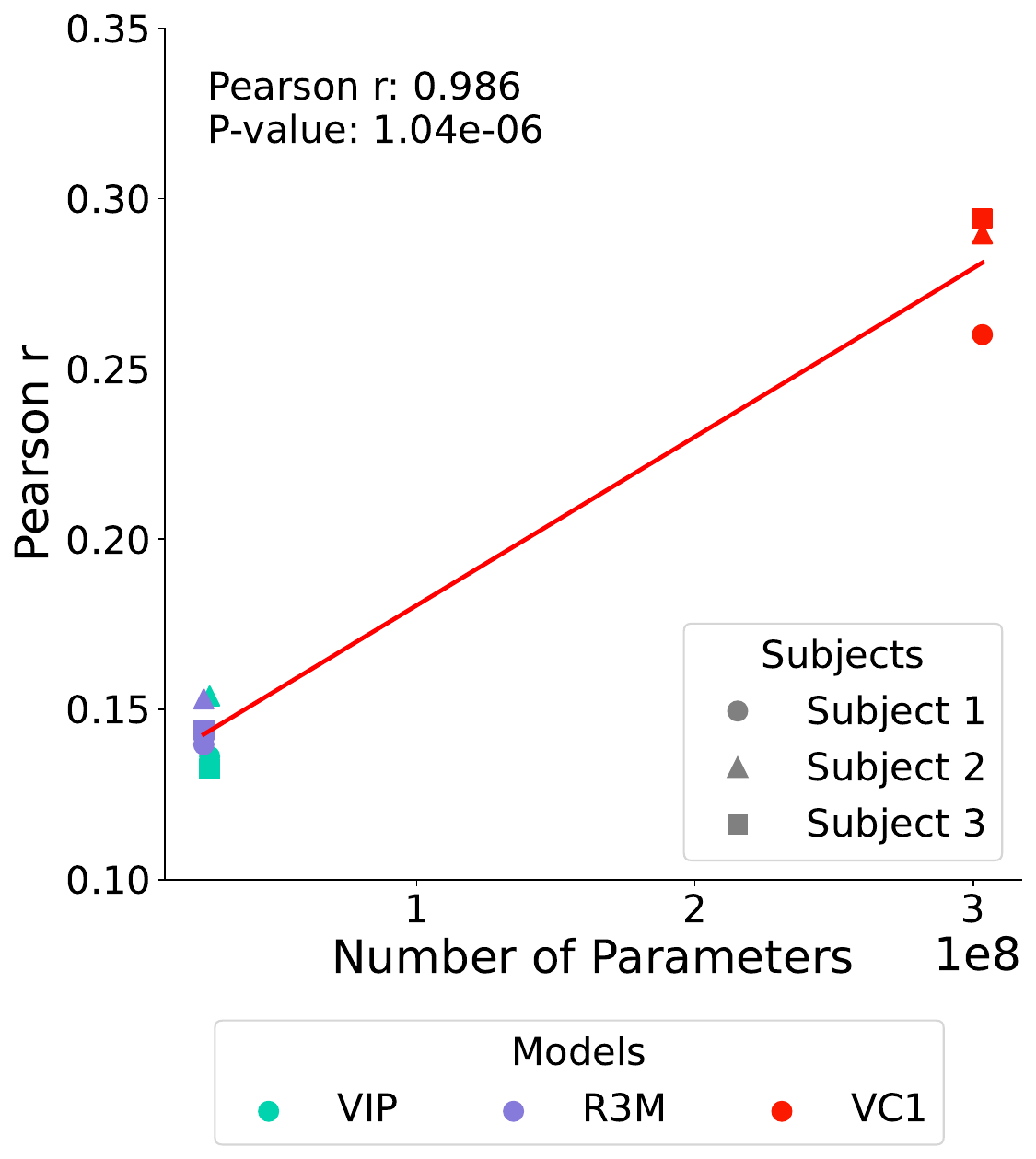}
        \caption{\textbf{Embodied AI Models}}
        \label{fig:decoding_epe}
    \end{subfigure}
    \caption{\textbf{Model size and encoding prediction performance.} Encoding model features of the viewed stimuli are used to predict voxel-wise fMRI brain activity~\cite{naselaris2011encoding}. The average Pearson-$r$ is plotted across all voxels for each model and with respect to the number of parameters in each model. \textbf{(a)} Encoding performance for models trained on static images.  \textbf{(b)} Encoding performance for models trained on videos. \textbf{(c)} Encoding performance for models trained to align representations of single frames across time for embodied AI visuomotor manipulation. 
    Model size and encoding prediction performance are not significantly correlated (by statistical test) for image and video models. This indicates that the model size is not a confound for the encoding performance of image and video models. In contrast, model size and encoding performance are significantly correlated ($p < 0.001$) for the embodied AI visual models. Encoding performance per model is plotted for each of the three participants -- the circle marker refers to S1, the triangle marker refers to S2, and the square marker refers to S3. Comparisons between models and model sizes are a critical step in building better and more interpretable models for understanding the human brain. Model architectures and training tasks, as well as the number of parameters and other model characteristics, instantiate varying constraints that have functional implications for brain prediction~\cite{Wang2023machineintelligence,conwell_large-scale_2024}.}
    \label{fig:supplemental-param-control}
\end{figure}

\begin{figure}[h]
    \centering
    \begin{subfigure}[b]{0.56\textwidth}
        \centering
        \includegraphics[width=1.0\textwidth]{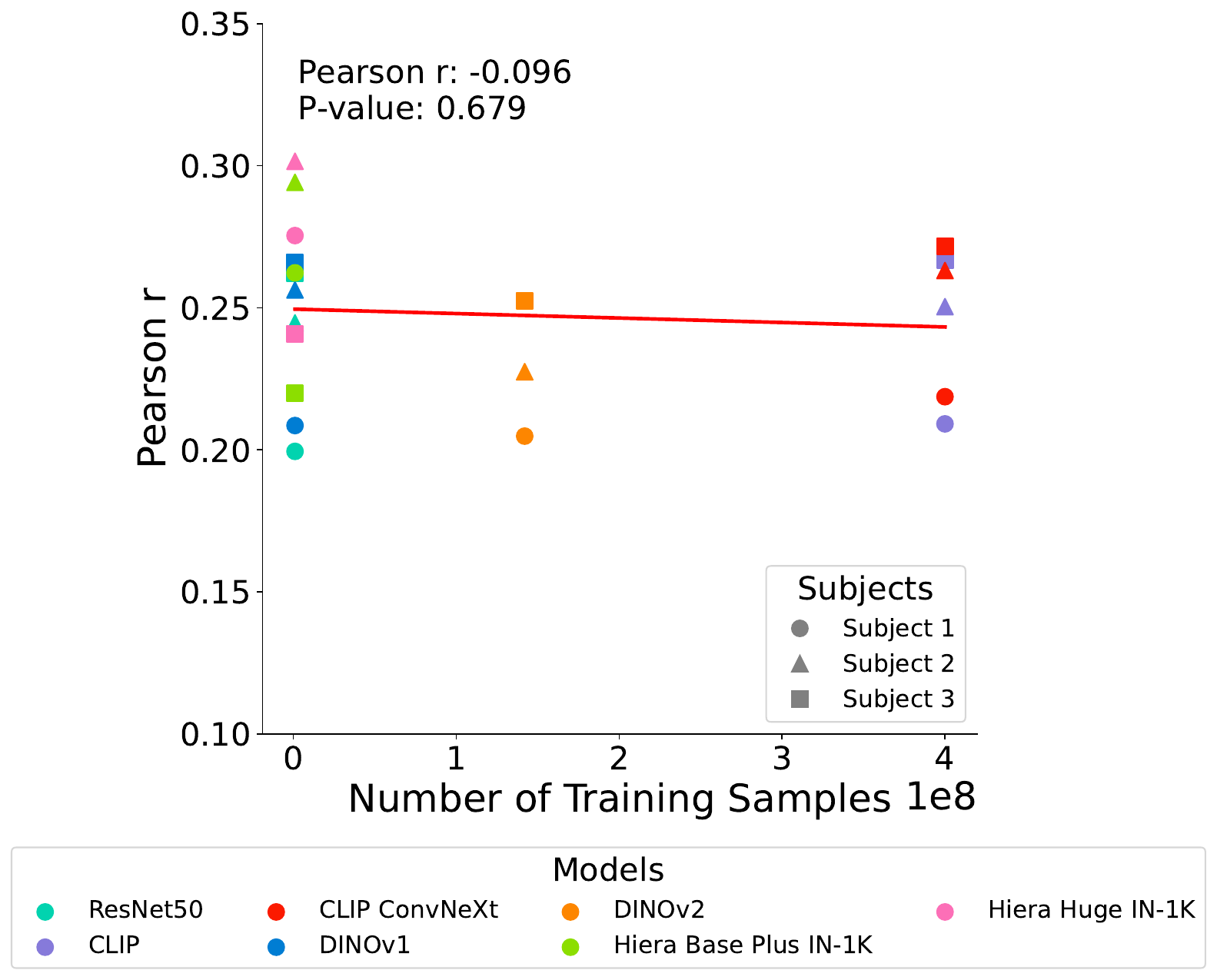}
        \caption{\textbf{Image Models} }
        \label{fig:decoding_cossim}
    \end{subfigure}\hfill
    
    \vspace{1em}
    
    \begin{subfigure}[b]{0.53\textwidth}
        \centering
        \includegraphics[width=1.\textwidth]{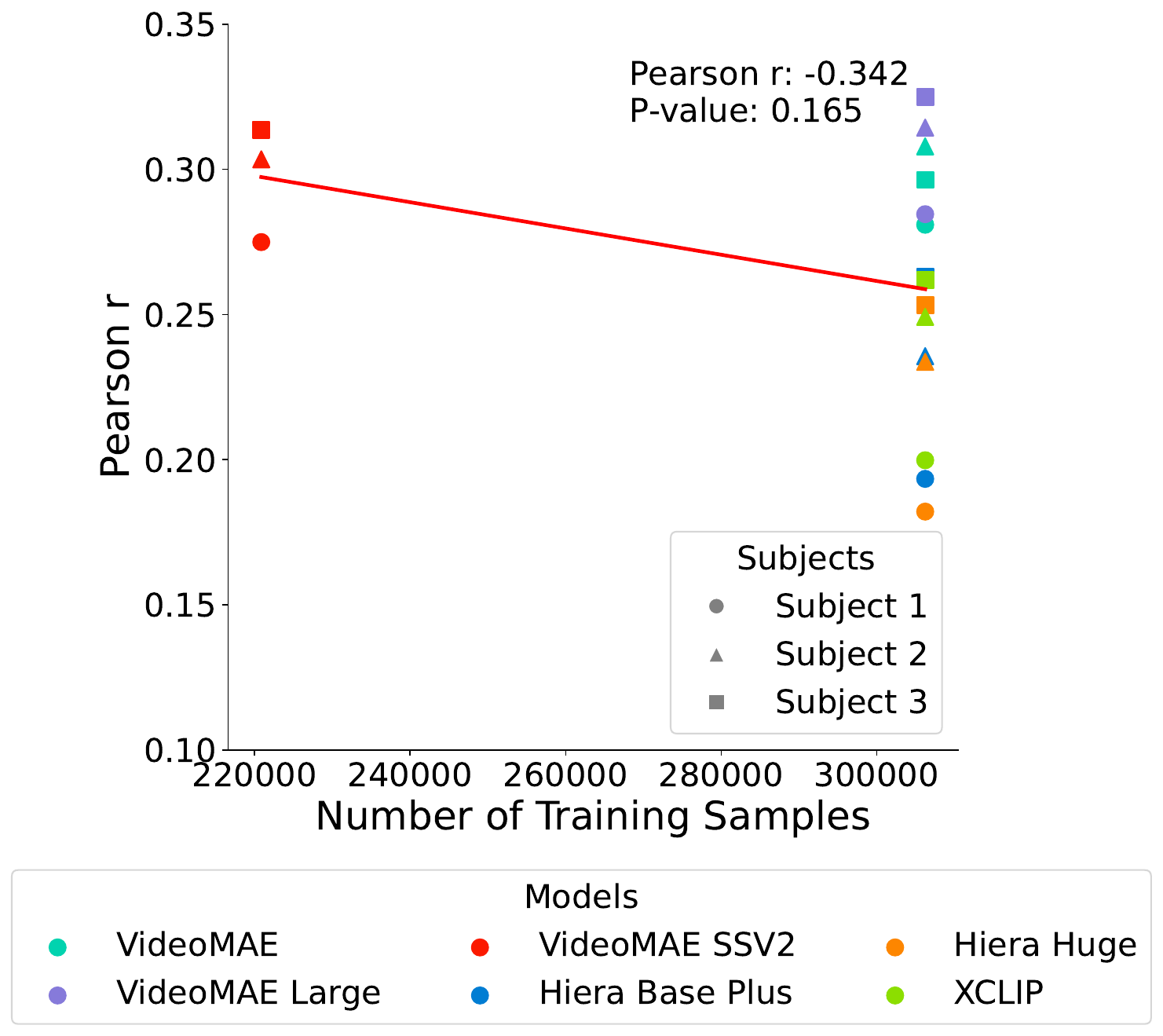}
        \caption{\textbf{Video Models} }
        \label{fig:decoding_cossim}
    \end{subfigure}\hfill
    \begin{subfigure}[b]{0.45\textwidth}
        \centering
        \includegraphics[width=0.9\textwidth]{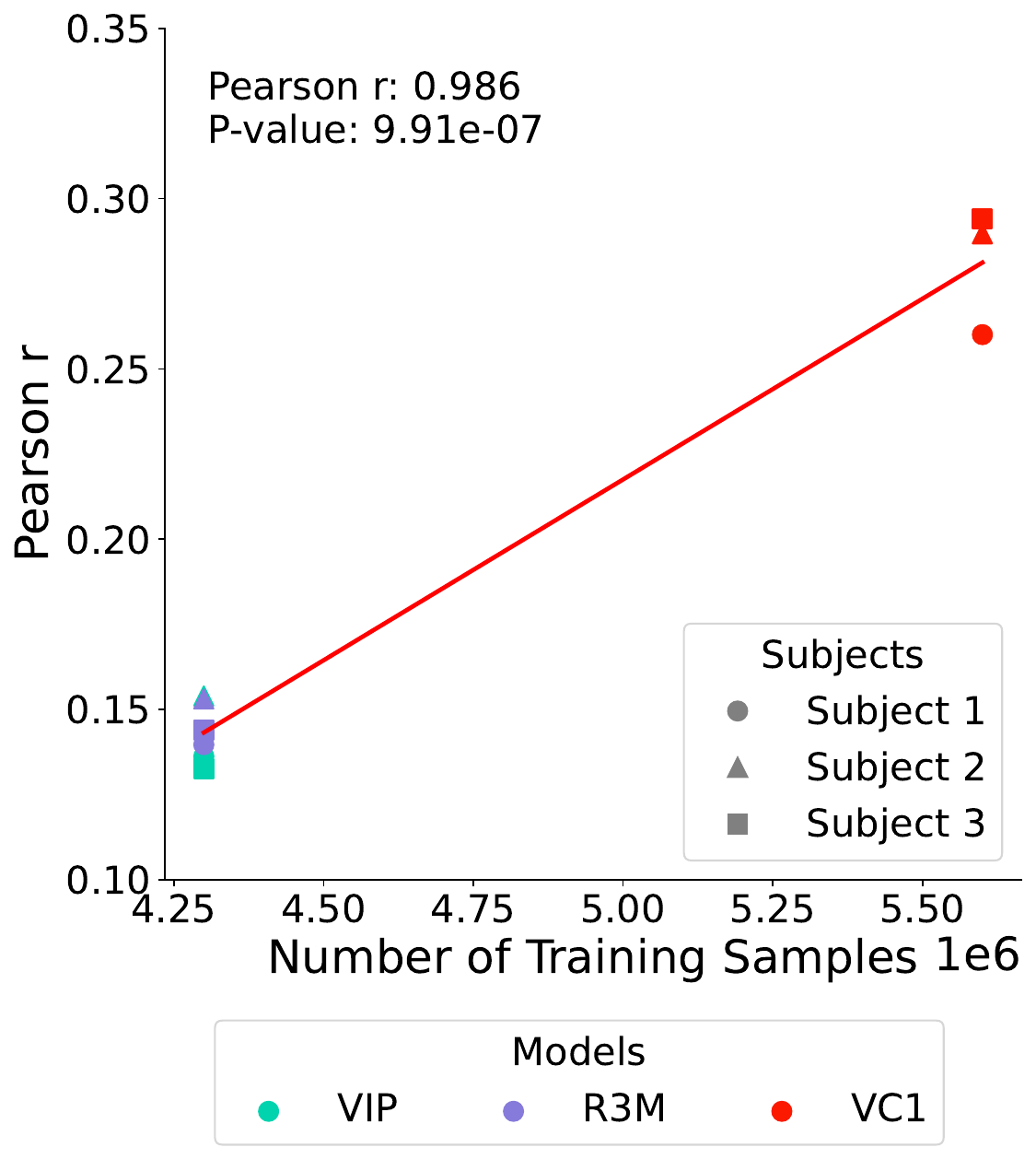}
        \caption{\textbf{Embodied AI Models}}
        \label{fig:decoding_epe}
    \end{subfigure}
    \caption{\textbf{Training data size and encoding prediction performance.} Encoding model features of the viewed stimuli are used to predict voxel-wise fMRI brain activity~\cite{naselaris2011encoding}. The average Pearson-$r$ is plotted across all voxels for each model and with respect to the number of samples each model is trained on. \textbf{(a)} Encoding performance for models trained on static images.  \textbf{(b)} Encoding performance for models trained on videos. \textbf{(c)} Encoding performance for models trained to align representations of single frames across time for embodied AI visuomotor manipulation. 
    Training data size and encoding prediction performance are not significantly correlated (by statistical test) for image and video models. This indicates that the training data size size is not a confound for the encoding performance of image and video models. In contrast, model size and encoding performance are significantly correlated ($p < 0.001$) for the embodied AI visual models. Encoding performance per model is plotted for each of the three participants -- the circle marker refers to S1, the triangle marker refers to S2, and the square marker refers to S3. %
    }
    \label{fig:supplemental-train-size-control}
\end{figure}

\clearpage

\begin{table}[h]
\centering
\begin{minipage}{0.45\textwidth}
\centering
\begin{tabular}{|l|l|c|}
\hline
\multicolumn{3}{|c|}{\textbf{VideoMAE Models (Pearson $r$)}} \\ \hline
\textbf{Model} & \textbf{Dataset} & \textbf{\small Encoding Performance} \\ \hline
VideoMAE Base & K-400 & 0.2964 \\ \hline
VideoMAE Large & K-400 & 0.3248 \\ \hline
VideoMAE Base & SSV2 & 0.3137 \\ \hline
\end{tabular}
\caption{\textbf{Control for dataset distribution: encoding performance of VideoMAE.} VideoMAE trained on SSV2 predicts fMRI brain activity at the same level as VideoMAE trained on Kinetics-400. All three VideoMAE models are better at predicting fMRI brain activity as compared to all other tested models as listed in Figure~7 in the main text. Combined with the results in Figure~\ref{fig:supplemental-param-control}, these results indicate that model size and training data size are unlikely to be confounding factors. More generally, these results suggest that the architecture and training paradigm used in VideoMAE leads to better fMRI brain activity prediction.}
\label{tab:videomae_performance}
\end{minipage}
\hspace{0.05\textwidth}
\begin{minipage}{0.45\textwidth}
\vspace{-21.5pt}
\centering
\begin{tabular}{|l|l|c|}
\hline
\multicolumn{3}{|c|}{\textbf{Hiera Models (Pearson $r$)}} \\ \hline
\textbf{Model} & \textbf{Dataset} & \textbf{\small Encoding Performance} \\ \hline
Hiera Base Plus & K-400 & 0.2629 \\ \hline
Hiera Base Plus & IN-1K & 0.2198 \\ \hline
Hiera Huge & K-400 & 0.2532 \\ \hline
Hiera Huge & IN-1K & 0.2407 \\ \hline
\end{tabular}
\caption{\textbf{Control for dataset distribution: encoding performance of Hiera.} Hiera models of the same size trained on Kinetics-400 (K-400) are better at predicting fMRI brain activity as compared to Hiera models trained on Imagenet-1K (IN-1K). These results indicate that fMRI brain activations encode dynamic visual information, as modeling temporal dynamics improves fMRI brain activity prediction when architecture is held constant.}
\label{tab:hiera_performance}
\end{minipage}
\end{table}

\clearpage
\subsection{Encoding Models: Voxel-wise Prediction Performance on Inflated Cortical Maps of the Brain}
\label{section: supplemental encoding models}
Inflated brain maps showing voxel-wise fMRI prediction performance, quantified as the Pearson correlation ($r$) between measured and predicted responses, for all visual encoding models not shown in the main text. Maps for each model are shown in alphabetical order. `AF' in the model name denotes a model using \textit{A}verage \textit{F}rames. All maps are shown using the same scale for ease of comparison between models.

\begin{figure}[h]
    \centering
    \includegraphics[width=0.9\textwidth]{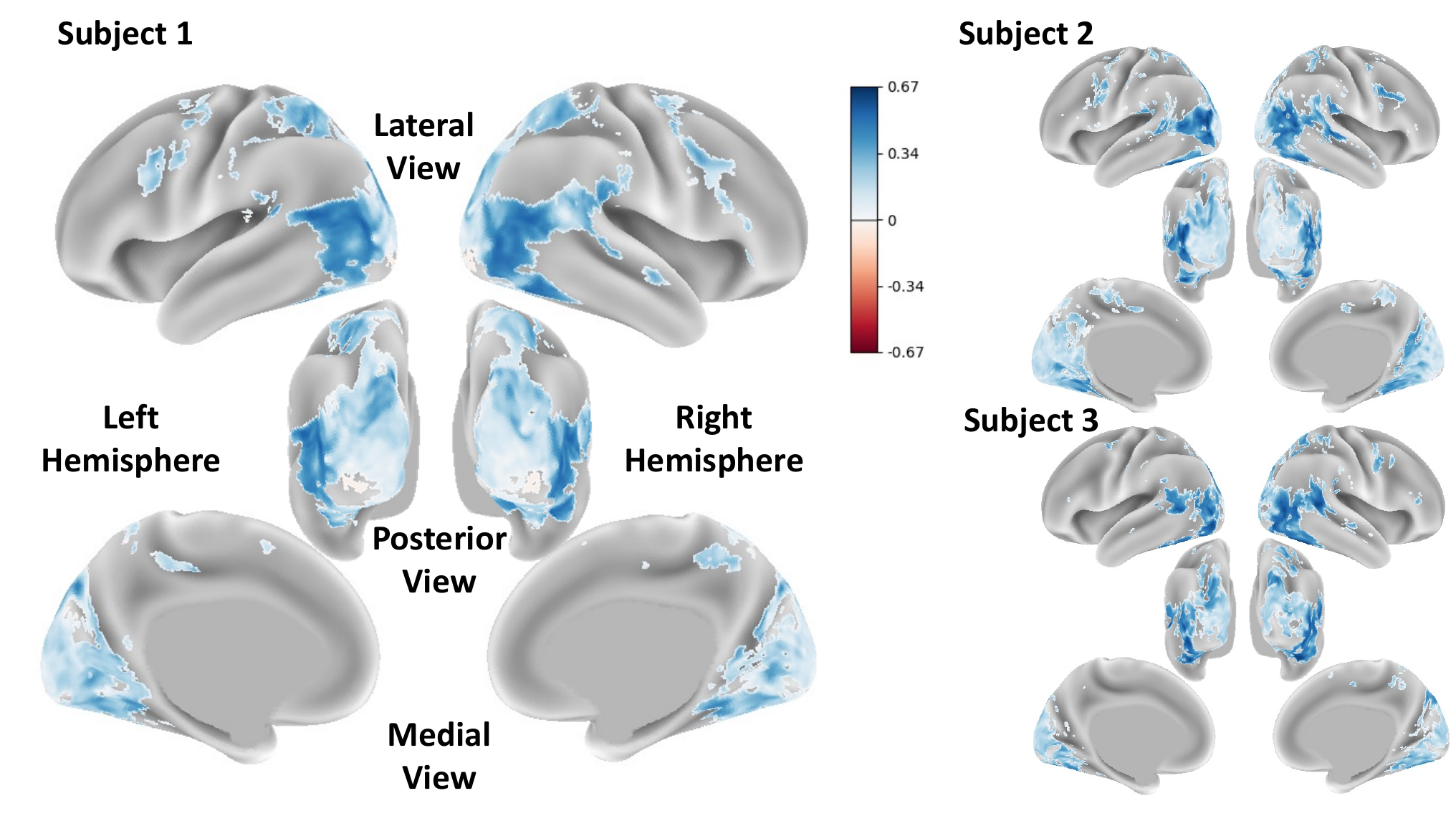}
    \caption{
    \brainmaptext{CLIP}
    }
    \label{fig:encoding brain maps clip}
\end{figure}

\begin{figure}[]
    \centering
    \includegraphics[width=0.9\textwidth]{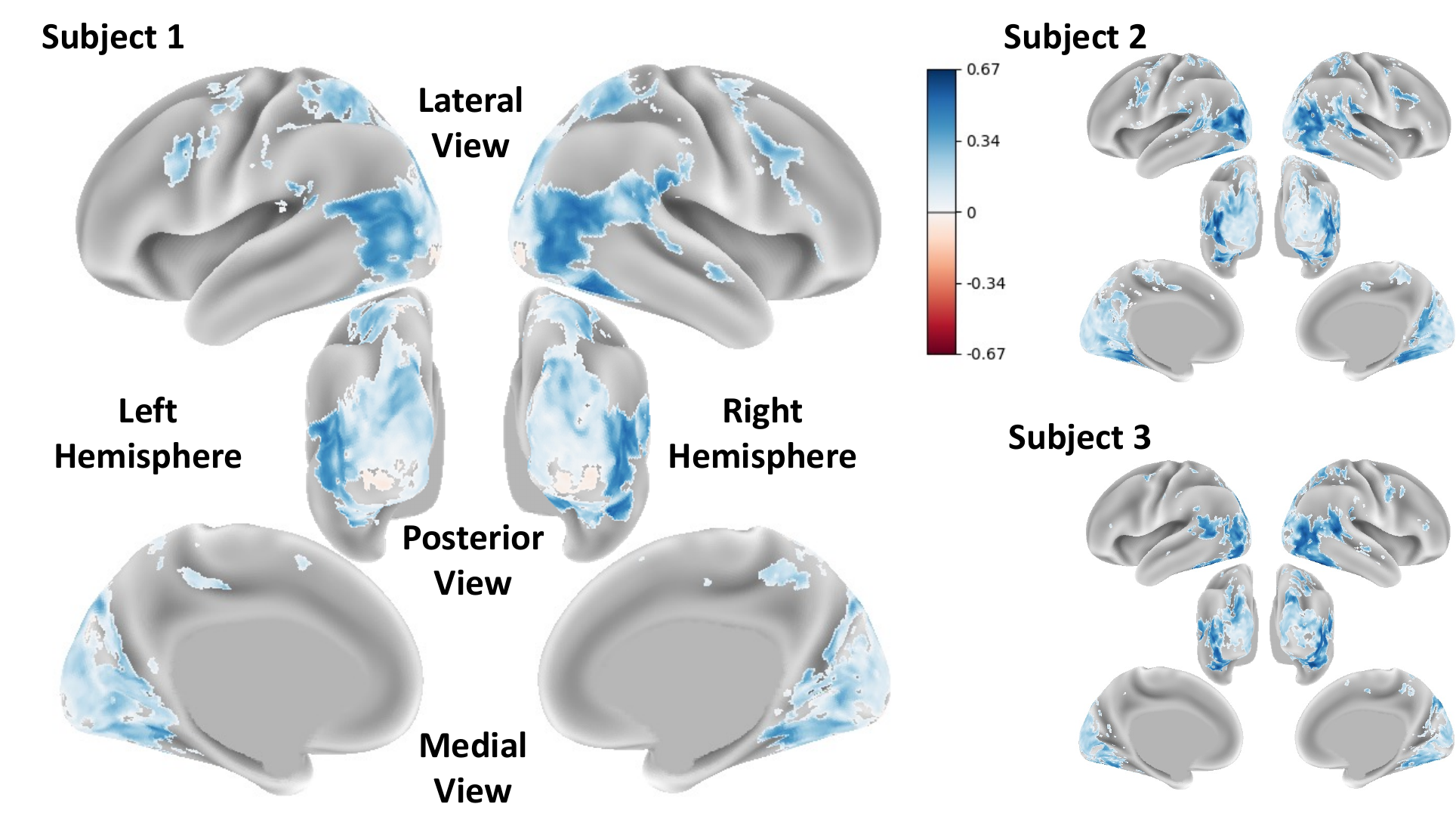}
    \caption{%
        \brainmaptext{CLIP AF}
        }
    \label{fig:encoding brain maps clip af}
\end{figure}

\begin{figure}[]
    \centering
    \includegraphics[width=0.9\textwidth]{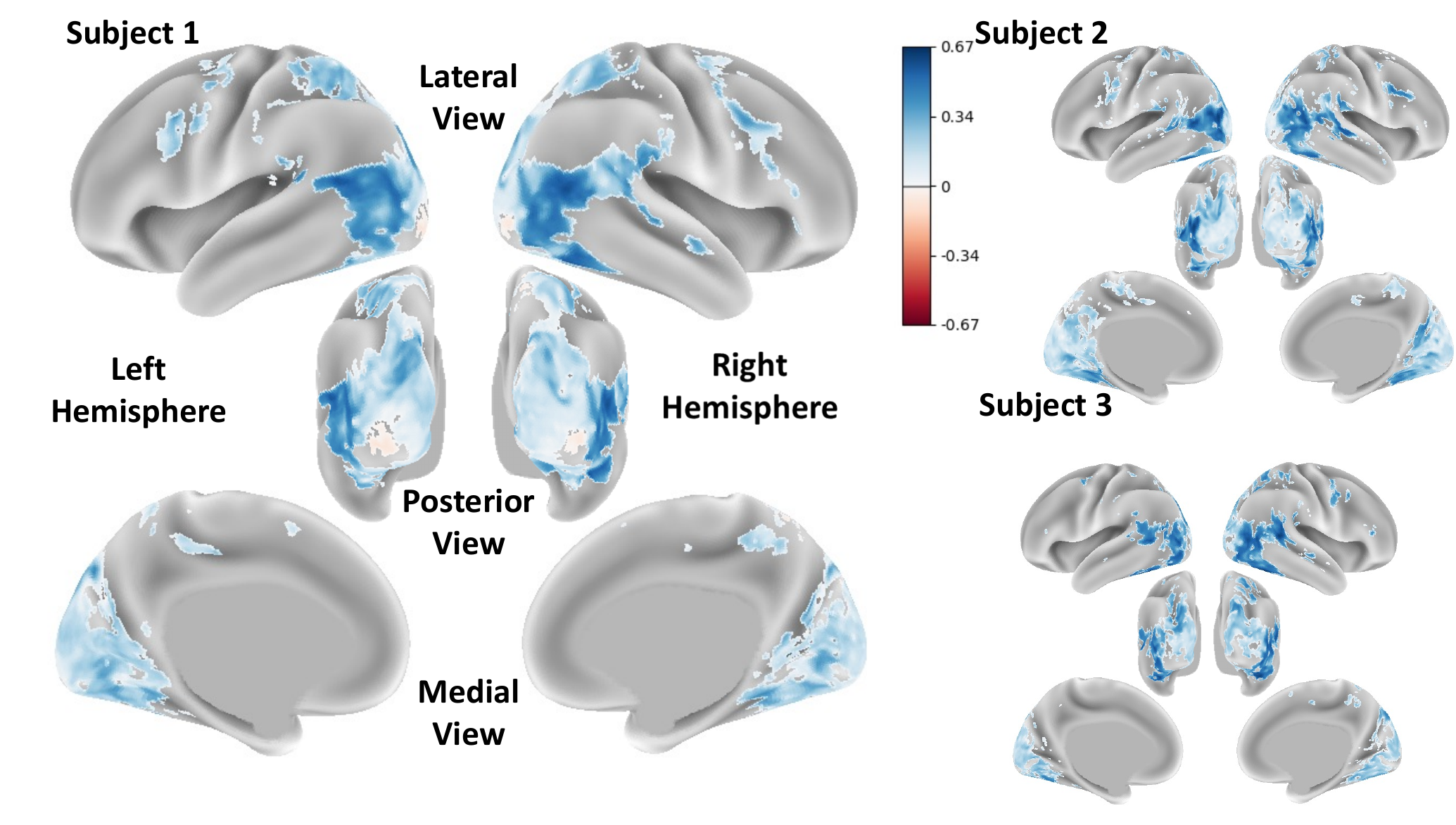}
    \caption{%
    \brainmaptext{CLIP ConvNeXt}
    }
    \label{fig:encoding brain maps clip convnext}
\end{figure}

\begin{figure}[]
    \centering
    \includegraphics[width=0.9\textwidth]{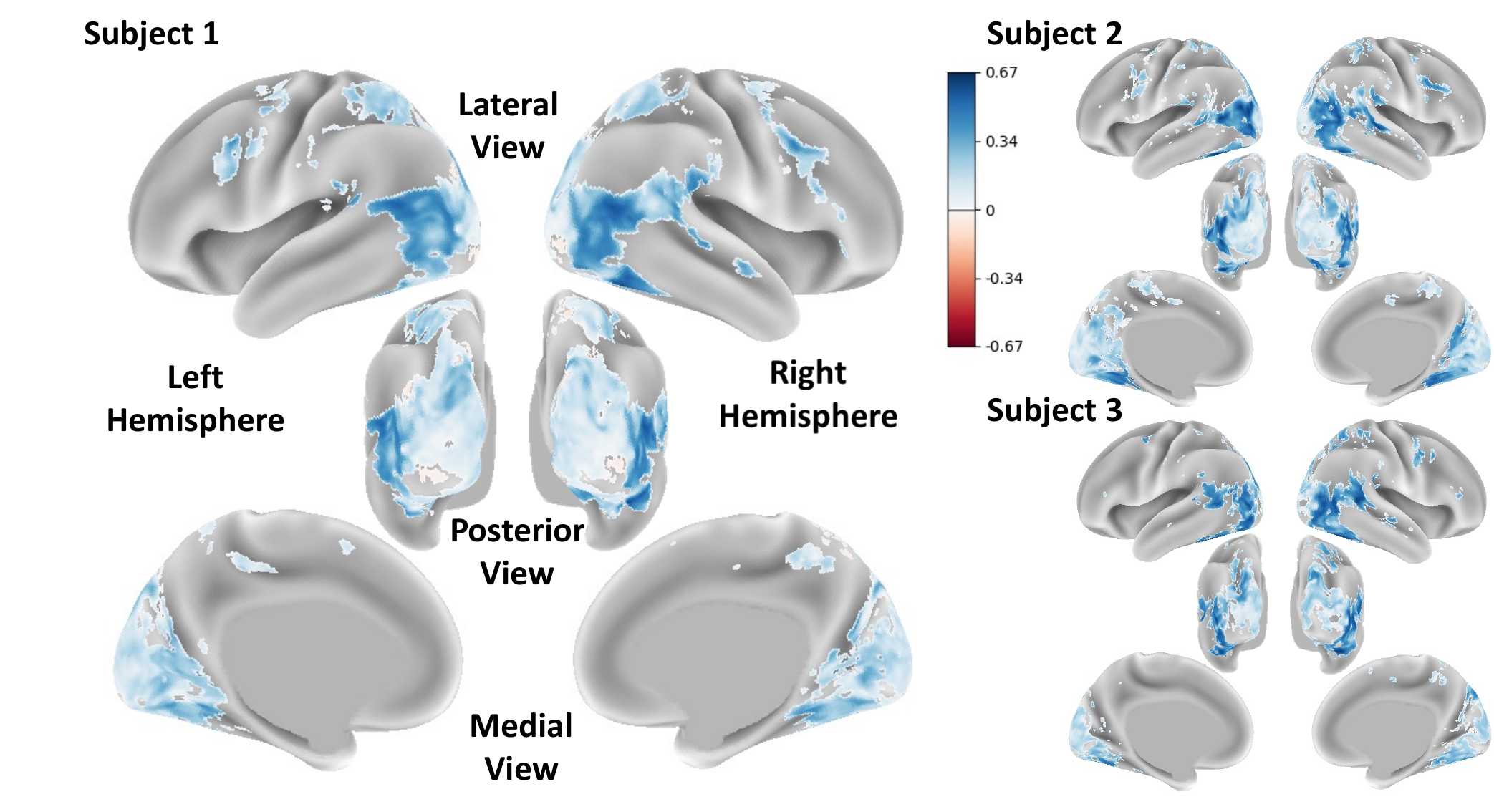}
    \caption{%
    \brainmaptext{CLIP ConvNeXt AF}
    }
    \label{fig:encoding brain maps clip convnext af}
\end{figure}

\begin{figure}[]
    \centering
    \includegraphics[width=0.9\textwidth]{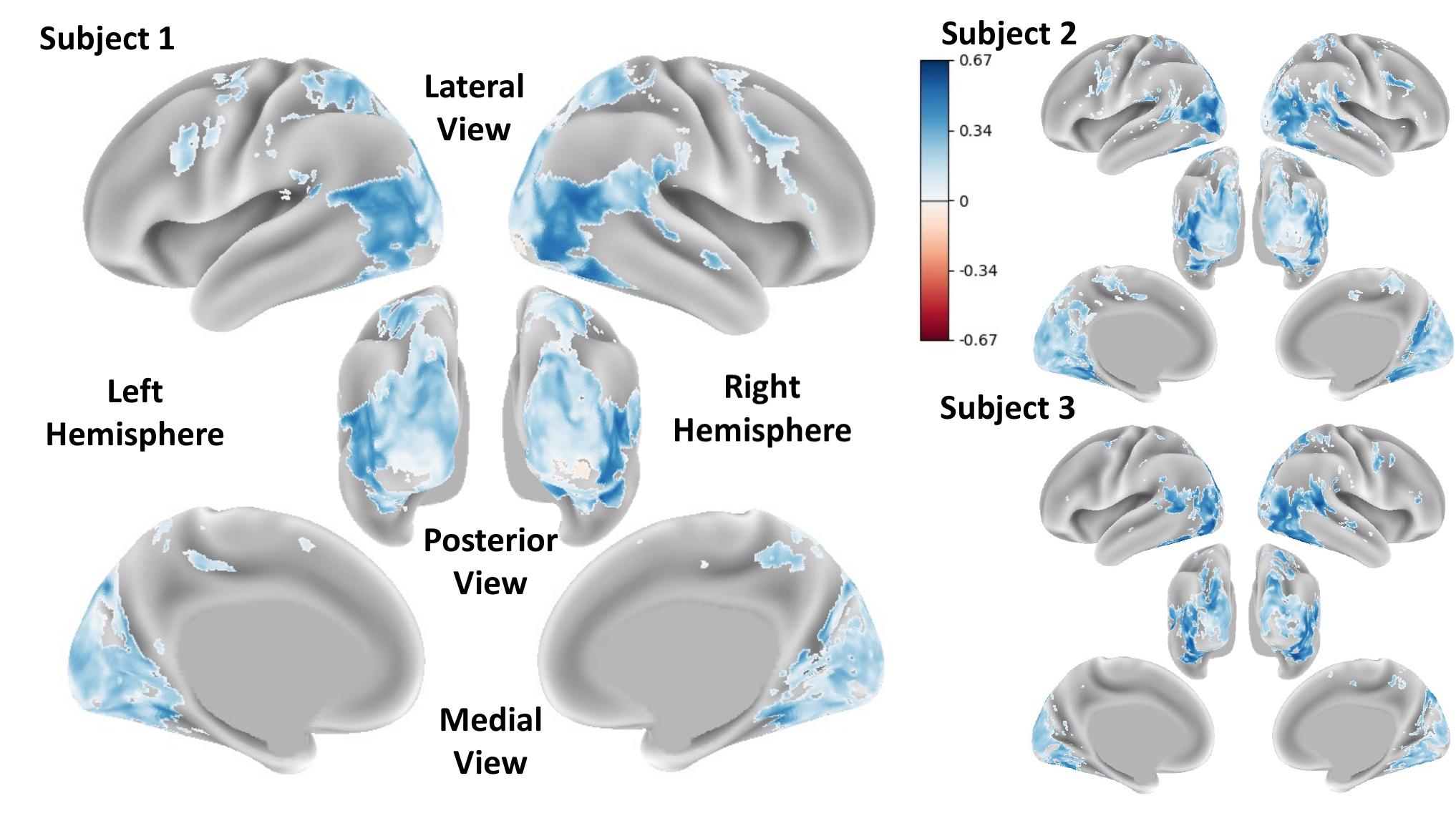}
    \caption{%
    \brainmaptext{DINOv1}
    }
    \label{fig:encoding brain maps dinov1}
\end{figure}

\begin{figure}[]
    \centering
    \includegraphics[width=0.9\textwidth]{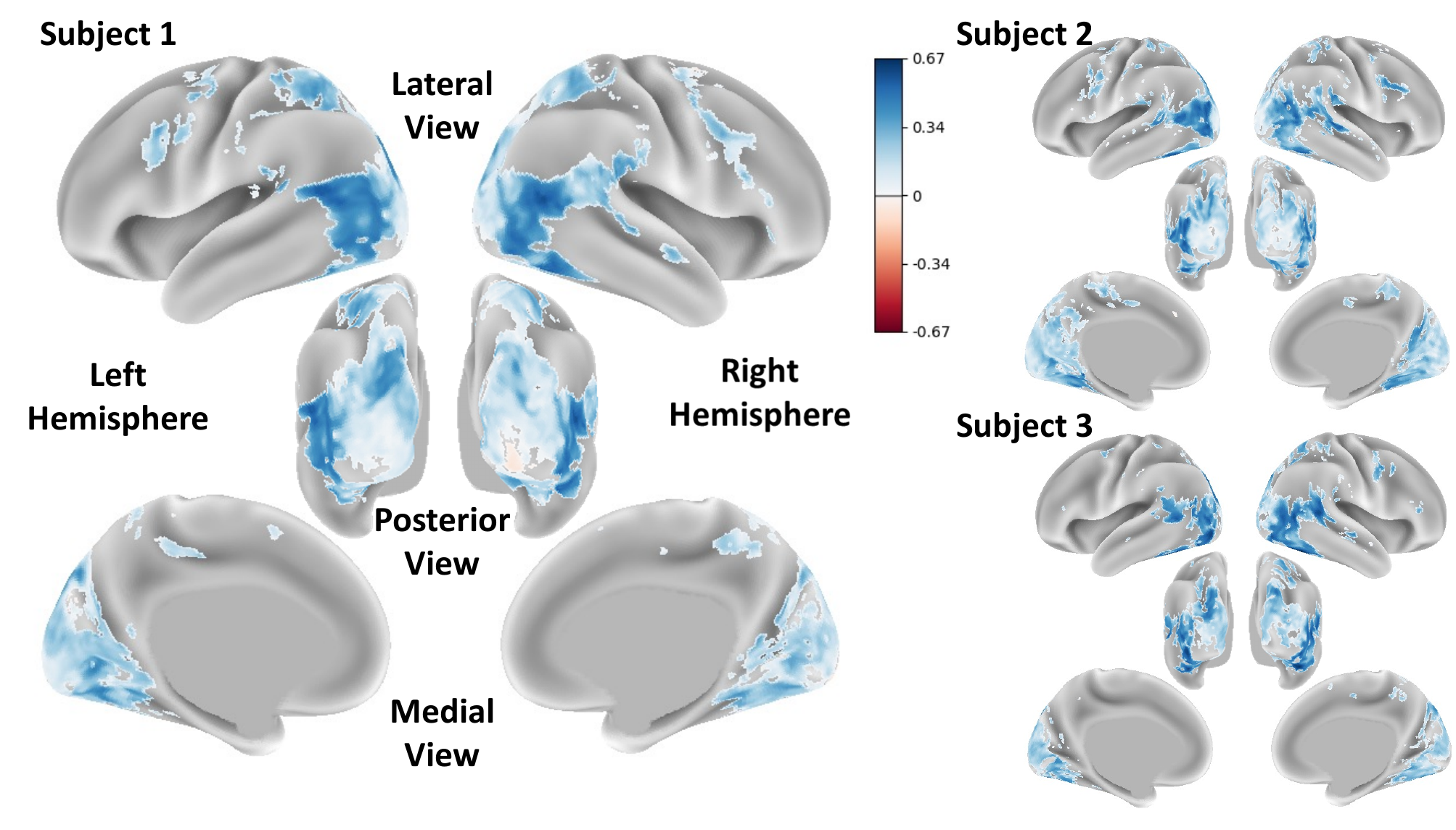}
    \caption{%
    \brainmaptext{DINOv2}
    }
    \label{fig:encoding brain maps dinov2}
\end{figure}

\begin{figure}[]
    \centering
    \includegraphics[width=0.9\textwidth]{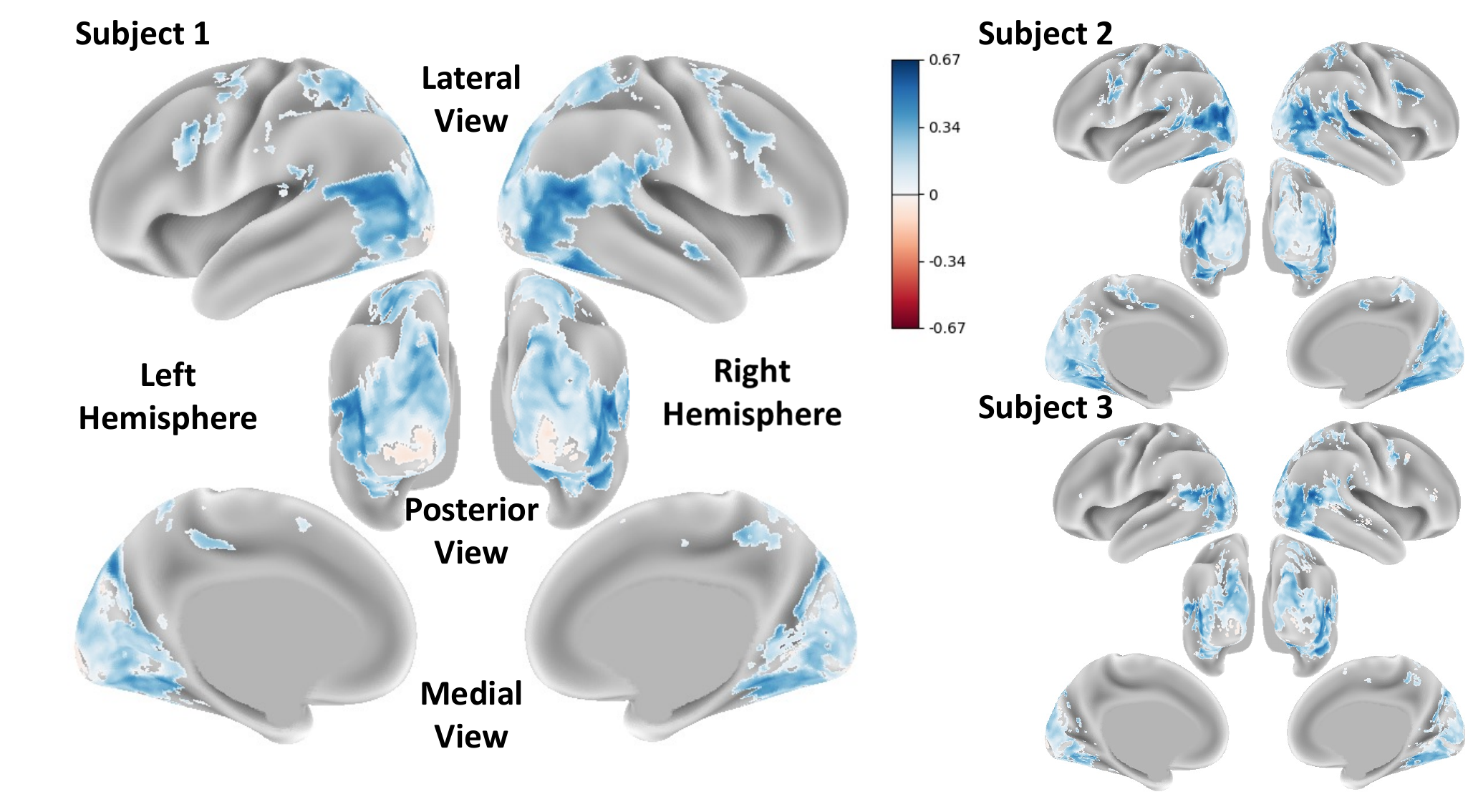}
    \caption{%
    \brainmaptext{Hiera Base Plus}
    }
    \label{fig:encoding brain maps hiera base plus}
\end{figure}

\begin{figure}[]
    \centering
    \includegraphics[width=0.9\textwidth]{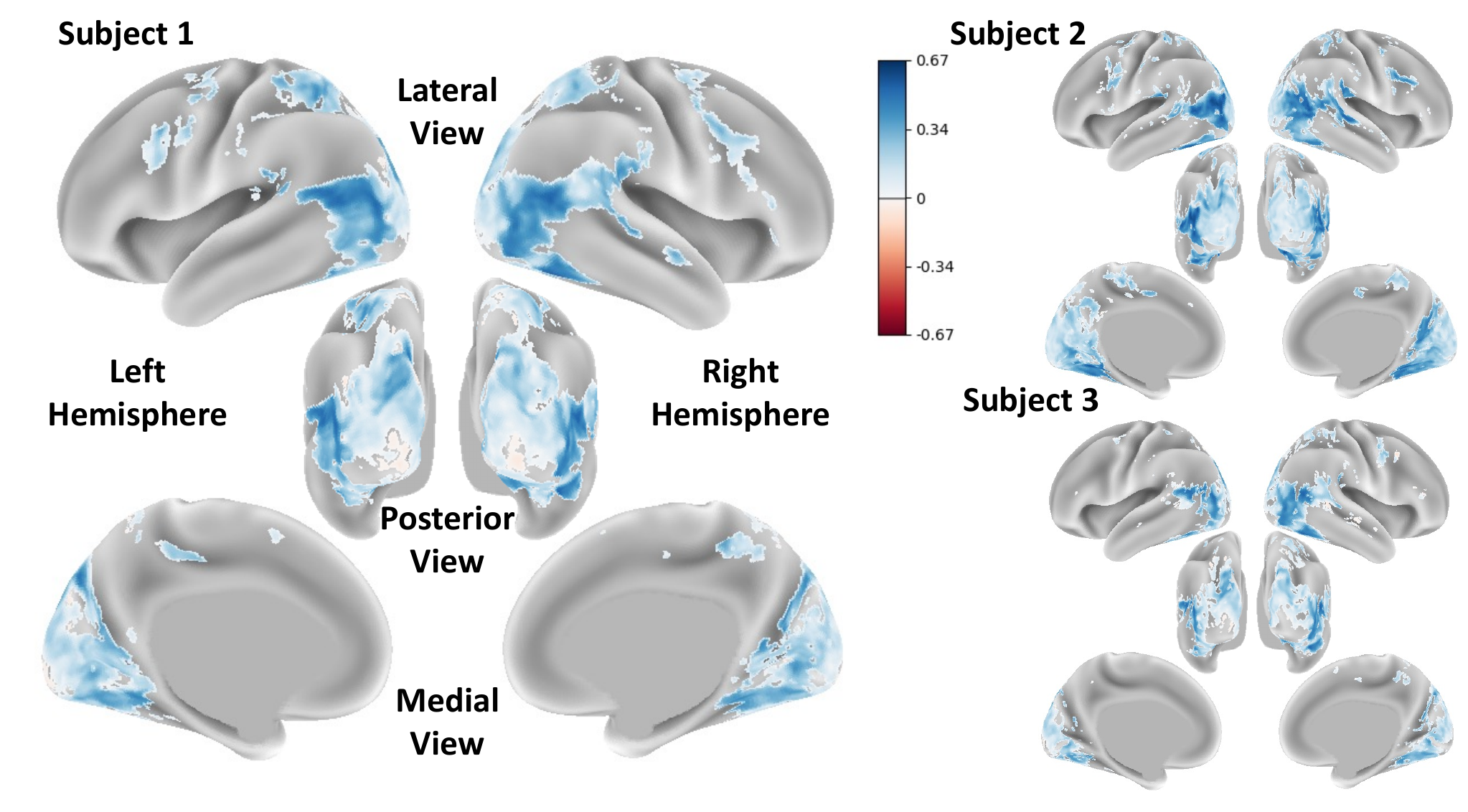}
    \caption{%
    \brainmaptext{Hiera Huge}
    }
    \label{fig:encoding brain maps hiera huge}
\end{figure}

\begin{figure}[]
    \centering
    \includegraphics[width=0.9\textwidth]{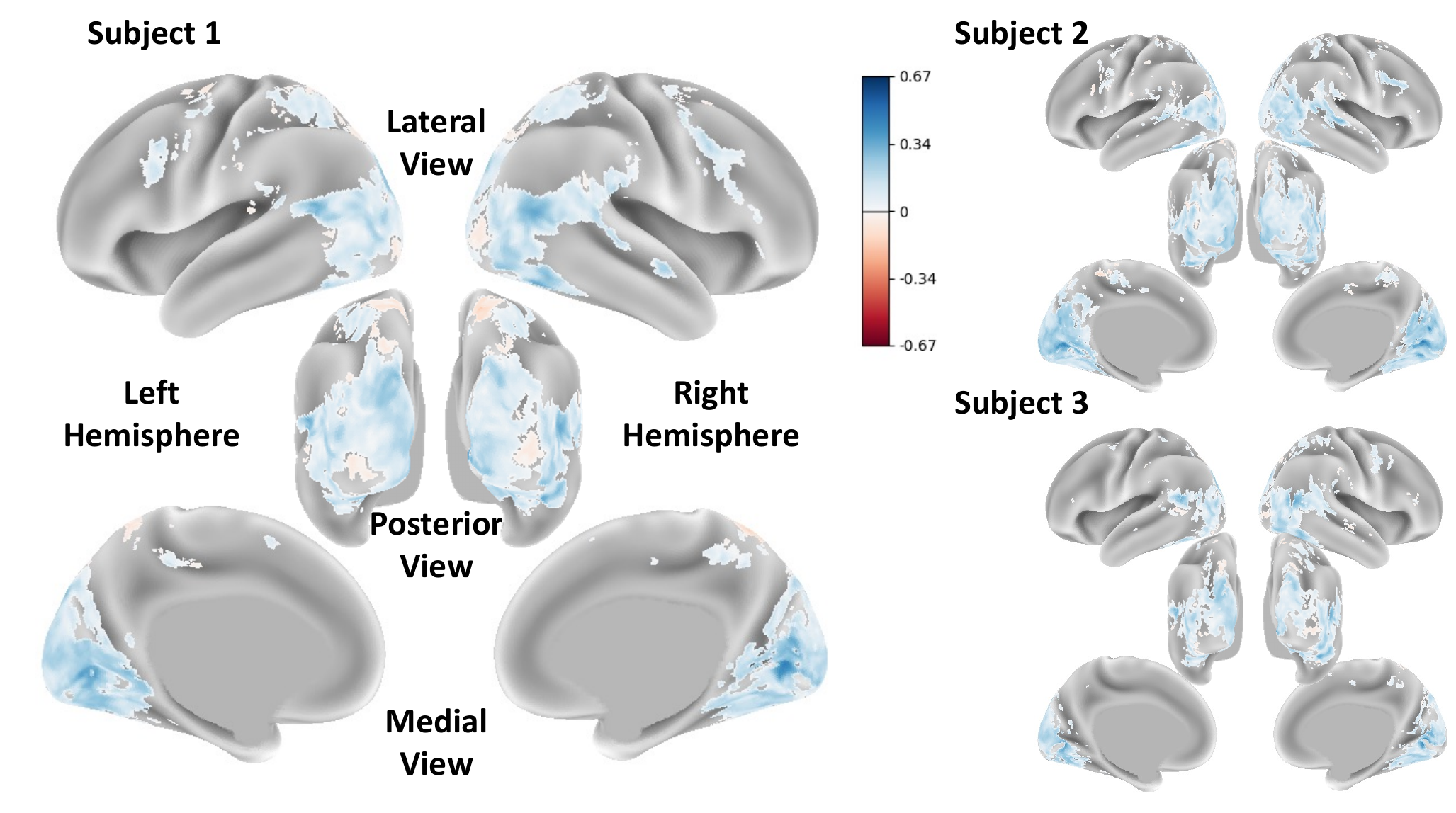}
    \caption{%
    \brainmaptext{R3M}
    }
    \label{fig:encoding brain maps r3m}
\end{figure}

\begin{figure}[]
    \centering
    \includegraphics[width=0.9\textwidth]{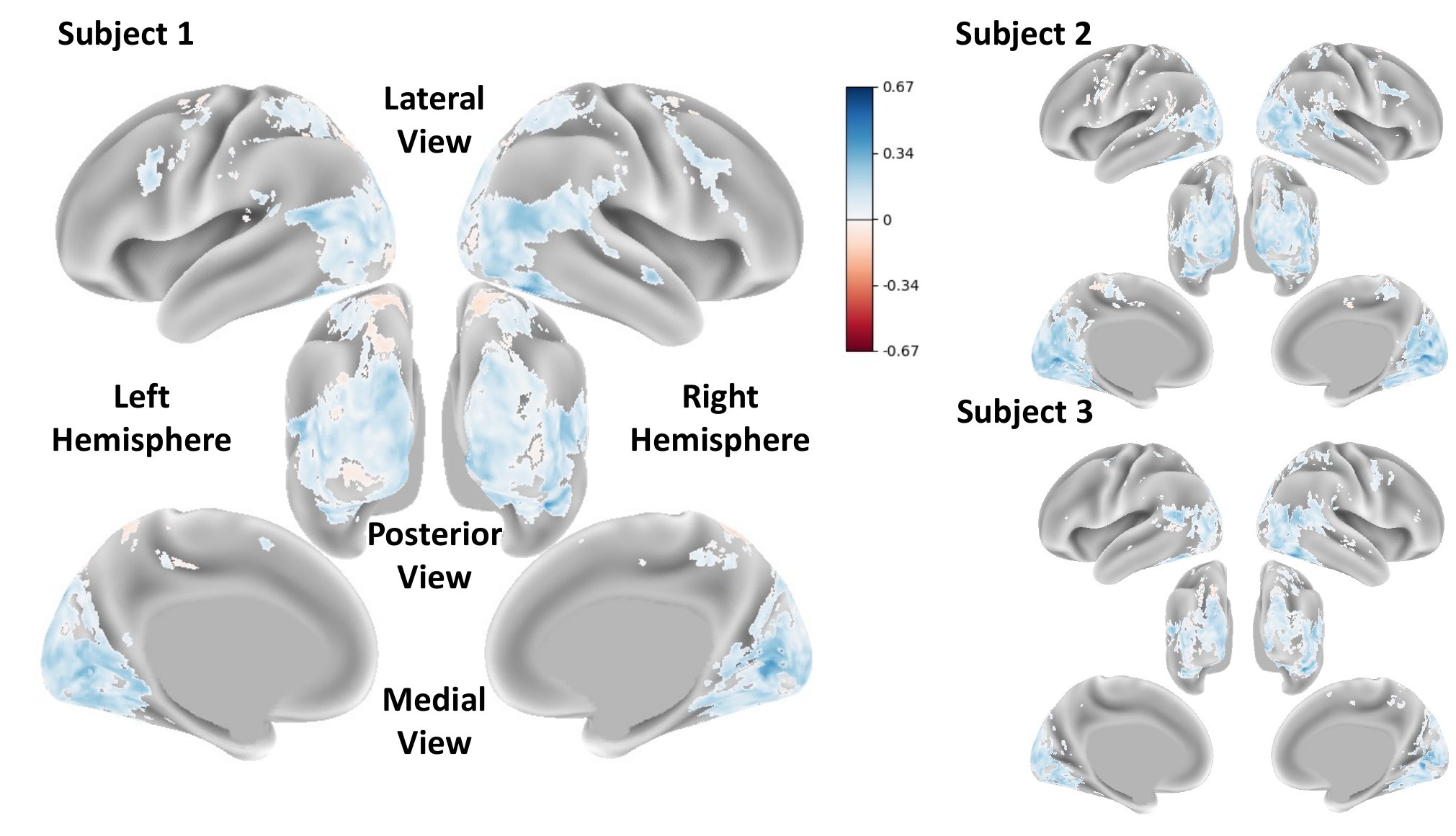}
    \caption{%
    \brainmaptext{R3M AF}
    }
    \label{fig:encoding brain maps r3m_af}
\end{figure}

\begin{figure}[]
    \centering
    \includegraphics[width=0.9\textwidth]{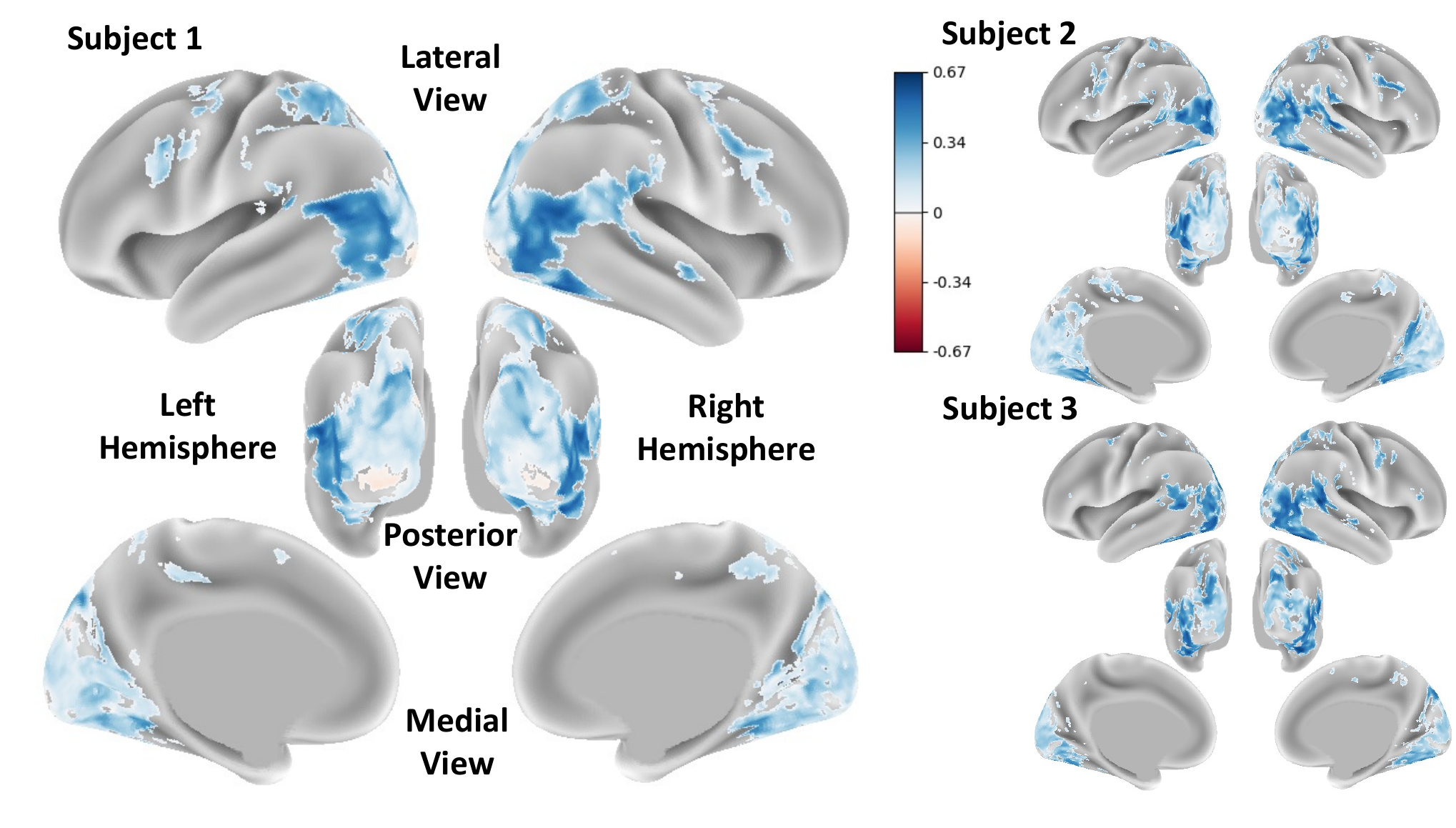}
    \caption{%
    \brainmaptext{ResNet50}
    }
    \label{fig:encoding brain maps resnet50}
\end{figure}

\begin{figure}[]
    \centering
    \includegraphics[width=0.9\textwidth]{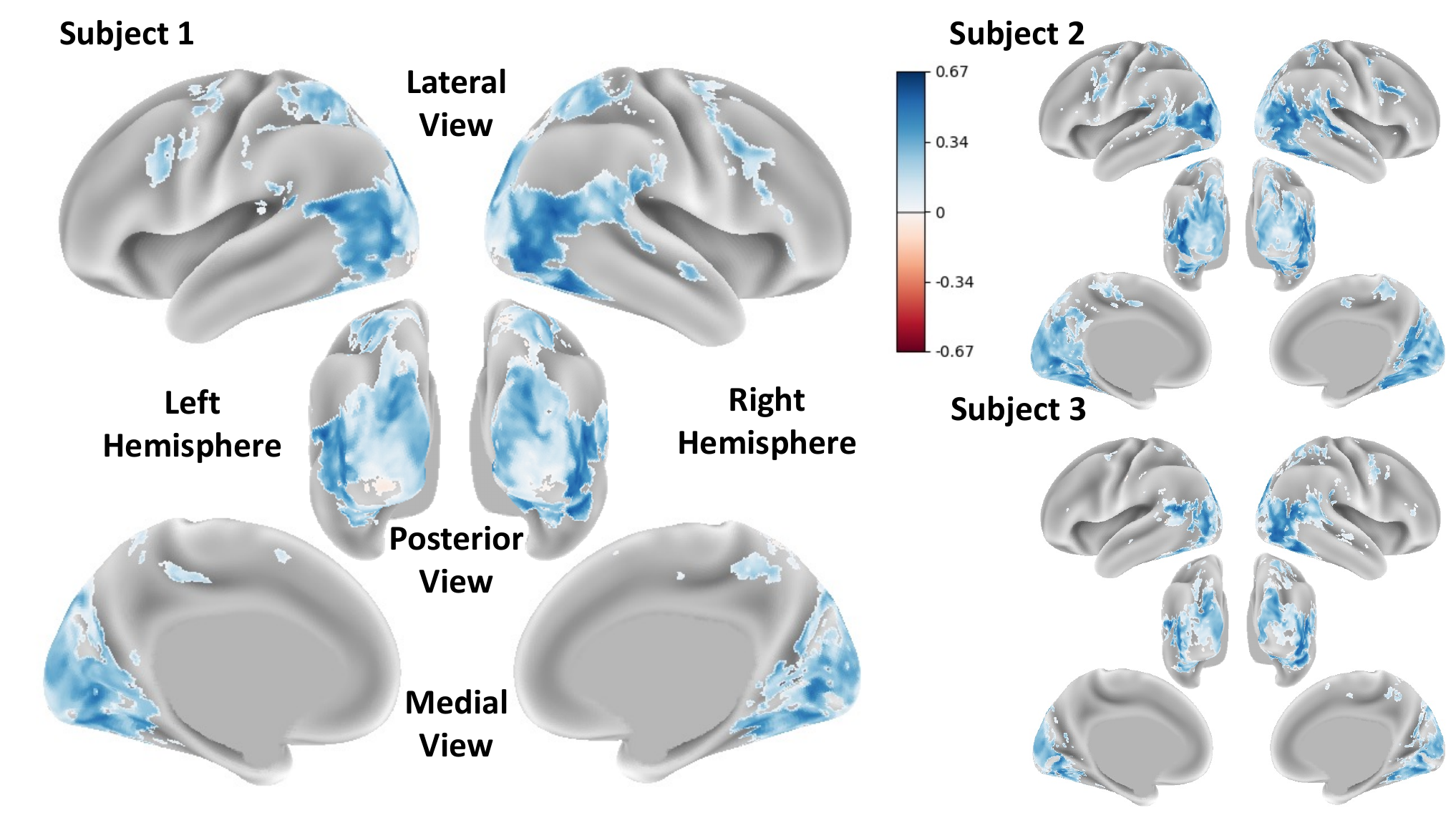}
    \caption{%
    \brainmaptext{VC1}
    }
    \label{fig:encoding brain maps vc1}
\end{figure}

\begin{figure}[]
    \centering
    \includegraphics[width=0.9\textwidth]{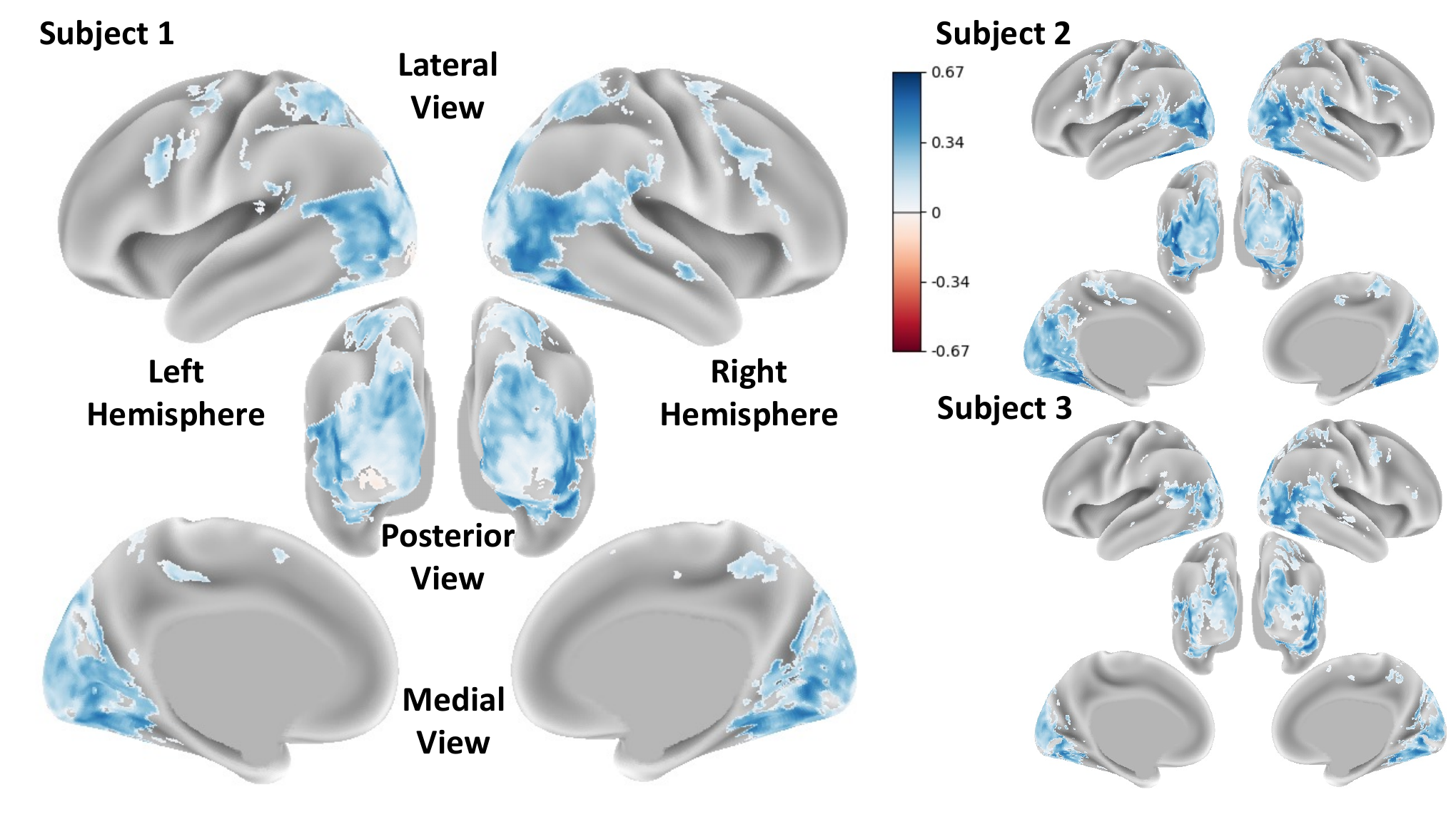}
    \caption{%
    \brainmaptext{VC1 AF}
    }
    \label{fig:encoding brain maps vc1_af}
\end{figure}

\begin{figure}[]
    \centering
    \includegraphics[width=0.9\textwidth]{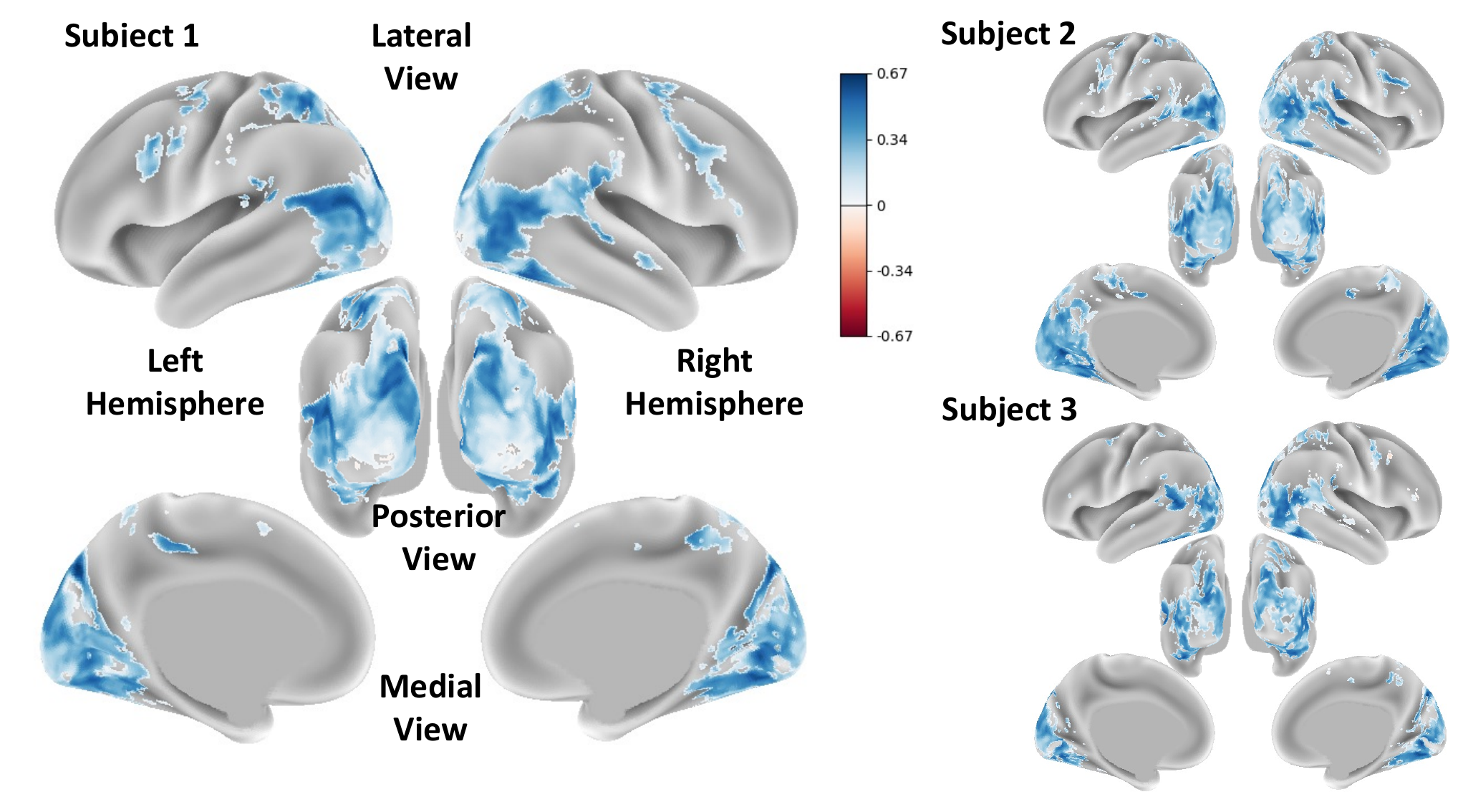}
    \caption{%
    \brainmaptext{VideoMAE}
    }
    \label{fig:encoding brain maps videomae}
\end{figure}

\begin{figure}[]
    \centering
    \includegraphics[width=0.9\textwidth]{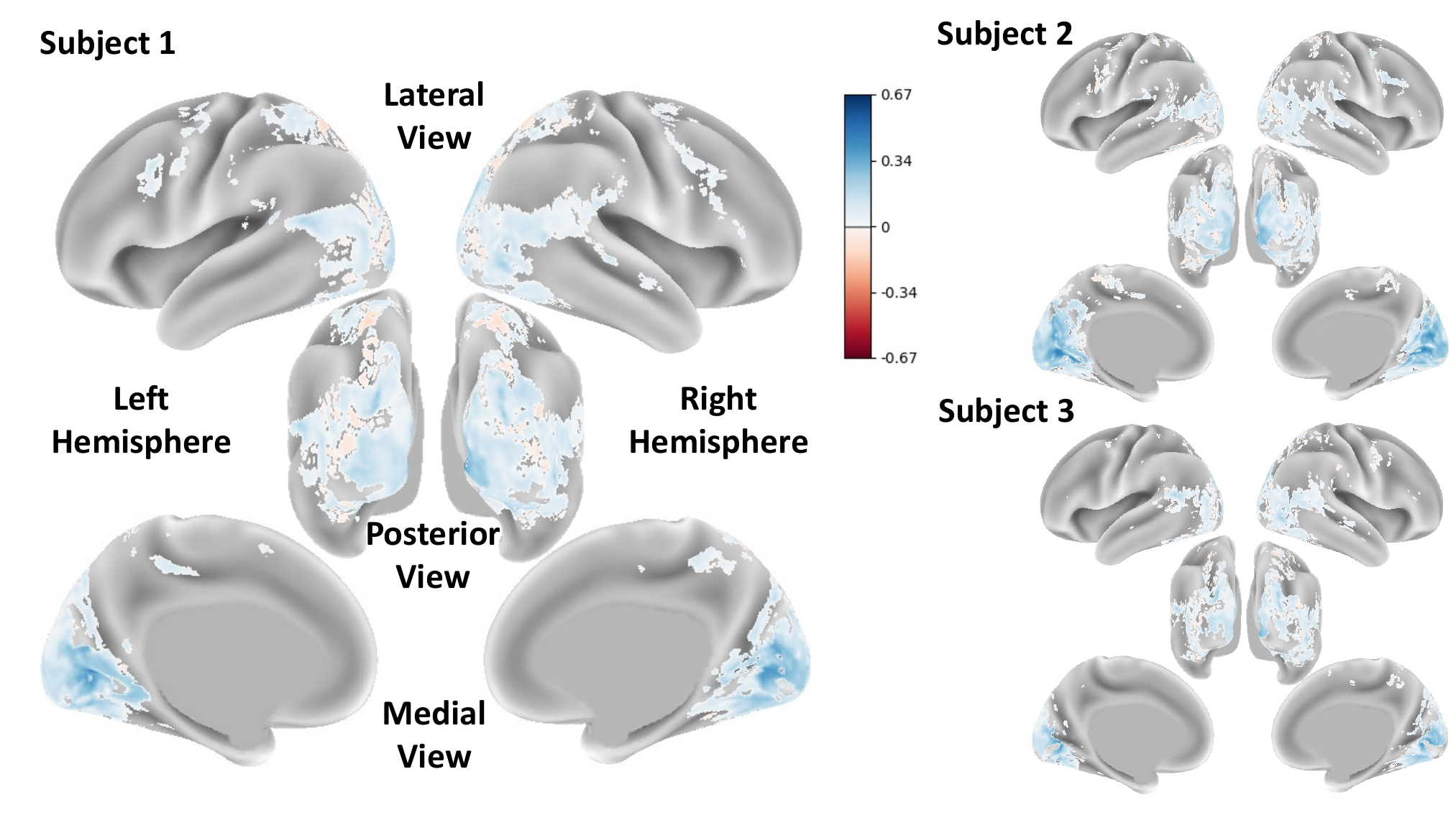}
    \caption{%
    \brainmaptext{VIP}
    }
    \label{fig:encoding brain maps vip}
\end{figure}

\begin{figure}[]
    \centering
    \includegraphics[width=0.9\textwidth]{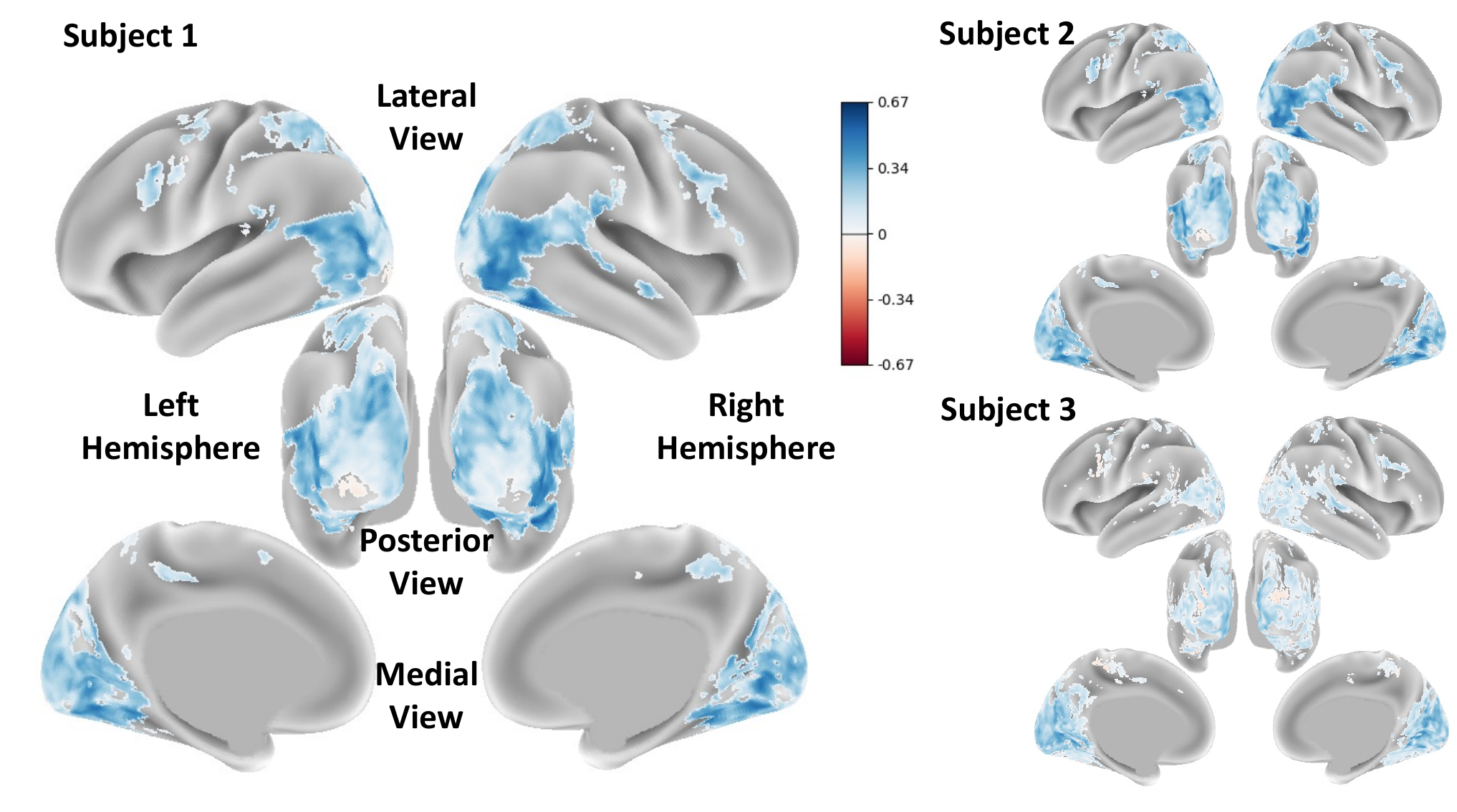}
    \caption{%
    \brainmaptext{VIP AF}
    }
    \label{fig:encoding brain maps vip_af}
\end{figure}

\begin{figure}[]
    \centering
    \includegraphics[width=0.9\textwidth]{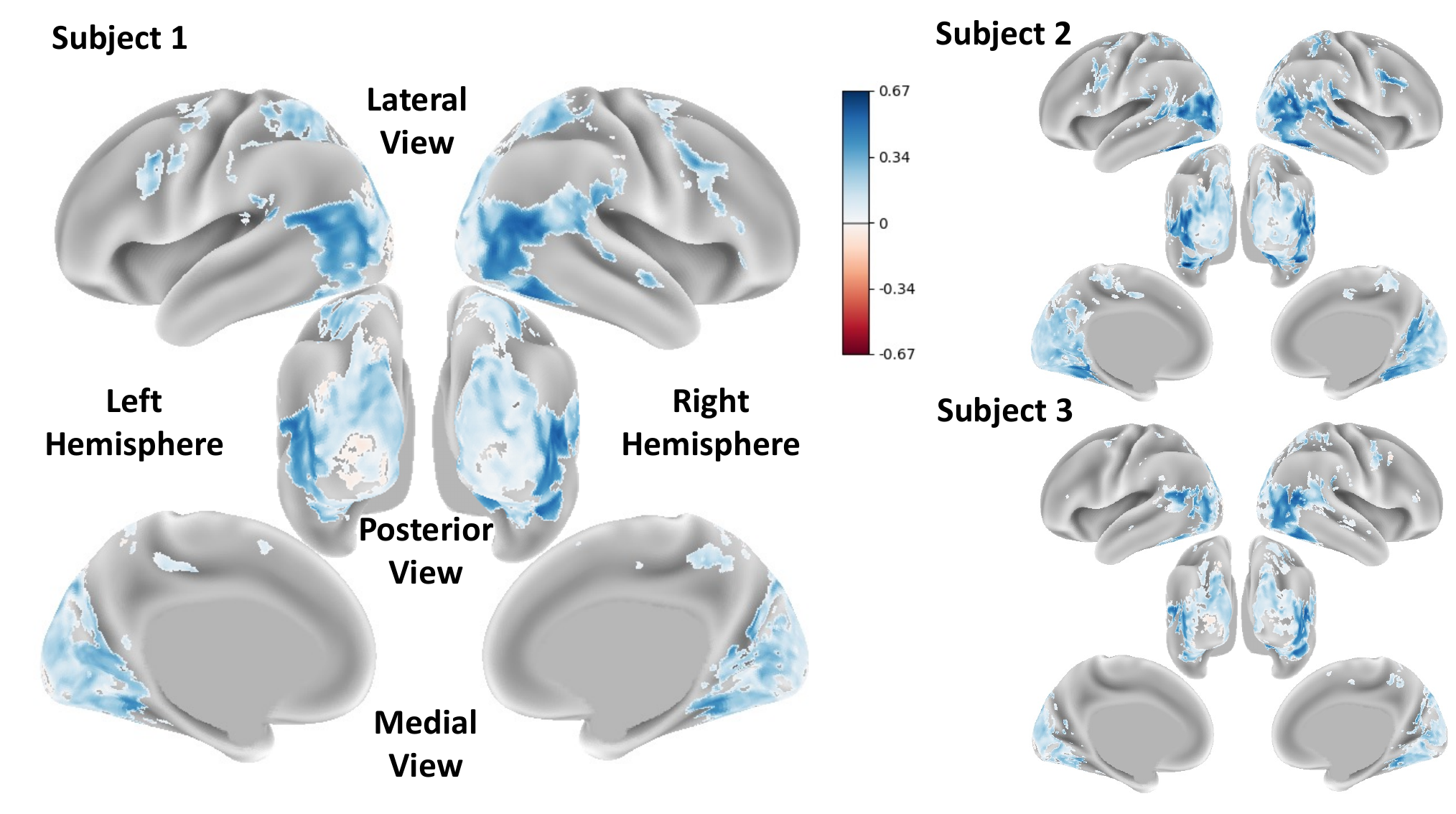}
    \caption{%
    \brainmaptext{XCLIP}
    }
    \label{fig:encoding brain maps xclip}
\end{figure}

\end{document}